\documentclass[11pt,a4paper]{article}
\usepackage[utf8]{inputenc}
\usepackage{a4wide}
\usepackage{amsmath,amssymb}
\usepackage{amsfonts}
\usepackage{cite}
\usepackage{color}
\usepackage{adjustbox}
\usepackage{multirow}
\usepackage{pdflscape}
\usepackage{textcomp}
\usepackage{gensymb}
\usepackage{soul}
\usepackage{graphicx}
\graphicspath{{figs/}}
\usepackage{diagbox}
\usepackage{pifont}
\usepackage{bigstrut}
\usepackage[small,bf]{caption}
\usepackage{tabulary}
\usepackage{booktabs}
\usepackage{stackengine}
\usepackage{mathtools}
\usepackage{enumerate}
\usepackage{makecell}
\usepackage{listings}
\usepackage{bbm}
\makeatletter
  
\newcommand{\Rmnum}[1]{\expandafter\@slowromancap\romannumeral #1@} 
\usepackage[unicode]{hyperref} 

\definecolor{greenLinks}{rgb}{0, 0.6, 0} 
\definecolor{blueLinks}{rgb}{0, 0, 0.6}
\definecolor{redLinks}{rgb}{0.6, 0, 0}
\definecolor{eprintLinks}{rgb}{0.4, 0.4, 0.4}
\definecolor{journalLinks}{rgb}{0.6, 0, 0}
\begin{document}

\pagenumbering{Alph}
\begin{titlepage}

\vspace*{15mm}

\begin{center}
{ \bf\LARGE {Neutrino mass and leptogenesis in the non‑SUSY \\[3mm] modular $A^{\prime}_5$ inverse seesaw model}}\\[8mm]
Xianshuo Zhang$^{\,a,}$\footnote{Email: \href{mailto:zxianshuo@yeah.net}{\texttt{zxianshuo@yeah.net}}},
Yakefu Reyimuaji$^{\,a,}$\footnote{Email: \href{mailto:yreyi@hotmail.com}{\texttt{yreyi@hotmail.com} }} \\
\vspace{8mm}
$^{a}$\,{\it School of Physical Science and Technology, Xinjiang University, Urumqi, Xinjiang 830017, China} \\

\end{center}
\vspace{8mm}

\begin{abstract}
\noindent A non-supersymmetric inverse seesaw model of neutrino mass based on the $A^{\prime}_5$ modular symmetry is presented. This framework provides a combined explanation for neutrino masses, mixing, and the cosmic baryon asymmetry through leptogenesis. Three concrete realizations are constructed, and their phenomenological predictions are analyzed. The results are not only compatible with the measured neutrino oscillation parameters within the current experimental 3$\sigma$ ranges, but also provide predictions for the neutrino mass ordering, Dirac and Majorana CP-violating phases, and the effective Majorana mass in neutrinoless double beta decay. The model further realizes TeV-scale leptogenesis consistent with the observed baryon asymmetry, rendering the scenario testable in both low-energy neutrino experiments and high-energy collider searches.

\end{abstract}

\end{titlepage}
\pagenumbering{arabic}



\section{Introduction}
\label{sec:intro}

The Standard Model (SM), completed by the discovery of the Higgs boson~\cite{ATLAS:2012yve,CMS:2012qbp}, stands as a cornerstone of modern particle physics, providing an extremely successful description of fundamental particles and their interactions. Nevertheless, the SM is unequivocally incomplete, as it fails to accommodate some observational evidence, notably the existence of nonzero neutrino masses and the observed matter–antimatter asymmetry of the Universe. These shortcomings provide compelling motivation for exploring physics beyond the SM (BSM).

The discovery of neutrino oscillations~\cite{Kajita:2016cak, McDonald:2016ixn} offers a definitive window into such BSM physics. In the SM, fermion masses arise via the Higgs mechanism, which requires both left- and right-handed chiral components. Because the SM does not include right-handed neutrinos, it predicts neutrinos to be massless. This prediction is ruled out by oscillation experiments, which imply that neutrinos possess small but non-zero masses. Resolving this discrepancy necessitates an extension of the SM, either through new particles or additional symmetries. Among the most natural and well-motivated frameworks for generating neutrino masses is the seesaw mechanism, which extends the particle content by introducing heavy mediators with SM-allowed interactions. While theoretically appealing, the conventional seesaw mechanism typically involves an energy scale far beyond the experimental reach, a notable shortcoming of an otherwise elegant paradigm. In this context, the inverse seesaw mechanism has emerged as a particularly attractive alternative, as it can accommodate the observed light neutrino masses while remaining testable at experimentally accessible energies~\cite{Mohapatra:1986aw,Mohapatra:1986bd,Deppisch:2004fa,Dev:2009aw,CentellesChulia:2020dfh}.

Explaining the matter–antimatter asymmetry of the Universe presents another significant challenge. Although the Universe began in a state of near-perfect symmetry between matter and antimatter, the observed baryon asymmetry, quantified by the ratio of the baryon number density to the photon number density, is precisely measured to be $\eta = (6.12 \pm 0.04) \times 10^{-10}$~\cite{Planck:2018vyg,ParticleDataGroup:2024cfk}. As emphasized by Sakharov~\cite{Sakharov:1967dj}, its dynamical generation requires three essential conditions: baryon number violation, C and CP violation, and departure from thermal equilibrium. Although the SM contains ingredients that partially satisfy these criteria, the CP violation it provides is many orders of magnitude small to generate the observed asymmetry. This is another indicator of the need for new physics.

A particularly elegant and economical solution is leptogenesis~\cite{Fukugita:1986hr}. In this scenario, a lepton asymmetry is generated dynamically through the out-of-equilibrium decays of heavy right-handed neutrinos in seesaw extensions of the SM, and is then partially converted into a baryon asymmetry via $B+L$-violating sphaleron processes. Beyond providing a compelling origin for matter-antimatter asymmetry, leptogenesis also establishes a theoretical link to neutrino mass generation within the seesaw framework. However, in conventional thermal leptogenesis, the lower bound on the lightest right-handed neutrino mass, $M_1 \gtrsim 10^9$ GeV, places the scale of new physics far beyond experimental reach. This constraint can be circumvented through resonant leptogenesis~\cite{Pilaftsis:1997jf, Pilaftsis:2003gt}, wherein the CP asymmetry is resonantly enhanced by nearly degenerate right-handed neutrinos with masses as low as the TeV scale. This energy range intriguingly coincides with the scale typical of the inverse seesaw mechanism. Moreover, the small lepton-number-violating parameter, essential for the inverse seesaw mechanism, also quantifies the mass splitting between the heavy singlet states. Its smallness naturally ensures a small mass degeneracy, providing the necessary condition for enhancing the CP asymmetry in leptogenesis (though not necessarily maximal enhancement). These intrinsic features of the inverse seesaw mechanism suggest the possibility of embedding both neutrino mass generation and leptogenesis within a common, experimentally testable framework.

While the inverse seesaw mechanism naturally explains the smallness of neutrino masses, it does not, by itself, address the flavor puzzle—the origin of the observed hierarchies in neutrino masses and the distinctive patterns of mixing angles. A widely adopted approach for constructing quantitative and predictive models of neutrino flavor is the framework of flavor symmetries. Within this framework, the flavor symmetry can be realized non-linearly by identifying it with a modular group. From a bottom-up perspective, modular symmetry has emerged as a powerful tool to explain the flavor structure of the SM~\cite{Feruglio:2017spp}. By constraining Yukawa couplings to transform as specific modular forms and eliminating the need for ad hoc flavon fields, this approach significantly reduces the number of free parameters in a model, thus enhancing its predictive power~\cite{Feruglio:2019ybq, Ding:2023htn}.

Traditionally, modular symmetry has been formulated within supersymmetric (SUSY) frameworks. However, the continued absence of experimental evidence for SUSY particles motivates a shift toward exploring modular symmetry in non-SUSY contexts~\cite{Qu:2024rns, Qu:2025ddz}. Recent work demonstrates the viability and fruitfulness of this direction, in which Yukawa couplings are constrained by more general classes of automorphic forms, such as polyharmonic  Maa{\ss} forms, which introduce new degrees of freedom and open up avenues for model building. Several successful models have been constructed within this non-SUSY paradigm, particularly those based on the finite modular group $A_4$~\cite{Qu:2024rns, Kumar:2024uxn, Nomura:2024atp, Nomura:2024nwh, Kobayashi:2025hnc, Loualidi:2025tgw, Kumar:2025bfe, Nomura:2025ovm, Nomura:2025raf, Zhang:2025dsa, Priya:2025wdm, Kumar:2025nut, Nanda:2025lem, Jangid:2025thp, Gao:2025jlw}. In contrast, the application of other finite groups and their double covers such as $S_3$~\cite{Okada:2025jjo}, $S_4$~\cite{Ding:2024inn}, and $A_5$~\cite{Li:2024svh, Li:2025kcr} remains relatively understudied. Investigating this broader landscape of symmetries holds the potential to yield new insights into fermion mass hierarchies and mixing patterns, while also offering a pathway toward unraveling the origin of the matter–antimatter asymmetry. This constitutes a promising direction for future research.

Within the non-SUSY framework, several recent studies~\cite{Nanda:2025lem, Priya:2025wdm, Kumar:2025bfe, Nasri:2026nbf} have proposed unified explanations for neutrino mass, mixing, and cosmic baryon asymmetry by invoking the $A_4$ modular symmetry. However, as noted above, research exploring neutrino mass generation and leptogenesis in the context of alternative modular groups remains limited. In this work, we construct an inverse seesaw model based on the non-holomorphic $A^{\prime}_5$ modular symmetry, which provides a combined description of neutrino mass, mixing, and the baryon asymmetry of the Universe. Compared with models based on other low-level modular groups, such as $S_3$, $A_4$, $S_4$, $A_5$ and their double covers, the $A_5^\prime$ group possesses the largest order and the richest set of irreducible representations. This feature allows for more flexible assignments of fermion and scalar fields, as well as a broader selection of modular forms, enabling a successful fit to neutrino oscillation data and a simultaneous explanation of the baryon asymmetry through leptogenesis. Moreover, existing studies based on these groups typically require higher energy scales to realize neutrino mass generation or leptogenesis. Our framework retains the predictive power inherent to modular symmetry while realizing neutrino mass via the low-scale seesaw mechanism. It predicts distinct collider and flavor physics signatures and successfully achieves leptogenesis at the TeV scale, with results consistent with current observational constraints.

The structure of this paper is as follows. Section~\ref{sec:model} presents the theoretical framework of $A^{\prime}_5$ modular symmetry within the inverse seesaw model, laying the foundation for subsequent analysis. Section~\ref{sec:b-model} elaborates on the construction and implementation of three specific models proposed in this work. In Sec.~\ref{sec:Num}, we perform a numerical analysis of the neutrino mass spectrum and flavor mixing patterns. Section~\ref{sec:lept} begins with an overview of the core aspects of the leptogenesis mechanism, followed by an analysis of the leptogenesis results for three specific models. Finally, Sec.~\ref{sec:Con} summarizes our main conclusions and outlines potential directions for future research. Appendix~\ref{app:Modular} provides a brief introduction to the basic concepts of modular symmetry relevant to this work.

\section{General setup}
\label{sec:model}

We extend the SM flavor sector by incorporating a non-holomorphic modular symmetry based on the group $A^{\prime}_5$, the double cover of the alternating group $A_5$. This choice is motivated by the fact that $A^{\prime}_5$ possesses irreducible representations of dimensions $\mathbf{1}$, $\mathbf{2}$, $\mathbf{3}$, $\mathbf{4}$, $\mathbf{5}$, and $\mathbf{6}$, offering sufficient flexibility to accommodate observed lepton mixing patterns while maintaining predictivity. The field content is enlarged by introducing three right-handed neutrinos $N$, three left-handed fermion singlets $S$, and a scalar singlet $\phi$ that couples $N$ and $S$. To further streamline construction and restrict allowed operators, we impose a global $U(1)_{B-L}$ symmetry under which the new fields carry appropriate charges~\footnote{The global $U(1)_{B-L}$ symmetry is spontaneously broken by $\langle\phi\rangle=v_\phi$, yielding a massless Majoron $J$. Its effective coupling to charged leptons is $g_{J\ell\ell}\sim y_N^2 M_\ell/(16\pi^2 v_\phi)$~\cite{Herrero-Brocal:2023czw,Vicente:2026zen}. Using our best-fit values $y_N^2\sim3\times10^{-5}$ (Sec.~\ref{sec:Num}) and $v_\phi\sim10\ \text{TeV}$ (Sec.~\ref{sec:lept}), we estimate $\mathcal{B}(\mu\to eJ)\sim1.4\times10^{-8}$.  This is three orders of magnitude below the TWIST limit $<5.8\times10^{-5}$ ~\cite{TWIST:2014ymv} and  lies within the reach of COMET Phase-II ($\mathcal{O}(10^{-8}){-}\mathcal{O}(10^{-9})$)~\cite{COMET:2018auw,Xing:2022rob}. Moreover, the Higgs invisible decay $h\to JJ$ is strongly suppressed due to $v_\phi\gg v_h$; as discussed in Ref.~\cite{Avila:2025qsc}, for $v_\phi>1\ \text{TeV}$ the limit on the scalar mixing angle is relaxed, and the branching ratio remains far below the LHC bound. Hence our scenario is safe, and testable in future experiments.}. Neutrino masses are generated via the inverse seesaw mechanism, which naturally explains their smallness through a spontaneously broken lepton number symmetry.

The transformation properties of the fields under the electroweak gauge symmetry, the $U(1)_{B-L}$ symmetry, and the modular $A^{\prime}_5$ symmetry are summarized in Table~\ref{tab:field1}. The modular weights $k$ assigned to each field control their coupling to modular forms. For the scalar sector, we take them as singlets, $\rho_H = \rho_\phi = \mathbf{1}$, ensuring that the modular symmetry is broken spontaneously not by these fields but solely by the modulus field $\tau$. No additional flavon fields are introduced to break the flavor symmetry $A^{\prime}_5$; instead, all flavor structure originates from the vacuum expectation value (VEV) of $\tau$ and the modular forms that depend on it. Once the modulus $\tau$ acquires a VEV, the Yukawa couplings become determined up to a small number of coupling constants, thereby generating the masses and mixing patterns of the leptons dynamically.
 \begin{table}[ht]
  \begin{center}
    \renewcommand{\arraystretch}{1.3}
    \setlength{\tabcolsep}{12pt}
   \begin{tabular}{|c||c|c|c|c||c|c|}
     \toprule
      \text{Fields}  &$L$&$E$&$N$&$S$&$H$&$\phi$\\
      \hline  
      $SU(2)_{L}$&$\mathbf{2}$&$\mathbf{1}$&$\mathbf{1}$&$\mathbf{1}$&$\mathbf{2}$&$\mathbf{1}$\\
      $U(1)_{Y}$&-$\frac{1}{2}$&-1&0&0&$\frac{1}{2}$& $0$\\
       $U(1)_{B-L}$&$1$&$1$&$1$&$0$&$0$&$-1$\\
      $A_5^{\prime}$&$\rho_L$&$\rho_E$&$\rho_N$&$\rho_S$&$\rho_H$&$\rho_{\phi}$\\
      $k$&$k_L$&$k_{E}$&$k_N$&$k_{S}$& $k_{H}$& $k_{\phi}$ \\ \bottomrule
      \end{tabular}
   \caption{Summary of field assignments under the electroweak gauge group $SU(2)_{L} \times U(1)_{Y}$, the $U(1)_{B-L}$ symmetry and the modular $A_5^{\prime}$ group. For each field, the table presents the assigned irreducible representations and modular weight ($k_L, k_E, \ldots, k_{\phi}$).}
 \label{tab:field1}  
   \end{center}   
\end{table}

With these assignments, the symmetries allow only the terms $LHE$, $LHN$, $SN\phi$, and $SS$ in the lepton sector. The Lagrangian, invariant under the modular $ A^{\prime}_5 $ symmetry, is given by
\begin{equation}
   \label{eq:lep-lag}
   -\mathcal{L}_{\ell}=\bar{L}EHf_1(Y)+\bar{L}N\tilde{H}f_2(Y)+\bar{S}N\phi f_3(Y)+\bar{S}Sf_4(Y)+\mathrm{H.c.},
\end{equation}
where $f_1(Y)$, $f_2(Y)$, $f_3(Y)$, and $f_4(Y)$ denote functions of the modular forms~\footnote{The Lagrangian is expressed in a general form, and specific models will be discussed later. Furthermore, since our focus lies on the mass matrix of leptons, the kinetic terms are omitted. For a brief discussion of the kinetic terms, refer to Ref.~\cite{Zhang:2025dsa}.}. Here, $\tilde{H} \equiv i\sigma^2 H^{\ast}$ is a conjugate of the Higgs doublet. The four terms in Eq.~(\ref{eq:lep-lag}) correspond, respectively, to the charged-lepton mass term, the Dirac mass term for $N$, the mixing term between $N$ and $S$, and the Majorana mass term for $S$. After the Higgs field $H$ and the scalar singlet $\phi$ acquire their VEVs, $\langle H \rangle = v_h$ and $\langle \phi \rangle = v_{\phi}$, the symmetries are broken and the lepton mass matrices can be derived.  The charged-lepton mass matrix, denoted by $M_{\mathrm{cl}}$, is given by $M_{\mathrm{cl}}=\langle H \rangle f_1\left( Y\left(\langle \tau \rangle\right)\right)$. In the neutrino sector, working in the flavor basis where the three left-handed neutrino fields are arranged as $(\nu_L, N^c, S)$, the full $9 \times 9$ neutrino mass matrix takes the following block form: 
\begin{equation}
M=\begin{pmatrix}
  0& M_D &0 \\
  M^T_D& 0 &M^T_{SN} \\
  0&  M_{SN}&M_S
\end{pmatrix}.
\label{eq:Block-mass-matrix}
\end{equation}
where the submatrices are defined as $ M_D = v_h f_2(Y)$, $M_{SN}= v_{\phi}f_3(Y)$, and $M_S = f_4(Y)$. 

In the inverse seesaw mechanism, a hierarchical structure satisfying $\mathcal{O}(M_S) \ll \mathcal{O}(M_D)$ $ \ll \mathcal{O}(M_{SN})$ is required for the matrices $M_D$, $M_{SN}$, and $M_S$ to naturally generate small masses for the light neutrinos. With this hierarchy and the block matrix form in Eq.~(\ref{eq:Block-mass-matrix}), the effective mass matrix for the light neutrinos is obtained by
\begin{equation}
    M_{\nu}=M_DM^{-1}_{SN}M_S(M^T_{SN})^{-1}M^T_D.
    \label{eq:nu-mass}
\end{equation}
This matrix is diagonalized by a unitary matrix $U_{\nu}$, such that $U_{\nu}^T M_{\nu} U_{\nu} = \mathrm{diag}(m_1, m_2, m_3)$. The leptonic mixing matrix $U_{\mathrm{PMNS}}$, is then given by $U_{\mathrm{PMNS}} = U_l^{\dagger} U_{\nu}$, where $U_l$ diagonalizes the charged-lepton mass matrix via $U_l^{\dagger} M_{\mathrm{cl}} M_{\mathrm{cl}}^{\dagger} U_l = \mathrm{diag}(m_e^2, m_{\mu}^2, m_{\tau}^2)$. Further details can be found, e.g., in~\cite{Wang:2024qhe}.

In this work, we adopt the standard parametrization of the PMNS matrix,
\begin{equation}  
	U_{\mathrm{PMNS}} = \begin{pmatrix}  
		c_{12}c_{13} & s_{12}c_{13} & s_{13}e^{-i\delta_{\mathrm{CP}}} \\  
		-s_{12}c_{23} - c_{12}s_{13}s_{23}e^{i\delta_{\mathrm{CP}}} & c_{12}c_{23} - s_{12}s_{13}s_{23}e^{i\delta_{\mathrm{CP}}} & c_{13}s_{23} \\  
		s_{12}s_{23} - c_{12}s_{13}c_{23}e^{i\delta_{\mathrm{CP}}} & -c_{12}s_{23} - s_{12}s_{13}c_{23}e^{i\delta_{\mathrm{CP}}} & c_{13}c_{23}  
	\end{pmatrix}  
	\begin{pmatrix}  
		e^{i\eta_1} & 0 & 0 \\  
		0 & e^{i\eta_2} & 0 \\  
		0 & 0 & 1  
	\end{pmatrix} ,
\end{equation}
where $s_{ij} \equiv \sin\theta_{ij}$ and $c_{ij} \equiv \cos\theta_{ij}$. The quantities $\eta_1$ and $\eta_2$ are the two Majorana phases, and $\delta_{\mathrm{CP}}$ denotes the Dirac CP-violating phase. This parametrization will be used in our numerical analysis to compare with experimental data.

\section{Specific model realizations}
\label{sec:b-model}

The modular invariance approach offers considerable flexibility in model building. Within the $A^{\prime}_5$ group, there exist numerous irreducible representations $\rho_{\mathbf{r}}$, and the modular weights of fields can in principle take any integer value. Consequently, the general Lagrangian in Eq.~\eqref{eq:lep-lag} admits a multitude of possible realizations, each characterized by specific choices of representations and modular weights for the fields. This part focuses on constructing concrete neutrino mass models based on $A^{\prime}_5$ modular symmetry, with the dual objectives of explaining the observed pattern of neutrino masses and mixing, and investigating the feasibility of achieving successful leptogenesis at the TeV scale. To this end, we introduce three specific benchmark models, labeled model A, model B, and model C, each capable of accommodating the current neutrino oscillation data while providing a framework for leptogenesis, which will be discussed in the next section.

The field assignments under the modular $A^{\prime}_5$ symmetry for these three models are listed in Table~\ref{tab:field2}. A common feature across all models is that the newly introduced singlet fermion $S$ transforms as a triplet $\mathbf{3}$ under $A^{\prime}_5$, while both scalar fields $H$ and $\phi$ are assigned to the trivial singlet representation $\mathbf{1}$. This choice ensures minimal complexity in the scalar sector, while allowing sufficient flexibility in the fermion sector to generate the desired mass textures.

The first column of Table~\ref{tab:field2} indicates the number of dimensionless real free parameters employed in the subsequent numerical analysis for each model. Models A and B share the same representation assignments for the fields $(\rho_L, \rho_E, \rho_N, \rho_S)$ but differ in their modular weight assignments $(k_L, k_E, k_N, k_S, k_H, k_{\phi})$. Both models contain 5 free parameters. Model C, with a different set of representation and weight assignments, contains 6 free parameters, offering slightly more flexibility in fitting the observational data.

A remark on notation is in order. The symbol $\mathbf{\hat{2}^{\prime}}$ (and analogously $\mathbf{\hat{2}}$) appearing in the table indicates that the corresponding field transforms as a doublet representation under $A^{\prime}_5$, but only its first two components are considered as active degrees of freedom in the low-energy theory, with the remaining component effectively decoupling or transforming as a singlet. Accordingly, the paired modular weights of these doublet and singlet constructions are written in parentheses, the first for the doublet and the second for the singlet. This constructive approach, sometimes referred to as representation reduction, allows for greater control over the resulting mass matrix textures while maintaining consistency with the modular symmetry framework.

The selection of modular weights in these models is restricted to integers in the range $-4\le k \le 3$. This choice serves two important purposes. First, it makes the model construction tractable by avoiding the increased complexity associated with higher-weight modular forms. Second, and more significantly, higher modular weights would induce a greater multiplicity of representations for the modular multiplets $Y^{(k)}_{\mathbf{r}}$ (where the subscript $\mathbf{r}$ denotes an irreducible representation of $A^{\prime}_5$), potentially introducing an undesirable proliferation of free parameters. By limiting the weights to this range, we strike a balance between predictive power and phenomenological viability.
\begin{table}[h!]
  \centering
  \renewcommand{\arraystretch}{2}
  \setlength{\tabcolsep}{2.5 pt}
  \begin{tabular}{||c||c||c|c|c||c|c|c||}
    \toprule
    \multirow{1}{*}{\makecell{Free\\Parameter}} &
    \text{Model} &
    $(\rho_L,k_L)$ &
    $(\rho_E,k_E)$ &
    $(\rho_N,k_N)$ &
    $(\rho_S,k_S)$ &
    $(\rho_H,k_H)$ &
    $(\rho_{\phi},k_{\phi})$ \\
    \cline{2-8}
    \hline  
    \multirow{2}{*}{$\mathbf{5}$} &
    $\mathbf{A}$ &
    $(\mathbf{3},-1)$ &
    $(\mathbf{\hat{2}^{\prime}}\oplus\mathbf{1},(4,3))$ &
    $(\mathbf{3^{\prime}},3)$ &
    $(\mathbf{3},1)$ &
    $(\mathbf{1},0)$ &
    $(\mathbf{1},0)$ \\
    \cline{2-8}
    &
    $\mathbf{B}$ &
    $(\mathbf{3},-2)$ &
    $(\mathbf{\hat{2}^{\prime}}\oplus\mathbf{1},(3,6))$ &
    $(\mathbf{3^{\prime}},4)$ &
    $(\mathbf{3},0)$ &
    $(\mathbf{1},0)$ &
    $(\mathbf{1},0)$ \\
    \hline 
    $\mathbf{6}$ &
    $\mathbf{C}$ &
    $(\mathbf{\hat{2}^{\prime}}\oplus\mathbf{1},(-2,1))$ &
    $(\mathbf{3},0)$ &
    $(\mathbf{3^{\prime}},-2)$ &
    $(\mathbf{3},1)$ &
    $(\mathbf{1},1)$ &
    $(\mathbf{1},3)$ \\
    \bottomrule
  \end{tabular}
  \caption{Specific representation and weight assignments of the three benchmark models considered in this work. For each model, the table specifies the irreducible representations $\rho$ under the $A^{\prime}_5$ group and the modular weights $k$ assigned to the fields $L$, $E$, $N$, $S$, $H$, and $\phi$. The notation $\mathbf{\hat{2}^{\prime}}\oplus\mathbf{1}$ indicates that the corresponding field transforms partly as a doublet (with only its first two components active) and partly as a singlet. Models A and B contain 5 free parameters each, while model C contains 6 free parameters.}
  \label{tab:field2}  
\end{table}

With these field assignments, the modular forms $Y^{(k)}_{\mathbf{r}}$ entering the functions $f_i(Y)$ in Eq.~\eqref{eq:lep-lag} are uniquely determined up to number of coupling constants. The explicit construction of the Yukawa couplings and the resulting mass matrices for each model will be presented in the following subsections, followed by a detailed numerical analysis of their phenomenological implications for neutrino oscillations and leptogenesis.

\subsection{Models A and B}
\label{sec:ma-mb}

Models A and B share a common structure in their field representation assignments under the $A^{\prime}_5$ modular symmetry, differing only in the specific modular weights assigned to each field (see Table~\ref{tab:field2}). This common structure allows us to present their Lagrangians and mass matrices simultaneously, while the distinct weight assignments lead to different phenomenological predictions that will be explored in the numerical analysis.

In both models, the left-handed lepton doublet $L = (L_1, L_2, L_3)$ transforms as the three-dimensional irreducible representation $\mathbf{3}$ of $A^{\prime}_5$. The right-handed charged lepton fields $E = (E_1, E_2, E_3)$ are assigned to decompose as the direct sum $\mathbf{\hat{2}^{\prime}} \oplus \mathbf{1}$. Specifically, the first two components $E_{1,2} \equiv (E_1, E_2)$ transform as the two-dimensional irreducible representation $\mathbf{\hat{2}^{\prime}}$, while the third component $E_3$ transforms as the trivial singlet $\mathbf{1}$.

The modular-invariant Lagrangian for the charged lepton sector is given by
\begin{equation}
  -\mathcal{L}_{\mathrm{CL}}=\alpha_{\mathrm{CL}}(\bar{L}E_{1,2})_{\mathbf{\hat{6}}}Y^{(k_1)}_{\mathbf{\hat{6}}\Rmnum{1}}H +\beta_{\mathrm{CL}}(\bar{L}E_{1,2})_{\mathbf{\hat{6}}}Y^{(k_1)}_{\mathbf{\hat{6}}\Rmnum{2}}H +\gamma_{\mathrm{CL}}(\bar{L}E_3)_{\mathbf{3}}Y^{(k_2)}_{\mathbf{3}}H +\mathrm{H.c.},
  \label{eq:cl-lag}
\end{equation}
where the subscripts indicate the $A^{\prime}_5$ representation into which the corresponding bilinear contracts, and $Y^{(k)}_{\mathbf{r}}$ denotes the modular multiplet of weight $k$ transforming in the representation $\mathbf{r}$. Modular weights are constrained by the requirement that the total weight of each term vanish, satisfying $ -k_1 = k_L + k_{E_{1,2}} + k_H $ and $ -k_2 = k_L + k_{E_3} + k_H $. Using the field weight assignments from Table~\ref{tab:field2}, the specific values of $k_1$ and $k_2$ for models A and B are computed and summarized in Table~\ref{tab:weight}. The coupling constants $\alpha_{\mathrm{CL}}$, $\beta_{\mathrm{CL}}$, and $\gamma_{\mathrm{CL}}$ are in general complex parameters. However, their phases can be absorbed through a redefinition of the charged lepton fields, allowing them to be treated as real parameters without loss of generality.
 \begin{table}[ht]
  \begin{center}
    \renewcommand{\arraystretch}{1.3}
\setlength{\tabcolsep}{12pt}
   \begin{tabular}{|c||c|c|c|c|c|}
     \toprule
      \text{Model}  &$k_1$&$k_2$&$k_3$&$k_4$&$k_5$\\
      \hline  
      A&$-3$&$-2$&$-2$&$-4$&$-2$\\

      B&$-1$&$-4$&$-2$&$-4$&$0$\\ \bottomrule
      \end{tabular}
   \caption{Modular weights $k_i$ for the modular forms appearing in the charged lepton and neutrino mass matrices of model A and model B, derived from the field assignments in Table~\ref{tab:field2} and the weight conservation conditions.}
 \label{tab:weight}  
   \end{center}   
\end{table}

From Eq.~(\ref{eq:cl-lag}), after electroweak symmetry breaking, we obtain the charged lepton mass matrix
\begin{equation}
  M_{CL}= \begin{pmatrix}
      \sqrt{2} \alpha_{\mathrm{CL}} Y^{(k_1)}_{\mathbf{\hat{6}}\Rmnum{1},6} + \sqrt{2} \beta_{\mathrm{CL}} Y^{(k_1)}_{\mathbf{\hat{6}}\Rmnum{2},6} & \sqrt{2} \alpha_{\mathrm{CL}} Y^{(k_1)}_{\mathbf{\hat{6}}\Rmnum{1},3} + \sqrt{2} \beta_{\mathrm{CL}} Y^{(k_1)}_{\mathbf{\hat{6}}\Rmnum{2},3} & \gamma_{\mathrm{CL}} Y^{(k_2)}_{\mathbf{3},1} \\
      \sqrt{2} \alpha_{\mathrm{CL}} Y^{(k_1)}_{\mathbf{\hat{6}}\Rmnum{1},5} + \sqrt{2} \beta_{\mathrm{CL}} Y^{(k_1)}_{\mathbf{\hat{6}}\Rmnum{2},5} & \alpha_{\mathrm{CL}} Y^{(k_1)}_{\mathbf{\hat{6}}\Rmnum{1},-} + \beta_{\mathrm{CL}} Y^{(k_2)}_{\mathbf{\hat{6}}\Rmnum{1},-} & \gamma_{\mathrm{CL}} Y^{(k_2)}_{\mathbf{3},3} \\
      \alpha_{\mathrm{CL}} Y^{(k_1)}_{\mathbf{\hat{6}}\Rmnum{1},+} + \beta_{\mathrm{CL}} Y^{(k_1)}_{\mathbf{\hat{6}}\Rmnum{2},+} & \sqrt{2} \alpha_{\mathrm{CL}} Y^{(k_1)}_{\mathbf{\hat{6}}\Rmnum{1},4} + \sqrt{2} \beta_{\mathrm{CL}} Y^{(k_1)}_{\mathbf{\hat{6}}\Rmnum{2},4} & \gamma_{\mathrm{CL}} Y^{(k_2)}_{\mathbf{3},2}
  \end{pmatrix}v_h,
  \label{eq:model_abCL}
  \end{equation}
where we have defined the combinations $ Y^{(k_1)}_{\mathbf{\hat{6}}A,\pm}= Y^{(k_1)}_{\mathbf{\hat{6}}A,1} \pm  Y^{(k_1)}_{\mathbf{\hat{6}}A,2}$ for $A\in \{I,~II\}$. The numerical subscripts on the modular forms denote their components according to the basis conventions given in Appendix~\ref{app:Modular}.

In the neutrino sector, the gauge singlet fermions $S = (S_1, S_2, S_3)$ and $N = (N_1, N_2, N_3)$ are both assigned to triplet representations of $A^{\prime}_5$, transforming as $\mathbf{3}$ and $\mathbf{3}^{\prime}$, respectively. Neutrino part of the Lagrangian consists of three parts,
\begin{equation}
  \mathcal{L}_{\nu}=\mathcal{L}_{D}+\mathcal{L}_{SN}+\mathcal{L}_{S},
\end{equation}
corresponding to the Dirac mass term, the $N$-$S$ mixing term, and the Majorana mass term for $S$, respectively. Their explicit forms, invariant under both the gauge symmetries and the modular $A^\prime_5$ symmetry, are given by
\begin{align}
  -\mathcal{L}_{D}& =\alpha_{D}(\bar{L}N)_{\mathbf{5}}Y^{(k_3)}_{\mathbf{5}}\tilde{H}+\mathrm{H.c.},\\
  -\mathcal{L}_{SN}&=\alpha_{SN}(\bar{S}N)_{\mathbf{5}}Y^{(k_4)}_{\mathbf{5}}\phi+\mathrm{H.c.},\\
  -\mathcal{L}_{SS}&=\alpha_S(\bar{S}S)_{\mathbf{1S}}Y^{(k_5)}_{\mathbf{1}}+\beta_S(\bar{S}S)_{\mathbf{5S}}Y^{(k_5)}_{\mathbf{5}}+\mathrm{H.c.}.
\end{align}
In these expressions, modular weights satisfy the constraints $-k_3 = k_L + k_N + k_H$, $-k_4 = k_S + k_N + k_{\phi}$, and $-k_5 = 2k_S$. The resulting values of $k_3$, $k_4$, and $k_5$ for models A and B, derived from the field assignments in Table~\ref{tab:field2}, are listed in Table~\ref{tab:weight}. 

After the scalar fields acquire their VEVs, the mass matrices take the following forms. The Dirac mass matrix $M_D$ connecting $\nu_L$ and $N$ is
\begin{equation}
 M_D= \alpha_{D}\begin{pmatrix}
    \sqrt{3}Y^{(k_3)}_{\mathbf{5},1} & Y^{(k_3)}_{\mathbf{5},4} & Y^{(k_3)}_{\mathbf{5},3} \\
    Y^{(k_3)}_{\mathbf{5},5} & -\sqrt{2}Y^{(k_3)}_{\mathbf{5},3} & -\sqrt{2}Y^{(k_3)}_{\mathbf{5},2} \\
    Y^{(k_3)}_{\mathbf{5},2} & -\sqrt{2}Y^{(k_3)}_{\mathbf{5},5} & -\sqrt{2}Y^{(k_3)}_{\mathbf{5},4}
    \end{pmatrix}v_h,
  \end{equation}
The $N$-$S$ mixing matrix $M_{SN}$ exhibits an identical structure, up to exchange $k_3 \leftrightarrow k_4$ and $v_h \leftrightarrow v_\phi$, 
\begin{equation}
  M_{SN}=\alpha_{SN}\begin{pmatrix}
    \sqrt{3}Y^{(k_4)}_{\mathbf{5},1} & Y^{(k_4)}_{\mathbf{5},4} & Y^{(k_4)}_{\mathbf{5},3} \\
    Y^{(k_4)}_{\mathbf{5},5} & -\sqrt{2}Y^{(k_4)}_{\mathbf{5},3} & -\sqrt{2}Y^{(k_4)}_{\mathbf{5},2} \\
    Y^{(k_4)}_{\mathbf{5},2} & -\sqrt{2}Y^{(k_4)}_{\mathbf{5},5} & -\sqrt{2}Y^{(k_4)}_{\mathbf{5},4}
    \end{pmatrix}v_{\phi},
  \end{equation}
Finally, the Majorana mass matrix $M_S$ for the singlet fermions $S$ is given by  
\begin{equation}
  M_S= \begin{pmatrix}
    \alpha_SY^{(k_5)}_{\mathbf{1}}+2\beta_SY^{(k_5)}_{\mathbf{5},1}&-\sqrt{3}\beta_SY^{(k_5)}_{\mathbf{5},5}&-\sqrt{3}\beta_SY^{(k_5)}_{\mathbf{5},2} \\
    -\sqrt{3}\beta_SY^{(k_5)}_{\mathbf{5},5}& \sqrt{6}\beta_SY^{(-2)}_{\mathbf{5},4} & \alpha_SY^{(-2)}_{\mathbf{1}}-\beta_SY^{(k_5)}_{\mathbf{5},1} \\
   -\sqrt{3}\beta_SY^{(k_5)}_{\mathbf{5},2} &\alpha_SY^{(k_5)}_{\mathbf{1}}-\beta_SY^{(k_5)}_{\mathbf{5},1}  &\sqrt{6}\beta_SY^{(k_5)}_{\mathbf{5},3}
  \end{pmatrix}.
\end{equation}

The parameters $\alpha_D$, $\alpha_{SN}$, $\alpha_S$, and $\beta_S$ appearing in the neutrino sector are, in general, complex. To reduce the number of free parameters and enhance the predictive power of the model, we impose a generalized CP symmetry (gCP)~\cite{Novichkov:2019sqv, Ding:2023htn}, which constrains these couplings to be real. Under this assumption, the only remaining complex quantity in the model is the modulus field $\tau$ itself. Consequently, the VEV $\langle \tau \rangle$ becomes the unique source of CP violation, determining both the magnitudes and phases of the modular forms $Y^{(k)}_{\mathbf{r}}(\tau)$ and, through them, the entire flavor structure of the lepton sector.

For the purpose of numerical fitting, it is convenient to redefine the parameters to isolate the overall mass scales and reduce the number of independent dimensionless quantities. We introduce the following reparametrizations: $ \tilde{\alpha}_{\mathrm{CL}}=\alpha_{\mathrm{CL}}v_h$, $\tilde{\beta}_{\mathrm{CL}} = \beta_{\mathrm{CL}}/\alpha_{\mathrm{CL}}$, $\tilde{\gamma}_{\mathrm{CL}} = \gamma_{\mathrm{CL}}/\alpha_{\mathrm{CL}}$, $\Lambda = \alpha_{SN} v_{\phi}$, and $\tilde{\beta}_{S} = \beta_{S}/\alpha_{S}$. The charged lepton mass matrix can then be written as $M_{\mathrm{CL}} = \tilde{\alpha}_{\mathrm{CL}} m_{\mathrm{CL}}$, where $m_{\mathrm{CL}}$ depends only on the dimensionless ratios $\tilde{\beta}_{\mathrm{CL}}$, $\tilde{\gamma}_{\mathrm{CL}}$, and the modulus $\tau$. Similarly, the effective light neutrino mass matrix from Eq.~(\ref{eq:nu-mass}) takes the form $M_{\nu} = n m_{\nu}$, with the overall scale factor $n = (\alpha_Dv_h  / \Lambda)^2 \alpha_{S}$. Thus, the lepton phenomenology in models A and B is governed by five dimensionless real free parameters: $\tilde{\beta}_{\mathrm{CL}}$, $\tilde{\gamma}_{\mathrm{CL}}$, $\tilde{\beta}_S$, $\mathrm{Re}(\tau)$, and $\mathrm{Im}(\tau)$. The overall mass scale $\tilde{\alpha}_{\mathrm{CL}}$ and factor $n$ are determined separately by the mass scale of charged leptons and requirement of reproducing the neutrino mass-squared differences, inferred from oscillation data.

\subsection{Model C}
\label{sec:mc}

Model C adopts a different assignment of modular representations compared to models A and B, interchanging the roles of the left-handed lepton doublet $L$ and the right-handed charged leptons $E$. This alternative construction explores a different region of the theory space and leads to distinct phenomenological predictions.

In this model, the right-handed charged leptons $E_i$ transform as the three-dimensional irreducible representation $\mathbf{3}$ of $A^\prime_5$. The right-handed neutrinos $N_i$ are assigned to another triplet $\mathbf{3^{\prime}}$, while the singlet fermions $S_i$ transform as $\mathbf{3}$. For the left-handed lepton doublets $L_i$, the first two components $L_{1,2}$ transform as the two-dimensional irreducible representation $\mathbf{\hat{2}^{\prime}}$ with modular weight of $-2$, and the third component $L_3$ transforms as the trivial singlet $\mathbf{1}$ with unit modular weight. The scalar fields $H$ and $\phi$ are assigned modular weights $k_H = 1$ and $k_{\phi} = 3$, respectively, as indicated in Table~\ref{tab:field2}.

The modular-invariant Lagrangian for the charged lepton sector takes the form
\begin{equation}
  -\mathcal{L}_{\mathrm{CL}}=\alpha_{\mathrm{CL}}(\bar{L}_{1,2}E)_{\mathbf{\hat{6}}}Y^{(1)}_{\mathbf{\hat{6}}\Rmnum{1}}H+\beta_{\mathrm{CL}}(\bar{L}_{1,2}E)_{\mathbf{\hat{6}}}Y^{(1)}_{\mathbf{\hat{6}}\Rmnum{2}}H+\gamma_{\mathrm{CL}}(\bar{L}_3E)_{\mathbf{3}}Y^{(-2)}_{\mathbf{3}}H  +\mathrm{H.c.}.
  \label{eq:cl-lag-c}
\end{equation}
Comparing with Eq.~\eqref{eq:cl-lag} for models A and B, we observe that the roles of $L$ and $E$ are interchanged. Consequently, the charged lepton mass matrix derived from Eq.~\eqref{eq:cl-lag-c} is essentially the transpose of the matrix given in Eq.~\eqref{eq:model_abCL}. After electroweak symmetry breaking, we obtain
\begin{equation}
\begin{pmatrix}
  \sqrt{2} \alpha_{\mathrm{CL}} Y^{(1)}_{\mathbf{\hat{6}}\Rmnum{1},6} + \sqrt{2} \beta_{\mathrm{CL}} Y^{(1)}_{\mathbf{\hat{6}}\Rmnum{2},6} &
  \sqrt{2} \alpha_{\mathrm{CL}} Y^{(1)}_{\mathbf{\hat{6}}\Rmnum{1},5} + \sqrt{2} \beta_{\mathrm{CL}} Y^{(1)}_{\mathbf{\hat{6}}\Rmnum{2},5} &
  \alpha_{\mathrm{CL}} Y^{(1)}_{\mathbf{\hat{6}}\Rmnum{1},+} + \beta_{\mathrm{CL}} Y^{(1)}_{\mathbf{\hat{6}}\Rmnum{2},+} \\
  \sqrt{2} \alpha_{\mathrm{CL}} Y^{(1)}_{\mathbf{\hat{6}}\Rmnum{1},3} + \sqrt{2} \beta_{\mathrm{CL}} Y^{(1)}_{\mathbf{\hat{6}}\Rmnum{2},3} &
  \alpha_{\mathrm{CL}} Y^{(1)}_{\mathbf{\hat{6}}\Rmnum{1},-} + \beta_{\mathrm{CL}} Y^{(1)}_{\mathbf{\hat{6}}\Rmnum{1},-} &
  \sqrt{2} \alpha_{\mathrm{CL}} Y^{(1)}_{\mathbf{\hat{6}}\Rmnum{1},4} + \sqrt{2} \beta_{\mathrm{CL}} Y^{(1)}_{\mathbf{\hat{6}}\Rmnum{2},4} \\
  \gamma_{\mathrm{CL}} Y^{(-2)}_{\mathbf{3},1} &
  \gamma_{\mathrm{CL}} Y^{(-2)}_{\mathbf{3},3} &
  \gamma_{\mathrm{CL}} Y^{(-2)}_{\mathbf{3},2}
  \end{pmatrix}v_h.
\end{equation}

For the neutrino sector, the $N$-$S$ mixing term and the Majorana mass term for $S$ retain the same forms as in models A and B, namely
\begin{align}
  -\mathcal{L}_{SN}& =\alpha_{SN}(\bar{S}N)_{\mathbf{5}}Y^{(-2)}_{\mathbf{5}}\phi+\mathrm{H.c.},\\
  -\mathcal{L}_{SS}&=\alpha_S(\bar{S}S)_{\mathbf{1S}}Y^{(0)}_{\mathbf{1}}+\beta_S(\bar{S}S)_{\mathbf{5S}}Y^{(0)}_{\mathbf{5}}+\mathrm{H.c.}.
\end{align}
The modular weights appearing in these expressions are determined by the field assignments in Table~\ref{tab:field2} and satisfy the conditions $-k_4 = k_S + k_N + k{\phi}$ and $-k_5 = 2k_S$, yielding $k_4 = -2$ and $k_5 = 0$ for model C. The Dirac mass term $\mathcal{L}_D$ in model C differs significantly from that of models A and B due to the different representation assignments for $L_i$. In the $A^{\prime}_5$ group, the tensor product decomposition relevant for the coupling between $L_{1,2}$ (transforming as $\mathbf{\hat{2}^{\prime}}$) and $N$ (transforming as $\mathbf{3^{\prime}}$) is $\mathbf{\hat{2}^{\prime}}\otimes \mathbf{3^{\prime}}=\mathbf{\hat{4}^{\prime}}\oplus \mathbf{\hat{2}^{\prime}}$. However, the modular form multiplet $Y^{(3)}_{\mathbf{r}}$ of weight $k=3$ does not contain a component transforming as $\mathbf{\hat{2}^{\prime}}$. Consequently, only the contraction into the $\mathbf{\hat{4}^{\prime}}$ representation is allowed. For the singlet component $L_3$, the coupling with $N$ proceeds via the $\mathbf{3^{\prime}}$ representation. The complete modular-invariant Dirac Lagrangian therefore consists of two terms,
\begin{equation}
\label{eq:nd-lag-c}
-\mathcal{L}_{D}=\alpha_{D}(\bar{L}_{1,2}N)_{\mathbf{\hat{4^{\prime}}}}Y^{(3)}_{\mathbf{\hat{4}^{\prime}}}\tilde{H}+\beta_{D}(\bar{L}_3N)_{\mathbf{3^{\prime}}}Y^{(0)}_{\mathbf{3^{\prime}}}\tilde{H}+\mathrm{H.c.}.
\end{equation}
The modular weights are constrained by $-k_{D_1} = k_{L_{1,2}} + k_N + k_H$ and $-k_{D_2} = k_{L_3} + k_N + k_H$, which from Table~\ref{tab:field2} give $k_{D_1} = 3$ and $k_{D_2} = 0$, consistent with the superscripts appearing in Eq.~\eqref{eq:nd-lag-c}.

After electroweak symmetry breaking, the Dirac mass matrix $M_D$ takes the form
\begin{equation}
\label{eq:md-c}
  M_D=\begin{pmatrix}
    -\sqrt{2}\alpha_{D}Y^{(3)}_{\mathbf{\hat{4}^{\prime},4}} & -\sqrt{3}\alpha_{D}Y^{(3)}_{\mathbf{\hat{4}^{\prime},2}} & \alpha_{D}Y^{(3)}_{\mathbf{\hat{4}^{\prime},1}} \\
    \sqrt{2}\alpha_{D}Y^{(3)}_{\mathbf{\hat{4}^{\prime},1}} & \alpha_{D}Y^{(3)}_{\mathbf{\hat{4}^{\prime},4}} & -\sqrt{3}\alpha_{D}Y^{(3)}_{\mathbf{\hat{4}^{\prime},3}}\\
    \beta_{D} Y^{(0)}_{\mathbf{3^{\prime}},1} &
    \beta_{D} Y^{(0)}_{\mathbf{3^{\prime}},3} &
    \beta_{D} Y^{(0)}_{\mathbf{3^{\prime}},2}
    \end{pmatrix}v_h.
\end{equation}
The $N$-$S$ mixing matrix $M_{SN}$ and the Majorana mass matrix $M_S$ for the singlet fermions $S$ are given by
\begin{align}
\label{eq:msn-c}
 M_{SN} & = \alpha_{SN}\begin{pmatrix}
    \sqrt{3}Y^{(-2)}_{\mathbf{5},1} & Y^{(-2)}_{\mathbf{5},4} & Y^{(-2)}_{\mathbf{5},3} \\
    Y^{(-2)}_{\mathbf{5},5} & -\sqrt{2}Y^{(-2)}_{\mathbf{5},3} & -\sqrt{2}Y^{(-2)}_{\mathbf{5},2} \\
    Y^{(-2)}_{\mathbf{5},2} & -\sqrt{2}Y^{(-2)}_{\mathbf{5},5} & -\sqrt{2}Y^{(-2)}_{\mathbf{5},4}
    \end{pmatrix}v_{\phi}, \\[6pt]
    \label{eq:ms-c}
  M_S & =\begin{pmatrix}
    \alpha_SY^{(-2)}_{\mathbf{1}}+2\beta_2Y^{(-2)}_{\mathbf{5},1}&-\sqrt{3}\beta_2Y^{(-2)}_{\mathbf{5},5}&-\sqrt{3}\beta_2Y^{(-2)}_{\mathbf{5},2} \\
    -\sqrt{3}\beta_2Y^{(-2)}_{\mathbf{5},5}& \sqrt{6}\beta_2Y^{(-2)}_{\mathbf{5},4} & \alpha_SY^{(-2)}_{\mathbf{1}}-\beta_2Y^{(-2)}_{\mathbf{5},1} \\
   -\sqrt{3}\beta_2Y^{(-2)}_{\mathbf{5},2} &\alpha_SY^{(-2)}_{\mathbf{1}}-\beta_2Y^{(-2)}_{\mathbf{5},1}  &\sqrt{6}\beta_2Y^{(-2)}_{\mathbf{5},3}
  \end{pmatrix}.
\end{align}

As in models A and B, we impose a gCP to restrict the coupling constants to real values, leaving the modulus field $\tau$ as the sole source of CP violation. For numerical analysis, we introduce the following reparametrization to isolate the overall mass scales and reduce the number of independent dimensionless quantities: $\tilde{\beta}_{\mathrm{CL}}$, $\tilde{\gamma}_{\mathrm{CL}}$, $\tilde{\beta}_{D}$, $\tilde{\beta}_{S}$, $\mathrm{Re}(\tau)$ and $\mathrm{Im}(\tau)$. This corresponds to four real parameters and one complex parameter $\tau$. The charged lepton mass matrix is then expressed as $M_{\mathrm{CL}} = \tilde{\alpha}_{\mathrm{CL}} m_{\mathrm{CL}}$, where $m_{\mathrm{CL}}$ depends on $\tilde{\beta}_{\mathrm{CL}}$, $\tilde{\gamma}_{\mathrm{CL}}$, and the modulus $\tau$. The effective light neutrino mass matrix from Eq.~\eqref{eq:nu-mass} takes the form $M_{\nu} = n m_{\nu}$, with the overall scale factor $n = \left(\alpha_D v_h/\Lambda \right)^2 \alpha_S$ and $m_{\nu}$ depending on $\tilde{\beta}_D$, $\tilde{\beta}_S$, and $\tau$.

Having fully specified the three benchmark models, we now proceed in the following two sections to a systematic numerical analysis of their parameter spaces, searching for regions that simultaneously satisfy the current neutrino oscillation data and yield successful low-scale leptogenesis.

\section{Lepton masses and mixing  }
\label{sec:Num}

In this section, we perform a phenomenological study by numerical analysis of the three benchmark models A, B and C, introduced in Sec.~\ref{sec:b-model}, to assess their viability in reproducing the observed neutrino oscillation data and give predictions on parameters that have not yet been determined experimentally. For each model, we scan the relevant parameter space to identify regions that yield predictions consistent with experimental measurements within the $3\sigma$ confidence level. The analysis focuses on the dimensionless free parameters introduced in the previous section, as well as the modulus field $\tau$, which determines the flavor structure through its VEV.

To implement parameter scan, the dimensionless real parameters appearing in the mass matrices—specifically the coupling ratios $\tilde{\beta}_{\mathrm{CL}}$, $\tilde{\gamma}_{\mathrm{CL}}$, $\tilde{\beta}_{D}$ (where applicable) and $\tilde{\beta}_{S}$—are each varied over the range $(0, 10^3)$. This interval is chosen to be sufficiently broad to encompass natural values of the couplings while avoiding unnaturally large fine-tuning. The modulus field $\tau = \mathrm{Re}(\tau) + i\,\mathrm{Im}(\tau)$ is the only complex parameter in the model after the imposition of generalized gCP symmetry. Its VEV is constrained to lie within the fundamental domain of the modular group, defined by conditions $\mathrm{Im}(\tau)> 0$, $\left| \mathrm{Re}(\tau)\right| \le \frac{1}{2}$, and $\left| \tau\right| \ge 1$. The introduction of gCP symmetry further imposes a identification $\tau \sim -\tau^{\ast}$~\cite{Novichkov:2018ovf}, which maps points with negative $\mathrm{Re}(\tau)$ to positive values. This allows us to restrict our scan to $\mathrm{Re}(\tau) > 0$ without loss of generality, specifically considering the interval $\mathrm{Re}(\tau) \in (0, 0.5)$.

To quantify the agreement between model predictions and experimental data, we employ a chi-squared ($\chi^2$) function, defined as
\begin{equation}
\label{eq:chi2}
\chi^2=\sum_{i}\left( \frac{p_i-q_i}{\sigma_i} \right)^2,
\end{equation}
where $p_i$ denotes the value predicted by the model for a given observable, $q_i$ is the corresponding central value from experimental measurements, and $\sigma_i$ represents the associated $1\sigma$ uncertainty. The observables included in the fit are the charged lepton mass ratios $m_e/m_\mu$ and $m_\mu/m_\tau$, the neutrino mixing angles $\sin^2\theta_{12}$, $\sin^2\theta_{13}$, and $\sin^2\theta_{23}$, and the ratio of the solar to atmospheric mass squared differences $ \Delta m^2_{21}/\left|\Delta m^2_{3l}\right| $, where $l=1$ for normal ordering (NO) and $l=2$ for inverted ordering (IO).

The experimental values adopted in this analysis are listed in Table~\ref{tab:data}, which contains the recent global fits of neutrino masses and mixing parameters from Refs.~\cite{Esteban:2020cvm, Esteban:2024eli}. For the charged lepton mass ratios, we take the values at $M_Z$ scale from Ref.~\cite{Xing:2007fb}, with a conservative $1\sigma$ uncertainty of $0.1\%$ of the central value to account for radiative corrections and theoretical uncertainties. The scan over the parameter space and the minimization of the $\chi^2$ function are performed using the FlavorPy package~\cite{FlavorPy}, a dedicated computational tool for flavor model analysis.
\begin{table}[ht!]
	\begin{center}
		\renewcommand{\arraystretch}{1.5}
		\setlength{\tabcolsep}{15pt}
		\begin{tabular}{|c|c|c|}\toprule
			\text{Observable}& $\text{bfp} \pm 1\sigma$&$3\sigma$ range\\
			\hline  
			$\frac{m_e}{m_\mu } $&$0.004737$&$-$\\
			$\frac{m_\mu}{m_\tau }$&$0.05882$&$-$\\
			\hline  
			$\sin^2\theta_{12}$ (NO/IO)&$0.308^{+0.012}_{-0.011}$&0.275$\to$0.345\\
			$\sin^2\theta_{13}$ (NO)&$0.02215^{+0.00056}_{-0.00058}$&0.02030$\to$0.02388\\
			$\sin^2\theta_{13}$ (IO)&$0.02231^{+0.00056}_{-0.00056}$&0.02060$\to$0.02409\\
			$\sin^2\theta_{23}$ (NO)&$0.470^{+0.017}_{-0.013}$&0.435$\to$0.585\\
			$\sin^2\theta_{23}$ (IO)&$0.550^{+0.012}_{-0.015}$&0.440$\to$0.584\\
			\hline 
			$\Delta m^2_{21}/10^{-5}~$eV$^2$ &$7.49^{+0.19}_{-0.19}$&6.92$\to$8.05\\
			$\Delta m^2_{31}/10^{-3}~$eV$^2$ (NO)&$2.513^{+0.021}_{-0.019}$&2.451$\to$2.578\\
			$\Delta m^2_{32}/10^{-3}~$eV$^2$ (IO)&$-2.484^{+0.020}_{-0.0.020}$ &-2.547$\to$-2.421\\
			\bottomrule
		\end{tabular}   
	\end{center}   
	\caption{Experimental input values used in numerical analysis. The neutrino oscillation parameters are taken from the global fit results of Refs.~\cite{Esteban:2020cvm, Esteban:2024eli}, with a separate column specified for the NO and IO cases where applicable. Charged lepton mass ratios are evaluated on the scale of $M_Z$ following~\cite{Xing:2007fb}, with a $1\sigma$ uncertainty of $0.1\%$ assigned for the calculation of the $\chi^2$ function.}
	\label{tab:data}
\end{table}

With the parameter scan and goodness-of-fit method established, we now proceed to a detailed analysis of each benchmark model. In the following subsections, we present the results for model A, model B, and model C, identifying the best-fit points and characterizing the regions of parameter space that provide the most favorable agreement with experimental data.

\subsection{Phenomenology of model A}
\label{sec:p-ma}

We now present the results of our numerical analysis of model A. Through a scan of the parameter space defined in Sec.~\ref{sec:Num}, we have identified the phenomenologically viable regions of the modulus field $\tau$ and determined the predictions of model A for the relevant physical observables. This model is capable of accommodating both NO and IO cases of neutrino masses. In the following, we focus primarily on the NO scenario, which yields a better fit to the data.

For NO, the minimum of the $\chi^2$ function is found to be $\chi^2_{\text{min}} = 0.3758$, indicating an agreement with the experimental data. The best-fit values of the model's input parameters and the corresponding physical observables are
{\allowdisplaybreaks
  \begin{align}
    &\langle \tau \rangle=0.019+1.9268i,\quad \tilde{\beta}_{\mathrm{CL}} = 1.2995,\quad \tilde{\gamma}_{\mathrm{CL}}=0.0002,\nonumber\\ 
    &\tilde{\beta}_{S} = 2.9998,\quad n=0.01247~\mathrm{eV},\nonumber\\
    &m_e/m_{\mu}=0.004737,\quad m_{\mu}/m_{\tau}=0.05883, \nonumber\\
    &\sin^2\theta _{12}=0.3051,\quad \sin^2\theta _{13}=0.02185,\quad \sin^2\theta _{23}=0.4645,\label{eq:bf-modelA-NO}\\
    &\delta_{\mathrm{CP}}=0.7406\pi,\quad  \eta_1=1.232\pi,\quad  \eta_2=1.701\pi,\nonumber\\
    &m_1=0.018962~\mathrm{eV},\quad  m_2=0.02084~ \mathrm{eV},\quad  m_3=0.05362 ~\mathrm{eV},\nonumber\\
    &\sum m_i=0.09342~\mathrm{eV},\quad  m_{\beta}=0.02094~ \mathrm{eV},\quad  m_{\beta\beta}=0.008069~\mathrm{eV}.\nonumber
    \end{align}
}
Here, $m_{\beta}$ denotes the effective electron neutrino mass accessible in kinematic experiments such as KATRIN, which currently provides an upper limit of $m_{\beta} < 0.45$ eV at $90\%$ confidence level (C.L.)~\cite{Katrin:2024tvg}. The quantity $m_{\beta\beta}$ represents the effective Majorana neutrino mass that contributes to neutrinoless double-beta decay ($0\nu\beta\beta$); the KamLAND-Zen experiment reports a range of $m_{\beta\beta} < 0.028$–$0.122$ eV depending on the results from calculation of the nuclear matrix element~\cite{KamLAND-Zen:2024eml}. Moreover, cosmological observations from the Planck Collaboration impose an upper bound on the sum of neutrino masses, $\sum m_i < 0.12$ eV at $95\%$ confidence level~\cite{Vagnozzi:2017ovm,Planck:2018vyg}. All predictions at the best-fit point comfortably satisfy these experimental constraints. The combination of DESI baryon acoustic oscillation (BAO) measurements with CMB data from the Planck satellite and the Atacama Cosmology Telescope set a particularly tight limit on the sum of neutrino masses, $\sum m_\nu < 0.072$ eV at 95\% C.L. within the $\Lambda$CDM framework~\cite{DESI:2024mwx}, which lies below our best-fit values. Nonetheless, for some models, such as model B (NO) and model C (NO), the predicted values also fall below this most stringent limit. However, this bound is model dependent. In extended cosmologies, such as the dynamical dark energy model $w_0w_a$CDM, the limit relaxes to approximately $0.177$ eV~\cite{Chebat:2025kes}, well above our predictions.
\begin{figure}[h!]
	\centering
	\includegraphics[width=1\textwidth]{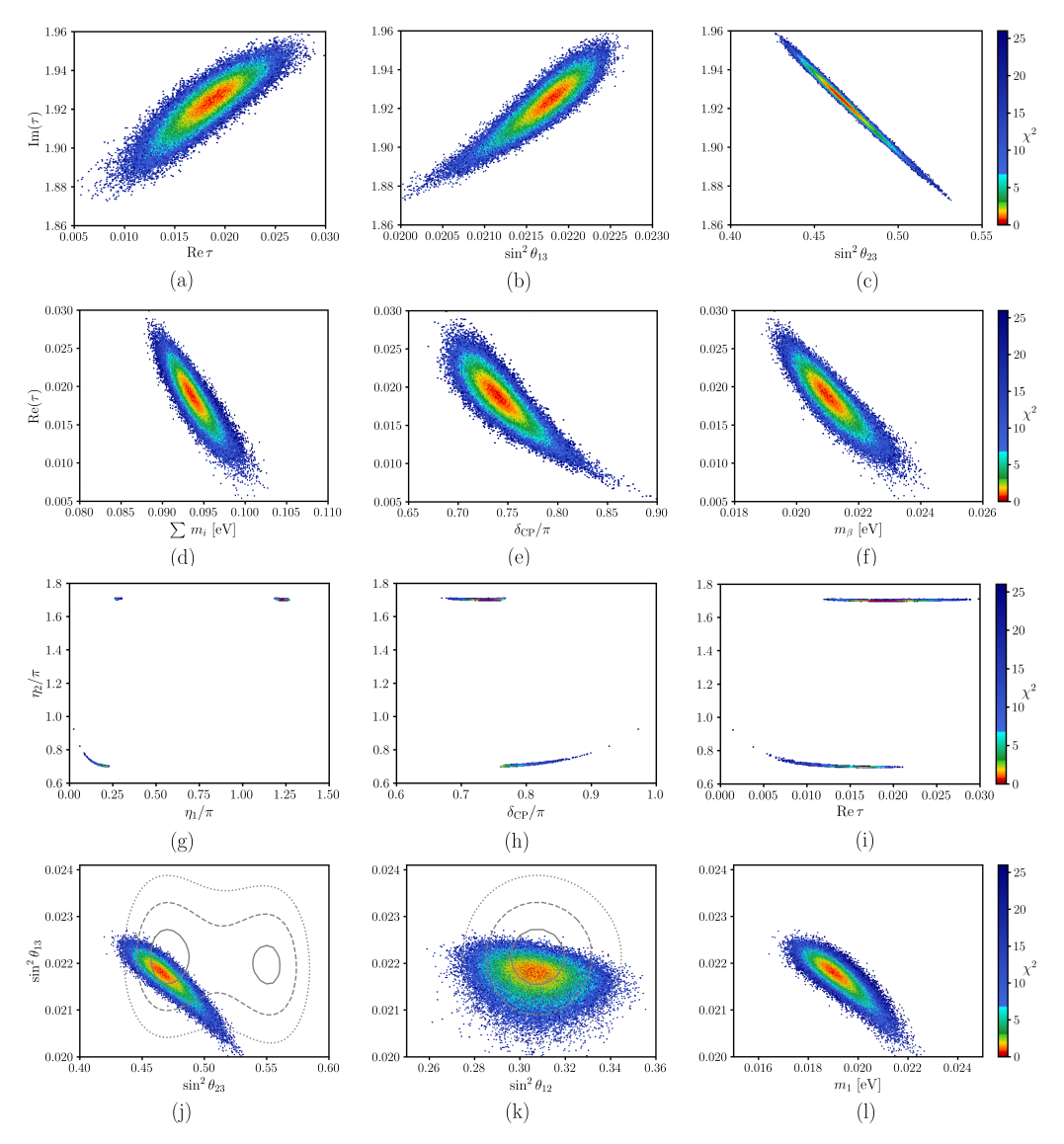}
	\caption{Results of the parameter scan for model A in NO scenario. The panels show: (a) the allowed region of the modulus field $\tau$; correlations between $\mathrm{Im}(\tau)$ and (b) $\sin^2\theta_{13}$, (c) $\sin^2\theta_{23}$; correlations between $\mathrm{Re}(\tau)$ and (d) the total neutrino mass $\sum m_i$, (e) the Dirac CP phase $\delta_{\mathrm{CP}}$, (f) the effective electron neutrino mass $m_{\beta}$; correlations involving the CP-violating phases: (g) $\eta_2$ vs. $\eta_1$, (h) $\eta_2$ vs. $\delta_{\mathrm{CP}}$, (i) $\eta_2$ vs. $\mathrm{Re}(\tau)$; and correlations among observables: (j) $\sin^2\theta_{13}$ vs. $\sin^2\theta_{23}$, (k) $\sin^2\theta_{13}$ vs. $\sin^2\theta_{12}$, (l) $\sin^2\theta_{13}$ vs. $m_1$. The color scale indicates the $\chi^2$ value, with brighter (red) regions corresponding to better agreement with experimental data.}
	\label{fig:m1no}
\end{figure}

Figure~\ref{fig:m1no} presents the results of the parameter scan for model A in the NO scenario, illustrating the correlations among various physical quantities and the modulus field $\tau$. Figure~\ref{fig:m1no}(a) shows the allowed region of $\langle \tau \rangle$ in the complex plane. The imaginary part $\mathrm{Im}(\tau)$ is preferentially large, concentrated around $1.87$–$1.96$, while the real part $\mathrm{Re}(\tau)$ remains close to zero, typically below $0.03$. This indicates that the model favors modulus values within the fundamental domain but away from the boundary where CP would be preserved.

Figures~\ref{fig:m1no}(b) and~\ref{fig:m1no}(c)  depict correlations between $\mathrm{Im}(\tau)$ and the neutrino mixing angles. There is a clear positive correlation between $\mathrm{Im}(\tau)$ and $\sin^2\theta_{13}$, with predicted values of $\sin^2\theta_{13}$ spanning the range $0.0200$–$0.0228$, fully consistent with the current $3\sigma$ experimental range. More pronounced is the strong anticorrelation between $\mathrm{Im}(\tau)$ and $\sin^2\theta_{23}$: as $\mathrm{Im}(\tau)$ increases, $\sin^2\theta_{23}$ decreases monotonically. The predicted range for $\sin^2\theta_{23}$ is $0.43$–$0.53$, covering the entire $3\sigma$ allowed interval. Notably, the majority of the viable parameter space yields $\sin^2\theta_{23} < 0.5$ (i.e., $\theta_{23} < 45^\circ$), with only a small region corresponding to larger $\chi^2$ values giving $\theta_{23} > 45^\circ$. This indicates that model A favors the lower octant of the atmospheric mixing angle. These predictions may provide useful references for the ongoing neutrino experiments.

Regarding mass scales and CP phases, Figs.~\ref{fig:m1no}(d)–~\ref{fig:m1no}(f) demonstrate that the real part of the modulus, $\mathrm{Re}(\tau)$, plays an important role in determining the absolute neutrino mass scale and CP-violating observables. All three quantities—the total neutrino mass $\sum m_i$, the Dirac CP phase $\delta_{\mathrm{CP}}$, and the effective electron neutrino mass $m_{\beta}$—exhibit a similar decreasing trend as $\mathrm{Re}(\tau)$ increases. The total neutrino mass varies within the narrow range $0.0875$–$0.1025$ eV, safely below the cosmological upper bound of $0.12$ eV. The Dirac CP phase is predicted to lie between $0.67\pi$ and $0.9\pi$, with the best-fit value at $0.74\pi$, consistent with current global fit indications. The effective electron neutrino mass $m_{\beta}$ ranges from $19.5$ meV to $24.5$ meV, well below the KATRIN sensitivity.

For Majorana phases, Figs.~\ref{fig:m1no}(g)–~\ref{fig:m1no}(i) show the correlations involving Majorana phases $\eta_1$ and $\eta_2$. A notable feature is the clustering of points in two distinct regions. When $\eta_2 \approx 1.7\pi$, the phase $\eta_1$ concentrates around $0.25\pi$ or $1.25\pi$. In contrast, when $\eta_1 < 0.25\pi$, the phase $\eta_2$ varies in the range $0.7\pi$–$0.8\pi$, with a distribution pattern similar to that of $\delta_{\mathrm{CP}}$. As $\mathrm{Re}(\tau)$ approaches zero, these points tend toward $\pi$. This feature comes from the constraints of gCP symmetry. When the modulus of $\tau$ takes  VEV near the boundary of the fundamental domain, symmetry is preserved and the CP phases approach $\pi$ or vanish.

Finally, Figs.~\ref{fig:m1no}(j)–~\ref{fig:m1no}(l) illustrate correlations among other directly measurable quantities. The correlation between $\sin^2\theta_{13}$ and $\sin^2\theta_{23}$ shows that the model predicts a slightly narrower range for $\sin^2\theta_{13}$ when $\sin^2\theta_{23}$ is in its lower octant. The correlation between $\sin^2\theta_{13}$ and $\sin^2\theta_{12}$ is relatively weak, with $\sin^2\theta_{12}$ confined to its experimentally allowed range. The panel showing $\sin^2\theta_{13}$ versus the lightest neutrino mass $m_1$ indicates that $m_1$ is typically around $0.019$ eV, with a slight negative correlation with $\sin^2\theta_{13}$.

\begin{figure}[h!]
	\centering
	\includegraphics[width=1\textwidth]{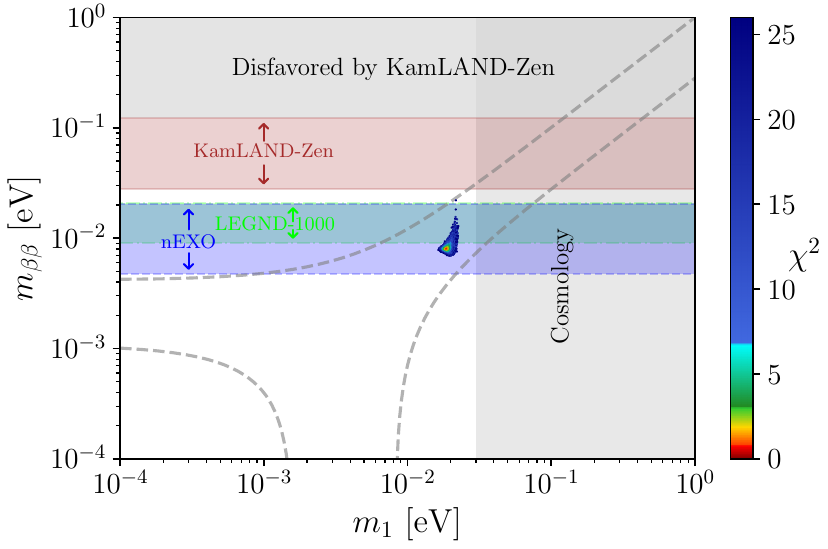}
	\caption{Prediction of model A for the effective Majorana neutrino mass $m_{\beta\beta}$ as a function of the lightest neutrino mass $m_{1}$ in NO scenario. The gray dashed area indicates parameter space favored by NO. Experimental constraints include the KamLAND-Zen exclusion limit (brown band, $m_{\beta\beta} < 0.028$–$0.122~\mathrm{eV}$~\cite{KamLAND-Zen:2024eml}), the target sensitivities of next-generation $0\nu\beta\beta$ experiments LEGEND-1000 (green band, $0.009$–$0.021~\mathrm{eV}$~\cite{LEGEND:2021bnm}) and nEXO (blue band, $0.0047$–$0.0203~\mathrm{eV}$~\cite{nEXO:2021ujk}), as well as the cosmological upper bound $m_{1} \gtrsim 0.037~\mathrm{eV}$  (vertical gray band) derived from $\sum m_i < 0.12$ eV~\cite{GAMBITCosmologyWorkgroup:2020rmf}.}
	\label{fig:m1no_0vbb}
\end{figure}

As for implications to $0\nu\beta\beta$ decay, Fig.~\ref{fig:m1no_0vbb} displays the prediction of model A for the effective Majorana neutrino mass $m_{\beta\beta}$ as a function of the lightest neutrino mass $m_1$ in NO scenario. The color-shaded region represents the model's allowed parameter space, which forms a well-defined band characteristic of the NO spectrum. The best-fit point, marked implicitly within this band, yields $m_{\beta\beta} = 0.00807$ eV, which is below the current KamLAND-Zen exclusion limit.
Model A predicts $m_{\beta\beta}$ to lie within the range $0.0118$–$0.0182$ eV across the viable parameter space. This interval is particularly significant as it falls entirely within the projected sensitivity reach of next-generation $0\nu\beta\beta$ experiments. The LEGEND-1000 Collaboration aims to probe effective Majorana masses down to $0.009$–$0.021$ eV~\cite{LEGEND:2021bnm}, while nEXO targets a sensitivity of $0.0047$–$0.0203$ eV~\cite{nEXO:2021ujk}. The predicted range of model A therefore lies squarely within the discovery potential of these forthcoming experiments, offering a concrete and testable target for experimental investigation.

Regarding the lightest neutrino mass, model A predicts that $m_1$ varies within $0.0164$–$0.0226$ eV. This range is certainly below the cosmological upper bound derived from Planck data, which excludes $m_1 \gtrsim 0.037$ eV at $95\%$ confidence level when translated from the constraint $\sum m_i < 0.12$ eV~\cite{Vagnozzi:2017ovm,Planck:2018vyg, GAMBITCosmologyWorkgroup:2020rmf}. All viable points in model A satisfy this constraint. 

As previously noted in the discussion of Fig.~\ref{fig:m1no}, there exists a clear correlation between the lightest neutrino mass $m_1$ and the mixing angle $\sin^2\theta_{13}$ (see Fig.~\ref{fig:m1no}(l)). This correlation arises from the underlying modular symmetry structure and provides an additional consistency check between different classes of observables. Specifically, points with smaller $m_1$ tend to correspond to slightly larger values of $\sin^2\theta_{13}$, though both quantities remain within their experimentally allowed ranges.

Based on these results of NO scenario, model A exhibits good agreement with current neutrino oscillation data while making precise and testable predictions for future experiments. The model favors the normal mass ordering and the lower octant of the atmospheric mixing angle, and predicts a range for the effective Majorana mass $m_{\beta\beta}$ that lies within the reach of next-generation $0\nu\beta\beta$ decay searches. These features position model A as a compelling and phenomenologically viable realization of the non-SUSY modular $A'_5$ inverse seesaw framework. We now turn to the phenomenological analysis of the IO option.

For the IO case, the minimum of the $\chi^2$ function is found to be $\chi^2_{\text{min}} = 2.8725$, which is notably larger than the value obtained for NO. The best-fit values of the model's input parameters and the corresponding physical observables are
{\allowdisplaybreaks
  \begin{align}
    &\langle \tau \rangle=0.48215+0.87702i,\quad \tilde{\beta}_{\mathrm{CL}} = 1.2995,\quad \tilde{\gamma}_{\mathrm{CL}}=65.941,\nonumber\\
    &\tilde{\beta}_{S} = 5.2598,\quad  n=0.001139~\mathrm{eV},\nonumber\\
    &m_e/m_{\mu}=0.004737,\quad m_{\mu}/m_{\tau}=0.05882,\nonumber\\
    &\sin^2\theta _{12}=0.3065,\quad \sin^2\theta _{13}=0.0223,\quad \sin^2\theta _{23}=0.4911,\label{eq:bf-modelA-IO}\\
    &\delta_{\mathrm{CP}}=0.3712\pi,\quad  \eta_1=0.445\pi,\quad  \eta_2=0.7746\pi,\nonumber\\
    &m_1=0.1251~\mathrm{eV},\quad  m_2=0.1254~ \mathrm{eV},\quad  m_3=0.115 ~\mathrm{eV},\nonumber\\
    &\sum m_i=0.3655~\mathrm{eV},\quad  m_{\beta}=0.1253~ \mathrm{eV},\quad  m_{\beta\beta}=0.07649~\mathrm{eV}.\nonumber
    \end{align}
}

The main reason for the larger $\chi^2$ value is due to the prediction of the atmospheric mixing angle. Although other physical quantities are in good agreement with the experimental results, it is not easy to achieve a viable range of $\sin^2\theta_{23}$, which satisfies $\sin^2\theta_{23} = 0.4911$ and lies within the $3\sigma$ experimental range. The model exhibits a preference for $\sin^2\theta_{23} < 0.5$ (i.e., $\theta_{23} < 45^\circ$), consistent with the trend observed in the NO scenario. The charged lepton mass ratios and the other mixing angles are in agreement with the experimental data.
  
A more serious concern is the predicted absolute neutrino mass scale. The sum of neutrino masses, $\sum m_i = 0.3655$ eV, significantly exceeds the stringent upper bound from cosmological observations, $\sum m_i < 0.12$ eV at the $95\%$ confidence level from Planck. But if one adopts a more conservative bound, the mass sum remains under the looser limit of $0.537$ eV sometimes quoted in older analyses. This tension indicates that the IO scenario of model A is disfavored by current cosmological data, and we therefore regard it as a less viable possibility compared to the NO case. The effective electron neutrino mass, accessible in kinematic experiments such as KATRIN, is predicted to be $m_{\beta} \in [0.121, 0.131]~\mathrm{eV}$, with a best-fit value of $0.1253$ eV. This range is below the current KATRIN upper limit of $m_{\beta} < 0.45$ eV~\cite{Katrin:2024tvg} and within the projected sensitivity of future tritium beta-decay experiments.
\begin{figure}[h!]
	\centering
	\includegraphics[width=1\textwidth]{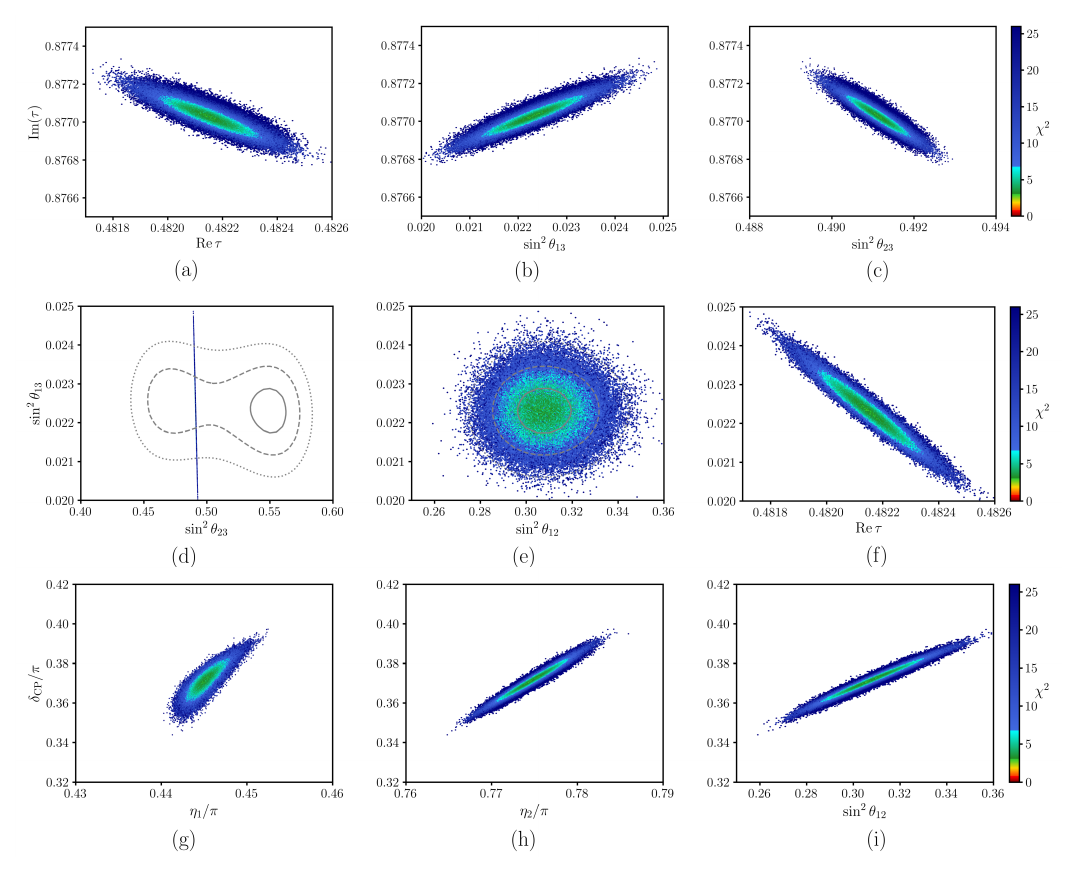}
	\caption{Results of the parameter scan for model A in the IO scenario. The panels show: (a) the allowed region of the modulus field $\tau$; correlations between $\mathrm{Im}(\tau)$ and (b) $\sin^2\theta_{13}$, (c) $\sin^2\theta_{23}$; correlations between $\sin^2\theta_{13}$ and (d) $\sin^2\theta_{23}$, (e) $\sin^2\theta_{12}$, (f) $\mathrm{Re}(\tau)$; and correlations involving the CP phases: (g) $\delta_{\mathrm{CP}}$ vs. $\eta_1$, (h) $\delta_{\mathrm{CP}}$ vs. $\eta_2$, (i) $\delta_{\mathrm{CP}}$ vs. $\sin^2\theta_{12}$. The color scale indicates the $\chi^2$ value, with brighter  regions corresponding to better agreement with experimental data.}
	\label{fig:m1io}
\end{figure}

As illustrated in Fig.~\ref{fig:m1io} the results of the parameter scan for model A in the IO scenario, the relations among the modulus field and mixing angles are in order. Figure~\ref{fig:m1io}(a) shows the allowed region of the modulus field $\tau$ in the complex plane. Unlike the NO case where $\mathrm{Re}(\tau)$ was near zero, here the modulus is concentrated around a fixed point near $\tau \approx \omega = e^{2\pi i/3}$, specifically $\mathrm{Re}(\tau) \approx 0.48$ and $\mathrm{Im}(\tau) \approx 0.88$. This indicates that the IO scenario selects a different region of the fundamental domain. Figures \ref{fig:m1io}(b) and \ref{fig:m1io}(c) show the correlations between $\mathrm{Im}(\tau)$ and the mixing angles. There is a positive correlation between $\mathrm{Im}(\tau)$ and $\sin^2\theta_{13}$, with predicted values that span the range $0.020$–$0.025$, consistent with the $3\sigma$ experimental interval. More notably, $\sin^2\theta_{23}$ exhibits a narrow allowed range, $0.489$–$0.493$, which lies entirely in the lower octant ($\theta_{23} < 45^\circ$). This sharp prediction is a distinctive feature of the IO scenario in model A and could be tested by future precision measurements of the atmospheric mixing angle.

For the purpose of showing the correlations among the mixing angles and $\mathrm{Re}(\tau)$, Figs.~\ref{fig:m1io}(d)–~\ref{fig:m1io}(f)  display the relationship between these parameters and the value of the reactor mixing angle $\sin^2\theta_{13}$. The correlation with $\sin^2\theta_{23}$ (see Fig~\ref{fig:m1io}(d)) is weak, which can be understood with the results in Figs~\ref{fig:m1io}(b),~\ref{fig:m1io}(c) and~\ref{fig:m1io}(f). As shown, there are opposite correlations of $\sin^2\theta_{13}$ and $\sin^2\theta_{23}$ with respect to $\tau$, and both angles are restricted to very small ranges. When these correlations are combined into Fig.~\ref{fig:m1io}(d), the opposite behavior almost cancels and appears as a sharp line shape that is shown there. The correlation of $\sin^2\theta_{13}$ with $\sin^2\theta_{12}$ (Fig.~\ref{fig:m1io}(e)) shows that the latter is also confined to a rather narrow band around $0.27$–$0.35$, consistent with the experimental data. Fig.~\ref{fig:m1io}(f) illustrates that $\sin^2\theta_{13}$ decreases slightly with $\mathrm{Re}(\tau)$, though the variation is small.

Given that CP phases, Figs.~\ref{fig:m1io}(g)–~\ref{fig:m1io}(i) illustrate the behavior of the CP phases and their correlations. Unlike the NO scenario, which exhibited two distinct clustered regions for the Majorana phases, the IO scenario yields a single region for each phase. The Dirac CP phase $\delta_{\mathrm{CP}}$ is predicted to be within the range $0.34\pi$–$0.40\pi$, with a best-fit value of $0.371\pi$. The Majorana phases are similarly constrained, $\eta_1$ varies between $0.44\pi$ and $0.453\pi$, while $\eta_2$ ranges from $0.765\pi$ to $0.786\pi$. All three CP phases exhibit positive correlations with each other and with $\sin^2\theta_{12}$, as shown in Figs.~\ref{fig:m1io}(g)–~\ref{fig:m1io}(i). These correlations reflect the underlying modular symmetry structure and provide a set of interconnected predictions that could be tested in future experiments capable of measuring leptonic CP violation.

\begin{figure}[h!]
	\centering
	\includegraphics[width=1\textwidth]{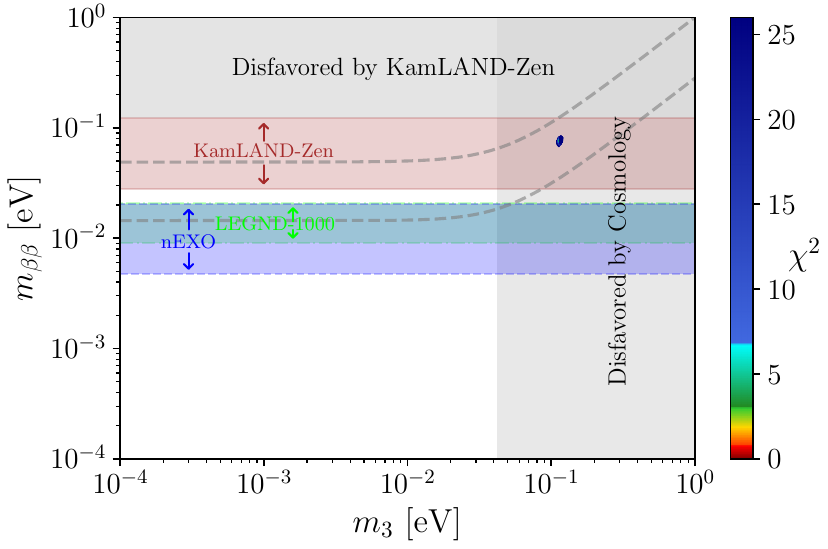}
	\caption{Prediction of model A for the effective Majorana neutrino mass $m_{\beta\beta}$ as a function of the lightest neutrino mass $m_3$ in the IO scenario. The gray dashed region indicates the entire parameter space, wherein the bold pointlike colored area is  the parameter space favored by the model. The vertical gray band shows the cosmological upper bound $m_3 \lesssim 0.042$ eV derived from $\sum m_i < 0.12$ eV~\cite{Vagnozzi:2017ovm,Planck:2018vyg, GAMBITCosmologyWorkgroup:2020rmf}. The predicted values of $m_3$ lie above this bound, indicating tension with cosmological data. Experimental constraints and sensitivities are as in Fig.~\ref{fig:m1no_0vbb}.}
	\label{fig:m1io-0vbb}
\end{figure}

In consideration of $0\nu\beta\beta$ decay, unlike the NO scenario, where the Majorana phases exhibited two distinct clustered regions, the IO case yields a single region for each phase, as discussed previously. This has implications for the predictions of $0\nu\beta\beta$ decay. Figure~\ref{fig:m1io-0vbb} presents the model's predictions for the effective Majorana mass $m_{\beta\beta}$ as a function of the lightest neutrino mass $m_3$. The gray dashed region indicates the parameter space for IO. The effective Majorana mass is predicted to be within the range $m_{\beta\beta} \in [0.068, 0.084]~\mathrm{eV}$, with a best-fit value of $0.0765$ eV. This interval falls entirely within the current exclusion limits of the KamLAND-Zen experiment, which limits $m_{\beta\beta} < 0.028$–$0.122$ eV depending on nuclear matrix element uncertainties. The entire predicted range exceeds the projected sensitivity of next-generation $0\nu\beta\beta$ experiments. The LEGEND-1000 aims to probe the effective Majorana masses to $0.009$–$0.021$ eV, while nEXO targets a sensitivity of $0.0047$–$0.0203$ eV. The predictions of model A in the IO scenario are well within the exclusion precisions of these forthcoming experiments, offering a testable target. 

More importantly, the predicted range lies within the disfavored region of cosmological observation. An observation from Fig.~\ref{fig:m1io-0vbb} is the predicted range of the lightest neutrino mass $m_3$. Model A predicts $m_3$ to lie in the range $0.111$–$0.121$ eV. However, as indicated by the vertical gray band, cosmological observations impose a strict upper bound on the sum of neutrino masses, $\sum m_i < 0.12$ eV, which translates to an upper bound on $m_3$ in the IO scenario of approximately $m_3 \lesssim 0.042$ eV when combined with the measured mass-squared differences. Clearly, the predicted value of $m_3$ exceeds this individual bound, but the larger issue, as noted above, is that the total mass sum $\sum m_i \approx 0.365$ eV exceeds the cosmological limit by a factor of three. This is a major factor that disfavors the IO scenario. Nevertheless, as noted previously, other cosmological bounds are model dependent. Extensions of the $\Lambda$CDM framework can relax the constraints on $\sum m_i$. For example, in $w_0w_a$CDM, the 95\% upper limit increases to $\sim 0.177$~eV, while in non-flat $\Lambda$CDM ($\Omega_K\neq0$) it rises to $\sim 0.085$~eV~\cite{Chebat:2025kes}. More exotic scenarios, such as interacting dark energy or neutrinos with time-varying masses~\cite{Lorenz:2018fzb}, can further weaken the bound (e.g., to several eV). Therefore, although the IO predictions in our models are disfavored by $\Lambda$CDM cosmology, they are not excluded in extended cosmological frameworks. We emphasize, however, that the NO remains preferred by the oscillation data and also satisfies the most stringent $\Lambda$CDM bounds.

To conclude the analysis of model A, we find that the model can accommodate both NO and IO cases, though with different levels of agreement with the experimental data. The IO case yields a somewhat poorer fit to the data compared to the NO scenario. The IO scenario predicts a much larger neutrino mass sum, which is strongly disfavored by cosmological observations. Nevertheless, its prediction for $m_{\beta\beta}$ is within the reach of future experiments. The model exhibits correlations among mixing angles and CP phases in both orderings, providing multiple interconnected predictions that can be tested in future neutrino facilities. Given the significant tension with cosmological data in the IO scenario, we consider the NO to be the phenomenologically preferred realization of model A. We now proceed to the analysis of model B.

\subsection{Phenomenology of model B}
\label{sec:p-mb}

Although the structure of model B is similar to that of model A, with the same representation assignments for the fields, the significant difference lies in the modular weights assigned to each field (see Tables~\ref{tab:field2} and~\ref{tab:weight}). This seemingly minor modification leads to significantly different phenomenological predictions, as we shall demonstrate. Similar to model A, the current experimental data favor the NO scenario, and we therefore focus our analysis on this case.

Taking into account the best-fit points, for model B with NO case, the minimum of $\chi^2_{\text{min}} = 0.8016$, indicating a good fit to the experimental data, albeit slightly larger than the value obtained for model A. The best-fit values of the model's input parameters and the derived physical observables are
{\allowdisplaybreaks
  \begin{align}
    &\langle \tau \rangle=0.03226+1.0416i,\quad \tilde{\beta}_{\mathrm{CL}} = 0.7696,\quad \tilde{\gamma}_{\mathrm{CL}}=0.000544,\nonumber\\
    &\tilde{\beta}_{S} = 0.02242,\quad n=0.43926~\mathrm{eV},\nonumber\\
    &m_e/m_{\mu}=0.004738,\quad m_{\mu}/m_{\tau}=0.05882,\nonumber\\
    &\sin^2\theta _{12}=0.3051,\quad \sin^2\theta _{13}=0.02222,\quad \sin^2\theta _{23}=0.5548, \label{eq:bf-modelB-NO}\\
    &\delta_{\mathrm{CP}}=0.6265\pi,\quad  \eta_1=1.594\pi,\quad  \eta_2=0.944\pi,\nonumber\\
    &m_1=0.001874~\mathrm{eV},\quad  m_2=0.008807~ \mathrm{eV},\quad  m_3=0.05012 ~\mathrm{eV},\nonumber\\
    &\sum m_i=0.0608~\mathrm{eV},\quad  m_{\beta}=0.00911~ \mathrm{eV},\quad  m_{\beta\beta}=0.001015~\mathrm{eV}.\nonumber
    \end{align}
}

A salient feature of model B is the very small values predicted for the absolute neutrino mass scale. The sum of neutrino masses, $\sum m_i = 0.0608$ eV, is well below the current cosmological upper bound of $0.12$ eV. The effective electron neutrino mass, $m_{\beta} = 0.00911$ eV, is much below the sensitivity of current kinematic experiments such as KATRIN. In particular, the effective Majorana mass for $0\nu\beta\beta$ reads $m_{\beta\beta} = 0.001015$ eV, which is more than an order of magnitude lower than the projected reach of next-generation experiments like LEGEND-1000 and nEXO. This presents a significant challenge for experimental verification.

\begin{figure}[h!]
	\centering
	\includegraphics[width=1\textwidth]{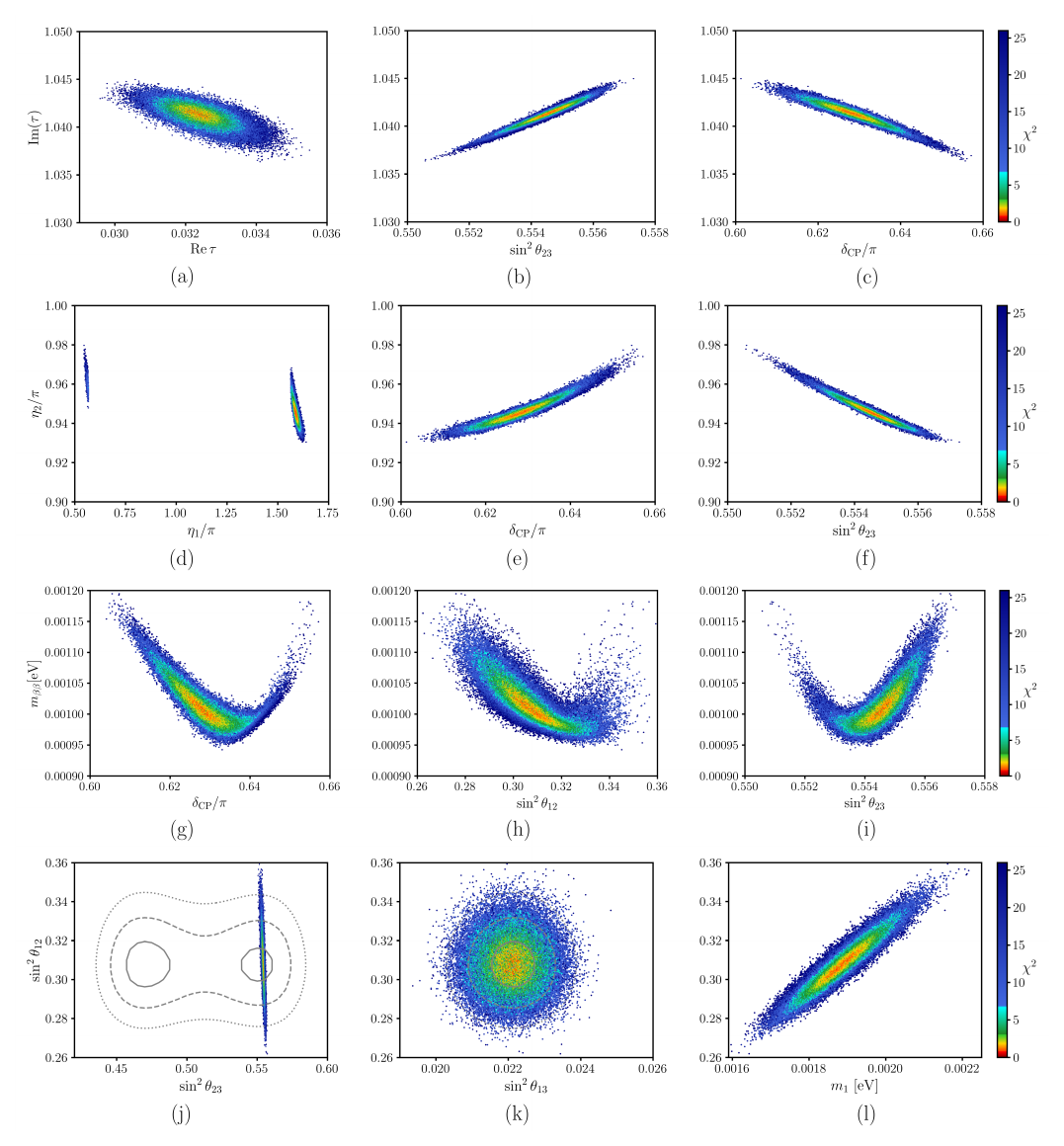}
	\caption{Results of the parameter scan for model B in the NO scenario. The panels show: (a) the allowed region of the modulus field $\tau$; correlations between $\mathrm{Im}(\tau)$ and (b) $\sin^2\theta_{23}$, (c) $\delta_{\mathrm{CP}}$; correlations involving the Majorana phase $\eta_2$ with (d) $\eta_1$, (e) $\delta_{\mathrm{CP}}$, (f) $\sin^2\theta_{23}$; correlations of the effective Majorana mass $m_{\beta\beta}$ with (g) $\delta_{\mathrm{CP}}$, (h) $\sin^2\theta_{12}$, (i) $\sin^2\theta_{23}$; and correlations among observables: (j) $\sin^2\theta_{12}$ vs. $\sin^2\theta_{23}$, (k) $\sin^2\theta_{12}$ vs. $\sin^2\theta_{13}$, (l) $\sin^2\theta_{12}$ vs. $m_1$. }
	\label{fig:m2no}
\end{figure}

Figure~\ref{fig:m2no} presents the results of the parameter scan for model B. Figure~\ref{fig:m2no}(a) shows the allowed region of the modulus $\tau$ in the complex plane. In contrast to model A, where $\mathrm{Im}(\tau)$ was concentrated around $1.9$–$2.0$, here $\mathrm{Im}(\tau)$ is close to unity, with the entire viable parameter space concentrated near the fixed point $\tau = i$. The real part $\mathrm{Re}(\tau)$ remains near zero, similar to model A. Figures~\ref{fig:m2no}(b) and~\ref{fig:m2no}(c) reveal correlations between $\mathrm{Im}(\tau)$ and the atmospheric mixing angle $\sin^2\theta_{23}$, as well as the Dirac CP phase $\delta_{\mathrm{CP}}$. As $\mathrm{Im}(\tau)$ increases, $\sin^2\theta_{23}$ also increases, while $\delta_{\mathrm{CP}}$ decreases. However, due to the narrow range of $\mathrm{Im}(\tau)$ (approximately $1.038$–$1.044$), the actual variation spans of these parameters are relatively limited. Specifically, $\sin^2\theta_{23}$ ranges from $0.55$ to $0.558$, placing it in the upper octant. This is in contrast to model A, which favored the lower octant. The Dirac CP phase varies within the interval $0.60\pi$ to $0.66\pi$, with the best-fit value at $0.6265\pi$.

With a view to CP-violating phases, Figs~\ref{fig:m2no}(d)–~\ref{fig:m2no}(f) exhibit the correlations among them. The Majorana phase $\eta_2$ is confined to a narrow interval, $(0.93$–$0.98)\pi$, with the best-fit value near $0.944\pi$ . More interesting is the behavior of $\eta_1$, which shows a bimodal distribution: one cluster around $0.55\pi$ and another around $1.6\pi$. The region with the minimum $\chi^2$ value corresponds to the latter, near $1.6\pi$, which is where the best-fit point is located. As $\eta_2$ increases, a monotonic trend is observed, the Dirac CP phase $\delta_{\mathrm{CP}}$ increases, while $\sin^2\theta_{23}$ decreases, but still lies above the value at the octant boundary. These correlations provide related predictions that could, in principle, be tested if multiple CP-violating observables become accessible in future experiments.

\begin{figure}[h!]
	\centering
	\includegraphics[width=1\textwidth]{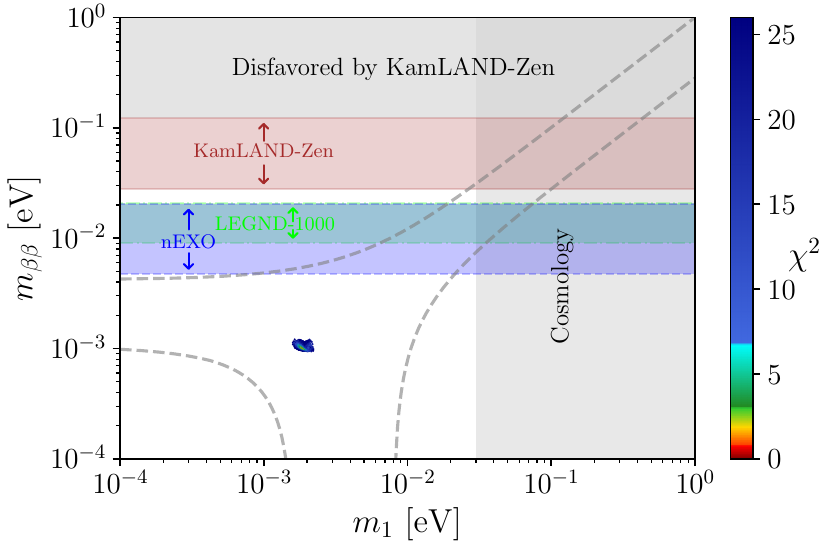}
	\caption{Prediction of model B for $m_{\beta\beta}$ as a function of the lightest neutrino mass $m_1$ in NO scenario. The predicted range, $m_{\beta\beta} \approx 0.95$–$1.20$ meV, lies well below the sensitivity thresholds of next-generation $0\nu\beta\beta$ experiments: LEGEND-1000 (green band, $0.009$–$0.021~\mathrm{eV}$~\cite{LEGEND:2021bnm}) and nEXO (blue band, $0.0047$–$0.0203$ eV~\cite{nEXO:2021ujk}), as well as the current KamLAND-Zen exclusion limit (brown band, $0.028$–$0.122$ eV~\cite{KamLAND-Zen:2024eml}). The vertical gray band indicates the cosmological upper bound $m_1 \lesssim 0.037$ eV derived from $\sum m_i < 0.12$ eV~\cite{Vagnozzi:2017ovm,Planck:2018vyg, GAMBITCosmologyWorkgroup:2020rmf}, which is satisfied by whole range of points.}
	\label{fig:m2no-0vbb}
\end{figure}

Projecting the effective Majorana mass $m_{\beta\beta}$ on the oscillation parameters,Figs.~\ref{fig:m2no}(g)–~\ref{fig:m2no}(i), in turn, illustrate its correlations with $\delta_{\mathrm{CP}}$, $\sin^2\theta_{12}$, and $\sin^2\theta_{23}$. The value of $m_{\beta\beta}$ is confined to very narrow range, $0.95$–$1.2$ meV, which is more than an order of magnitude smaller than the projected sensitivities of next-generation $0\nu\beta\beta$ experiments. This is further displayed in Fig.~\ref{fig:m2no-0vbb}, where the predicted region for $m_{\beta\beta}$ lies far below the exclusion limits of KamLAND-Zen and the target sensitivities of LEGEND-1000 and nEXO. This imposes a stringent challenge for experimental detection: if model B is realized in nature, $0\nu\beta\beta$ decay will likely remain unobservable for the foreseeable future.

Moving to mixing angle correlations, Figs.~\ref{fig:m2no}(j)–~\ref{fig:m2no}(l) show distributions of points among the mixing angles and the lightest neutrino mass $m_1$. The solar mixing angle $\sin^2\theta_{12}$ is confined to a narrow range around $0.26$–$0.36$, consistent with experimental data. Figure~\ref{fig:m2no}(j) shows a weak positive correlation between $\sin^2\theta_{12}$ and $\sin^2\theta_{23}$, while Fig.~\ref{fig:m2no}(k) indicates that $\sin^2\theta_{13}$ varies across the allowed parameter space. Figure~\ref{fig:m2no}(l) demonstrates that $m_1$ is predicted to be extremely small, in the range $0.0016$–$0.0022$ eV, with positive correlation with $\sin^2\theta_{12}$.

We now examine the predictions of model B for $0\nu\beta\beta$, which reveal one of the most pronounced differences from model A. As discussed in the previous section, the effective Majorana mass $m_{\beta\beta}$ at the best-fit point is already extremely small, $m_{\beta\beta} = 0.001015$ eV. The full parameter scan, shown in Fig.~\ref{fig:m2no-0vbb}, confirms that this is a generic feature of the model. The figure displays the prediction of model B for $m_{\beta\beta}$ as a function of the lightest neutrino mass $m_1$. The range of $m_{\beta\beta}$ is $[0.95, 1.20]$ meV, with the best-fit value at $1.015$ meV. This range is more than an order of magnitude smaller than the predictions of model A ($11.8$–$18.2$ meV) and lies far below the reach of any $0\nu\beta\beta$ experiment.
The lightest neutrino mass $m_1$ is predicted to be in the range $1.6$–$2.2$ meV, consistent with the cosmological bound $m_1 \lesssim 37$ meV from the Planck constraint. The vertical gray band indicates this exclusion region, and all viable points in model B satisfy it comfortably. The prediction poses a formidable challenge for $0\nu\beta\beta$ experiments. This is in stark contrast to model A, which places $m_{\beta\beta}$ within the reach of next-generation experiments. 

Examining the IO scenario of model B, it yields a larger $\chi^2$ value compared to the NO case, indicating a poorer fit to the experimental data. The minimum value of the $\chi^2$ function is found to be $\chi^2_{\text{min}} = 2.4783$, which is comparable to the value obtained for model A in the IO scenario. The best-fit values of the model's input parameters and the corresponding physical observables for model B in the IO scenario are the following:
{
\allowdisplaybreaks
  \begin{align}
    &\langle \tau \rangle=0.49684+2.0694i,\quad \tilde{\beta}_{\mathrm{CL}} = 0.8779, \quad\tilde{\gamma}_{\mathrm{CL}}=0.00085,\nonumber\\
    &\tilde{\beta}_{S} = 0.6656,\quad n=0.05075~\mathrm{eV},\nonumber\\
    &m_e/m_{\mu}=0.004737,\quad m_{\mu}/m_{\tau}=0.05882,\nonumber\\
    &\sin^2\theta _{12}=0.3062,\quad \sin^2\theta _{13}=0.02224,\quad \sin^2\theta _{23}=0.5221, \label{eq:bf-modelB-IO}\\
    &\delta_{\mathrm{CP}}=1.542\pi,\quad  \eta_1=0.4033\pi,\quad  \eta_2=1.586\pi,\nonumber\\
    &m_1=0.05372~\mathrm{eV},\quad  m_2=0.05441~ \mathrm{eV},\quad  m_3=0.02195 ~\mathrm{eV},\nonumber\\
    &\sum m_i=0.1301~\mathrm{eV},\quad  m_{\beta}=0.05436~ \mathrm{eV},\quad  m_{\beta\beta}=0.04644~\mathrm{eV}.\nonumber
    \end{align}
}    

Several features are notable. First, the modulus field takes a value with a relatively large imaginary part, $\mathrm{Im}(\tau) \approx 2.07$, placing it near the boundary of the fundamental domain. The real part, $\mathrm{Re}(\tau) \approx 0.497$, is close to the maximum allowed value of $0.5$. Second, consistent with the NO scenario of model B, but in contrast to model A, the atmospheric mixing angle is predicted to lie in the upper octant, with $\sin^2\theta_{23} = 0.5221 > 0.5$. Third, the sum of neutrino masses, $\sum m_i = 0.1301$ eV, exceeds the most stringent cosmological bound of $0.12$ eV but remains within the $0.24$ eV upper limit obtained when BAO data are excluded from the Planck analysis~\cite{Planck:2018vyg}.

\begin{figure}[h!]
	\centering
	\includegraphics[width=1\textwidth]{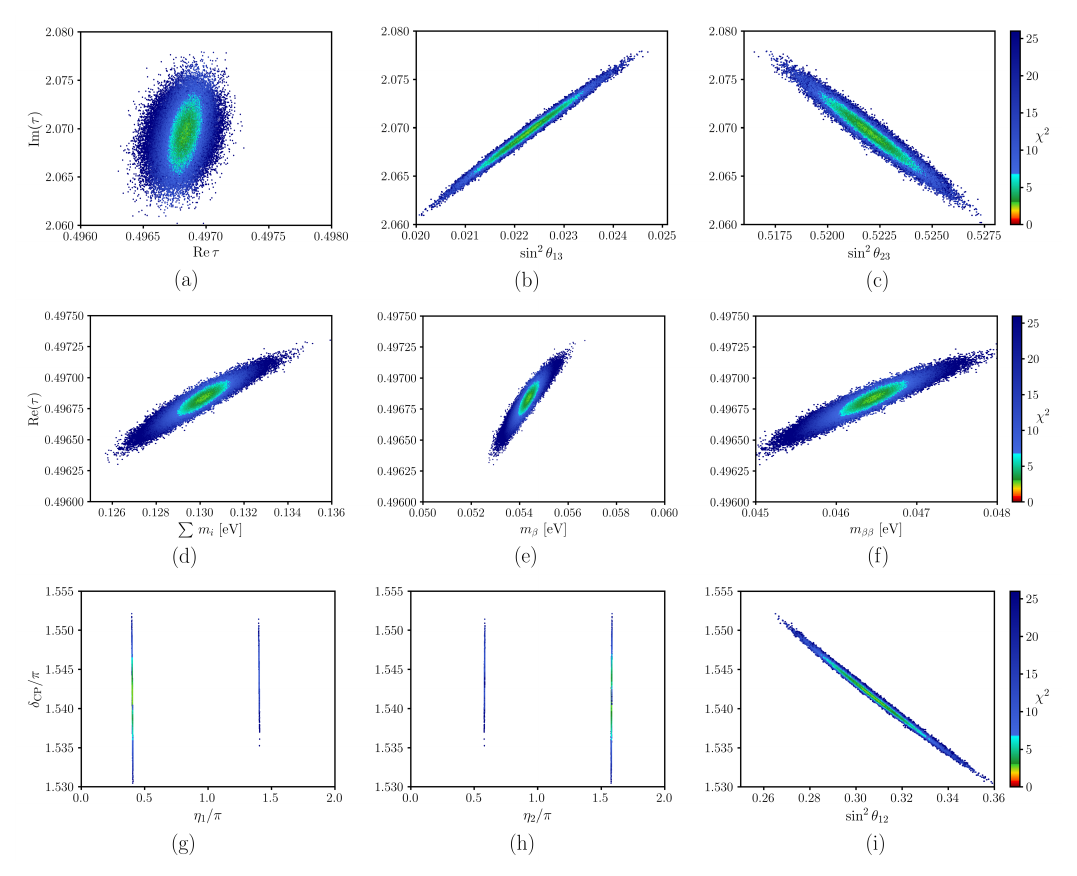}
	\caption{Results of the parameter scan for model B in the IO scenario. The panels show: (a) the allowed region of the modulus field $\tau$; correlations between $\mathrm{Im}(\tau)$ and (b) $\sin^2\theta_{13}$, (c) $\sin^2\theta_{23}$; correlations between $\mathrm{Im}(\tau)$ and (d) the total neutrino mass $\sum m_i$, (e) the effective electron neutrino mass $m_{\beta}$, (f) the effective Majorana mass $m_{\beta\beta}$; and correlations involving the CP phases: (g) $\delta_{\mathrm{CP}}$ vs. $\eta_1$, (h) $\delta_{\mathrm{CP}}$ vs. $\eta_2$, (i) $\delta_{\mathrm{CP}}$ vs. $\sin^2\theta_{12}$. The color scale indicates the $\chi^2$ value.}
	\label{fig:m2io}
\end{figure}

Figure~\ref{fig:m2io} presents the results of the parameter scan for model B in the IO scenario. Figure~\ref{fig:m2io}(a) shows the allowed region of the modulus field $\tau$ in the complex plane. Compared to model A in the IO scenario, the real part of $\tau$ is similar, but the imaginary part is significantly larger, concentrated around $2.06$–$2.08$. This places the modulus closer to the boundary of the fundamental domain. Figures~\ref{fig:m2io}(b) and~\ref{fig:m2io}(c) illustrate the correlations between $\mathrm{Im}(\tau)$ and the mixing angles $\sin^2\theta_{13}$ and $\sin^2\theta_{23}$, respectively. The behavior is qualitatively similar to that observed in model A (IO): $\sin^2\theta_{13}$ increases with $\mathrm{Im}(\tau)$, while $\sin^2\theta_{23}$ decreases. However, the predicted range for $\sin^2\theta_{23}$ is notably narrow, $0.515 \leq \sin^2\theta_{23} \leq 0.5275$, with the best-fit value at $0.5221$. This places the atmospheric mixing angle firmly in the upper octant, consistent with the NO scenario of model B but opposite to the preference of model A. The solar and reactor mixing angles, $\sin^2\theta_{12}$ and $\sin^2\theta_{13}$, span broader ranges but remain consistent within the $3\sigma$ experimental boundaries.

To show the correlation between $\mathrm{Re}(\tau)$ and three masses, $\sum m_i$, $m_{\beta}$ and $m_{\beta\beta}$, Figs.~\ref{fig:m2io}(d)–~\ref{fig:m2io}(f) demonstrate weak positive correlations. The total neutrino mass varies within the range $0.125$ eV $\leq \sum m_i \leq 0.135$ eV. As noted earlier, this exceeds the strictest cosmological bound of $0.12$ eV but remains below the less restrictive limit of $0.24$ eV obtained when BAO data are excluded. The effective electron neutrino mass $m_{\beta}$ ranges from $0.052$ eV to $0.057$ eV, well below the KATRIN upper limit of $0.45$ eV. The effective Majorana mass $m_{\beta\beta}$ is predicted to be in the interval $0.045$ eV $\leq m_{\beta\beta} \leq 0.048$ eV.

Correlations of the CP phases are illustrated in Figs.~\ref{fig:m2io}(g)–~\ref{fig:m2io}(i). The Dirac CP phase $\delta_{\mathrm{CP}}$ is confined to a narrow range, $1.53\pi$ to $1.553\pi$, with the best-fit value at $1.542\pi$. The Majorana phases exhibit highly concentrated distributions. The $\eta_1$ clusters around $0.4\pi$ or $1.4\pi$, while $\eta_2$ clusters around either $0.58\pi$ or $1.58\pi$. This bimodal structure is reminiscent of the behavior observed for $\eta_2$ in the NO scenario of model B, albeit the specific values differ. A negative correlation is observed between $\delta_{\mathrm{CP}}$ and $\sin^2\theta_{12}$, as $\delta_{\mathrm{CP}}$ increases, $\sin^2\theta_{12}$ decreases. 

The similarities between model B and model A in the IO scenario are noteworthy and likely stem from their analogous structural features under the $A^\prime_5$ modular group, particularly the identical representation assignments for the fields. Both models predict narrow ranges for $\sin^2\theta_{23}$, exhibit positive correlations between $\mathrm{Im}(\tau)$ and the mass parameters, and show bimodal distributions for some Majorana phases. However, the different modular weight assignments lead to quantitative differences. Model B predicts a larger $\mathrm{Im}(\tau)$, a slightly higher $\sin^2\theta_{23}$, and a somewhat smaller $m_{\beta\beta}$. These differences underscore the diversity of phenomenological predictions that can arise within the modular symmetry framework.

\begin{figure}[h!]
	\centering
	\includegraphics[width=1\textwidth]{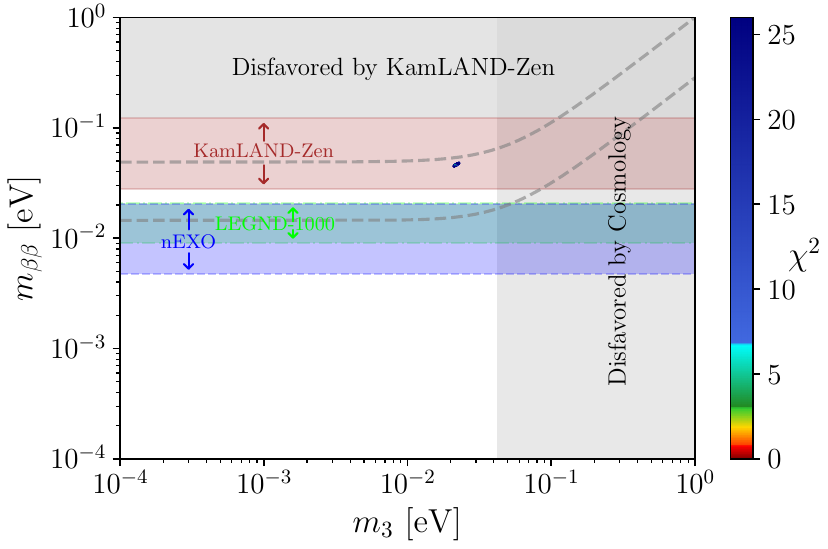}
	\caption{Prediction of model B for the effective Majorana neutrino mass $m_{\beta\beta}$ as a function of the lightest neutrino mass $m_3$ in the IO scenario. The predicted range, $m_{\beta\beta} \approx 0.045$–$0.048$ eV, lies within the current KamLAND-Zen exclusion limit (brown band, $0.028$–$0.122$ eV~\cite{KamLAND-Zen:2024eml}) and partially overlaps with the projected sensitivities of next-generation experiments LEGEND-1000 (green band, $0.009$–$0.021~\mathrm{eV}$~\cite{LEGEND:2021bnm}) and nEXO (blue band, $0.0047$–$0.0203$ eV~\cite{nEXO:2021ujk}). Unlike model A (IO), the lightest neutrino mass $m_3$ lies outside the gray cosmological exclusion band, satisfying the bound $m_3 \lesssim 0.042$ eV derived from $\sum m_i < 0.12$ eV.}
	\label{fig:m2io-0vbb}
\end{figure}
Regarding $0\nu\beta\beta$ decay, Fig.~\ref{fig:m2io-0vbb} displays the prediction of model B with IO case for the effective Majorana neutrino mass $m_{\beta\beta}$ as a function of the lightest neutrino mass $m_3$. The point-shape region represents the model's allowed parameter space, which forms a narrow band with $m_{\beta\beta}$ ranging from $0.045$ eV to $0.048$ eV. This interval falls entirely within the current exclusion limit of the KamLAND-Zen experiment ($0.028$–$0.122$ eV) and well above the projected sensitivities of next-generation $0\nu\beta\beta$ experiments. A difference from model A in the IO scenario concerns the lightest neutrino mass $m_3$. In model B, $m_3$ varies between approximately $0.021$ eV and $0.023$ eV, which lies outside the gray cosmological exclusion band from the Planck bound. This means that, unlike model A, the IO scenario of model B is not immediately ruled out by cosmology.

To close the analysis of model B with the IO scenario, it may be useful to highlight its some notable features. The model provides a moderate fit to the data, with a minimum $\chi^2$ of $2.4783$. The model predicts that the atmospheric mixing angle lies in the upper octant within a narrow range $\sin^2\theta_{23} \in [0.515, 0.5275]$. The modulus field is preferentially localized near the boundary of the fundamental domain. Although the total neutrino mass, $\sum m_i \approx 0.125$–$0.135$ eV, exceeds the most stringent cosmological upper limit, it remains compatible with the less restrictive limit of $0.24$ eV obtained when BAO data are excluded. The effective Majorana mass is predicted to be $m_{\beta\beta} \approx 0.045$–$0.048$ eV, a range that falls within the current KamLAND-Zen exclusion region. Having completed the phenomenological analysis of models A and B, we now turn to the analysis of model C.

\subsection{Phenomenology of the model C}
\label{sec:p-mc}

Unlike models A and B, which shared identical field representations but differed in modular weights, model C adopts a distinct assignment of representations for the lepton fields (see Table~\ref{tab:field2}). This leads to a different structure for the lepton mass matrices and, consequently, a set of phenomenological predictions that are different from those of the previous two models. Remarkably, model C provides an excellent fit to the current neutrino oscillation data for both mass orderings, with $\chi^2_{\text{min}} < 1$ in each case. We begin with the NO scenario.

For model C in the NO case, the minimum value of the $\chi^2$ function is very low, $\chi^2_{\text{min}} = 0.09214$, indicating an excellent agreement with the experimental data. The best-fit values of the model's input parameters and the corresponding physical observables are
{
\allowdisplaybreaks
  \begin{align}
    &\langle \tau \rangle=0.36932+1.1714i,\quad \tilde{\beta}_{\mathrm{CL}} = 2.2525,\quad \tilde{\gamma}_{\mathrm{CL}}=159.15,\nonumber\\
    &\tilde{\beta}_{D} =123.49,\quad \tilde{\beta}_{S} =4.7687,\quad n=1.3591\times 10^{-6}~\mathrm{eV},\nonumber\\
    &m_e/m_{\mu}=0.004737,\quad m_{\mu}/m_{\tau}=0.05883,\nonumber\\
    &\sin^2\theta _{12}=0.3052,\quad \sin^2\theta _{13}=0.02224,\quad \sin^2\theta _{23}=0.473, \label{eq:bf-modelC-NO}\\
    &\delta_{\mathrm{CP}}=0.1849\pi,\quad  \eta_1=0.1373\pi,\quad  \eta_2=1.111\pi,\nonumber\\
    &m_1=0.005344~\mathrm{eV},\quad  m_2=0.01016~ \mathrm{eV},\quad  m_3=0.05046 ~\mathrm{eV},\nonumber\\
    &\sum m_i=0.06596~\mathrm{eV},\quad  m_{\beta}=0.01042~ \mathrm{eV},\quad  m_{\beta\beta}=0.00631~\mathrm{eV}.\nonumber
    \end{align}
}

Several features are apparent. First, the absolute neutrino mass scale is very low, with the sum of neutrino masses $\sum m_i = 0.0660$ eV lying well below the stringent cosmological bound. Second, the effective mass $m_{\beta} = 0.01042$ eV is far below the KATRIN sensitivity of $0.45$ eV. Third, the effective Majorana mass $m_{\beta\beta} = 0.00631$ eV falls within the projected reach of next-generation $0\nu\beta\beta$ experiments such as nEXO.
\begin{figure}[h!]
	\centering
	\includegraphics[width=1\textwidth]{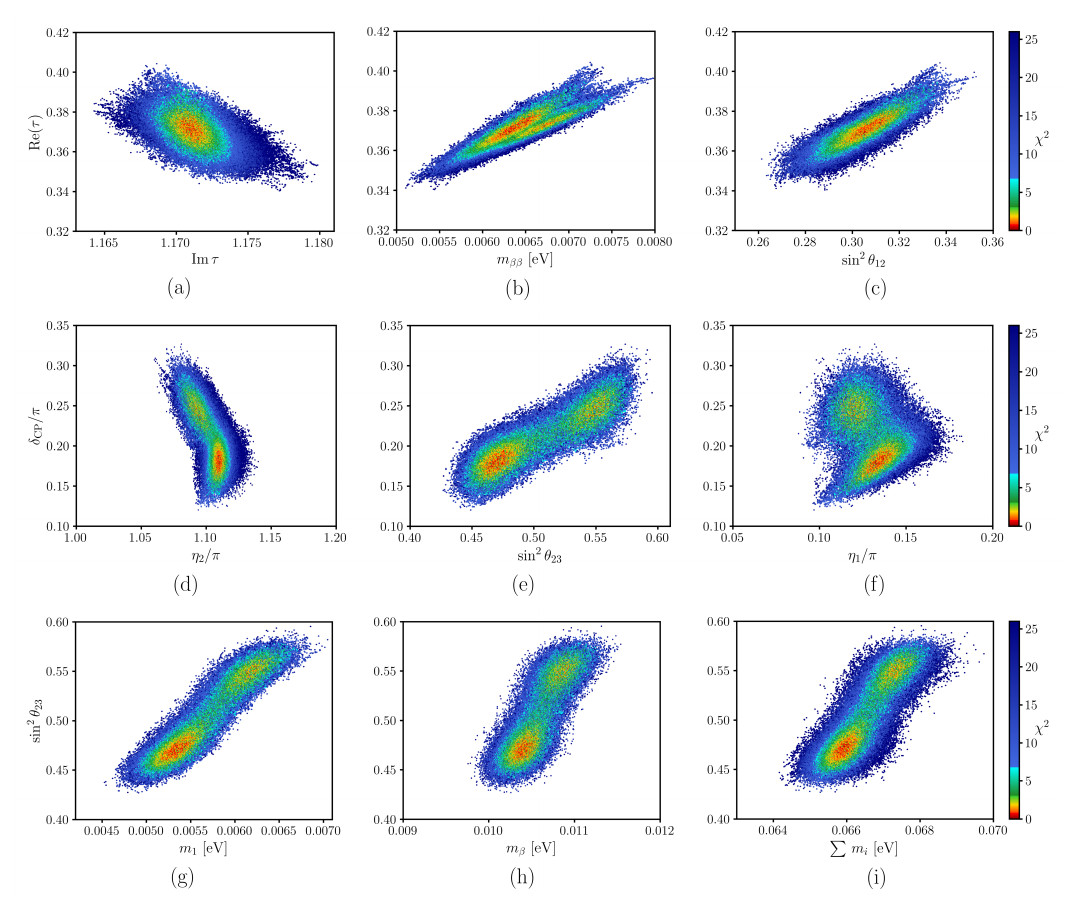}
	\caption{Results of the parameter scan for model C in the NO scenario. The panels show: (a) the allowed region of the modulus field $\tau$; correlations between $\mathrm{Re}(\tau)$ and (b) the effective Majorana mass $m_{\beta\beta}$, (c) the solar mixing angle $\sin^2\theta_{12}$; correlations involving the Dirac CP phase $\delta_{\mathrm{CP}}$ with (d) the Majorana phase $\eta_2$, (e) the atmospheric mixing angle $\sin^2\theta_{23}$, (f) the Majorana phase $\eta_1$; and correlations of the atmospheric mixing angle $\sin^2\theta_{23}$ with (g) the lightest neutrino mass $m_1$, (h) the effective electron neutrino mass $m_{\beta}$, and (i) the total neutrino mass $\sum m_i$.}
	\label{fig:m3no}
\end{figure}

Based on our numerical analysis, Fig.~\ref{fig:m3no} presents the results of the parameter correlations of model C with NO. Figure~\ref{fig:m3no}(a) shows the allowed region of the modulus field $\tau$. The viable region is concentrated around $\mathrm{Re}(\tau) \approx 0.37$ and $\mathrm{Im}(\tau) \approx 1.17$, which is different from the regions favored by models A and B. Figures~\ref{fig:m3no}(b) and~\ref{fig:m3no}(c) exhibit positive correlations between the real part of the modulus, $\mathrm{Re}(\tau)$, and two important quantities: the effective Majorana mass $m_{\beta\beta}$ and the solar mixing angle $\sin^2\theta_{12}$. Overall, as $\mathrm{Re}(\tau)$ increases, both $m_{\beta\beta}$ and $\sin^2\theta_{12}$ increase. This provides a link between the geometry of the modulus space and the observable parameters.

Unlike models A and B, where certain mixing angles were highly constrained, model C predicts broader ranges for mixing angles that fully cover the $3\sigma$ experimental intervals. However, interesting correlations emerge. Figures~\ref{fig:m3no}(d)–~\ref{fig:m3no}(f)  illustrate the behavior of the Dirac CP phase $\delta_{\mathrm{CP}}$. Its allowed range is $0.11\pi$ to $0.33\pi$, with the best-fit value at $0.185\pi$. In particular, $\delta_{\mathrm{CP}}$ exhibits a  connected bimodal distribution, with two low $\chi^2$ regions, visible in Figs.~\ref{fig:m3no}(d) and~\ref{fig:m3no}(f). This feature distinguishes model C from models A and B, where the CP phases were either unimodal or exhibited different bimodal patterns. Majorana phases are constrained to relatively narrow intervals $\eta_1 \in (0.088\pi, 0.183\pi)$ and $\eta_2 \in (1.06\pi, 1.14\pi)$. Figure~\ref{fig:m3no}(d) shows a positive correlation between $\delta_{\mathrm{CP}}$ and $\eta_2$, while Fig.~\ref{fig:m3no}(f) reveals a more complex relationship with $\eta_1$.

The atmospheric mixing angle $\sin^2\theta_{23}$ plays an out of the ordinary role in model C, exhibiting positive correlations with several physical quantities. Figures~\ref{fig:m3no}(e) and~\ref{fig:m3no}(g)–~\ref{fig:m3no}(i) of demonstrate that $\sin^2\theta_{23}$ increases with the Dirac CP phase $\delta_{\mathrm{CP}}$ (Fig.~\ref{fig:m3no}(e)), the lightest neutrino mass $m_1$ (Fig.~\ref{fig:m3no}(g)), the effective electron neutrino mass $m_{\beta}$ (Fig.~\ref{fig:m3no}(h)), and the total neutrino mass $\sum m_i$ (Fig.~\ref{fig:m3no}(i)). The effective electron neutrino mass $ m_{\beta} $ is constrained in $(9.76, 11.5)$ meV, significantly lower than the $0.45$ eV limit from the KATRIN experiment. These correlations are close to monotonic and relatively tight, meaning that a precise measurement of $\sin^2\theta_{23}$ in future experiments would provide indirect constraints on the absolute neutrino mass scale and the CP-violating phases. The predicted range for $\sin^2\theta_{23}$ itself is approximately $0.45$–$0.50$, which is well aligned with $3\sigma$ experimental range.

\begin{figure}[h!]
	\centering
	\includegraphics[width=1\textwidth]{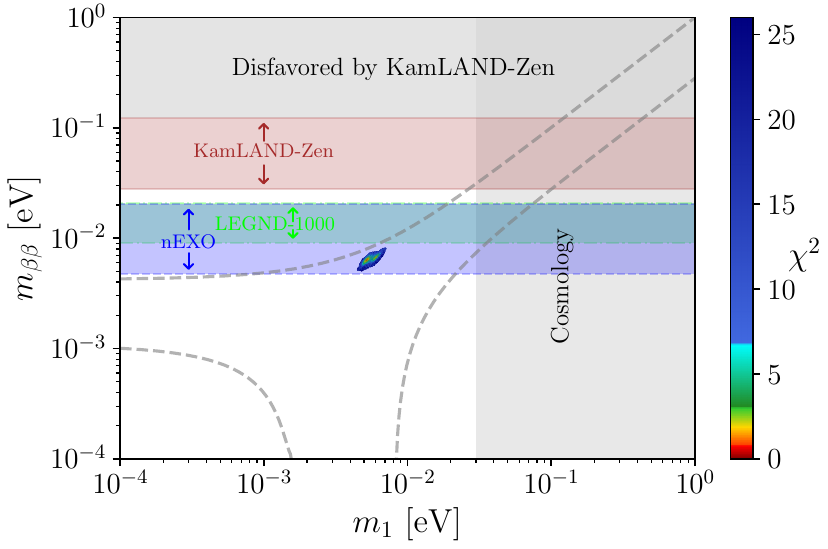}
	\caption{Prediction of model C for the effective Majorana neutrino mass $m_{\beta\beta}$ as a function of the lightest neutrino mass $m_1$ in the NO scenario. The colored region indicates the parameter space favored by the model, with $m_1$ ranging from approximately $0.0045$ to $0.0065$ eV and $m_{\beta\beta}$ ranging from $5.1$ to $8.1$ meV. Experimental and cosmological constraints are the same as such plots in analyses of models A and B. As shown, the entire $m_{\beta\beta}$ range falls within the projected sensitivity of nEXO, offering a clear target for experimental verification.}
	\label{fig:m3no-0vbb}
\end{figure}

We now examine the predictions of the NO case of model C for $0\nu\beta\beta$, which represent one of the most distinctive and experimentally promising features of this framework. As noted previously, the effective Majorana mass at the best-fit point is $m_{\beta\beta} = 0.00631$ eV. The complete parameter analysis, shown in Fig.~\ref{fig:m3no-0vbb}, shows that this is characteristic of the viable region of the model.
Figure~\ref{fig:m3no-0vbb} displays our result. The colored region is the allowed parameter space, which forms a band. Across this region, the lightest neutrino mass varies within the range $m_1 \in [0.0045, 0.0068]$ eV, while $m_{\beta\beta} \in [5.1, 8.1]$ meV. These values are consistent with the cosmological bound on the sum of neutrino masses, $\sum m_i < 0.12$ eV, which translates to an upper limit on $m_1$ of approximately $0.037$ eV in the NO scenario. The vertical gray band indicates this cosmological exclusion region, and all viable points in model C lie outside it. The current leading constraint on $m_{\beta\beta}$ comes from the KamLAND-Zen experiment, which excludes values in the range $0.028$–$0.122$ eV depending on the uncertainties of elements of the nuclear matrix~\cite{KamLAND-Zen:2024eml}. The narrow range of $m_{\beta\beta}$ in model C is a consequence of the parameter correlations observed in Fig.~\ref{fig:m3no}. Figure~\ref{fig:m3no}(b) shows a positive correlation between $\mathrm{Re}(\tau)$ and $m_{\beta\beta}$, which means that the modulus value controls the rate of $0\nu\beta\beta$ decay. Figures~\ref{fig:m3no}(g)–~\ref{fig:m3no}(i) further reveal that $m_{\beta\beta}$ is correlated with the atmospheric mixing angle $\sin^2\theta_{23}$ and the absolute mass scales. These relations may provide multiple channels for cross-checking the model. A measurement of $\sin^2\theta_{23}$ or $m_{\beta}$ would provide indirect information on the expected rate of $0\nu\beta\beta$ decay, and vice versa.

The predictions of model C, with $m_{\beta\beta}$ below $8.1$ meV, are more than a factor of three below the KamLAND-Zen limit, comfortably avoiding all current constraints. Looking to the future, next-generation $0\nu\beta\beta$ experiments aim to probe much lower masses. The nEXO experiment targets a sensitivity of $0.0047$–$0.0203$ eV~\cite{nEXO:2021ujk}, while LEGEND-1000 aims for $0.009$–$0.021~\mathrm{eV}$~\cite{LEGEND:2021bnm}. The entire predicted range falls within the projected range of nEXO and partially overlaps with the lower end of the LEGEND-1000 sensitivity band. This places model C in a highly favorable position for experimental testability. If nature has realized this particular modular symmetry assignment, there is a chance that $0\nu\beta\beta$ decay will be observed in the next generation of experiments.

We now complete the numerical analysis of model C by examining its predictions for the IO scenario. As in the NO case, model C provides an excellent fit to the experimental data, with a minimum $\chi^2$ value that is low but slightly higher than that of the NO scenario. The minimum is found to be $\chi^2_{\text{min}} = 0.2267$, which is in good agreement with the current neutrino oscillation data. The best-fit values of the input parameters of the model  and the corresponding physical observables are
\begin{equation}
  \begin{aligned}
    &\langle \tau \rangle=0.49855+1.0775i,\quad \tilde{\beta}_{\mathrm{CL}} = 3.4252,\quad \tilde{\gamma}_{\mathrm{CL}}=229.73,\\
    &\tilde{\beta}_{D} =31.611,\quad \tilde{\beta}_{S} =0.34199,\quad n=7.0405\times 10^{-5}~\mathrm{eV},\\
    &m_e/m_{\mu}=0.004737,\quad m_{\mu}/m_{\tau}=0.05882,\\
    &\sin^2\theta _{12}=0.3063,\quad \sin^2\theta _{13}=0.02231,\quad \sin^2\theta _{23}=0.548,\\
    &\delta_{\mathrm{CP}}=0.8451\pi,\quad  \eta_1=1.162\pi,\quad  \eta_2=0.1398\pi,\\
    &m_1=0.04924~\mathrm{eV},\quad  m_2=0.04999~ \mathrm{eV},\quad  m_3=0.003875 ~\mathrm{eV},\\
    &\sum m_i=0.1031~\mathrm{eV},\quad  m_{\beta}=0.04993~ \mathrm{eV},\quad  m_{\beta\beta}=0.04835~\mathrm{eV}.
    \end{aligned}
  \end{equation}

Several features distinguish this scenario from the IO predictions of models A and B. First, the total neutrino mass, $\sum m_i = 0.1031$ eV, which is below the stringent cosmological bound of $0.12$ eV, unlike models A and B whose IO predictions exceeded this limit. Second, the atmospheric mixing angle is predicted in the upper octant, with $\sin^2\theta_{23} = 0.548$, consistent with the global fit preference for IO. Third, the modulus field takes a value very close to the boundary of the fundamental domain, with $\mathrm{Re}(\tau) \approx 0.4986$, while its imaginary part is approximately $1.078$.

\begin{figure}[h!]
	\centering
	\includegraphics[width=1\textwidth]{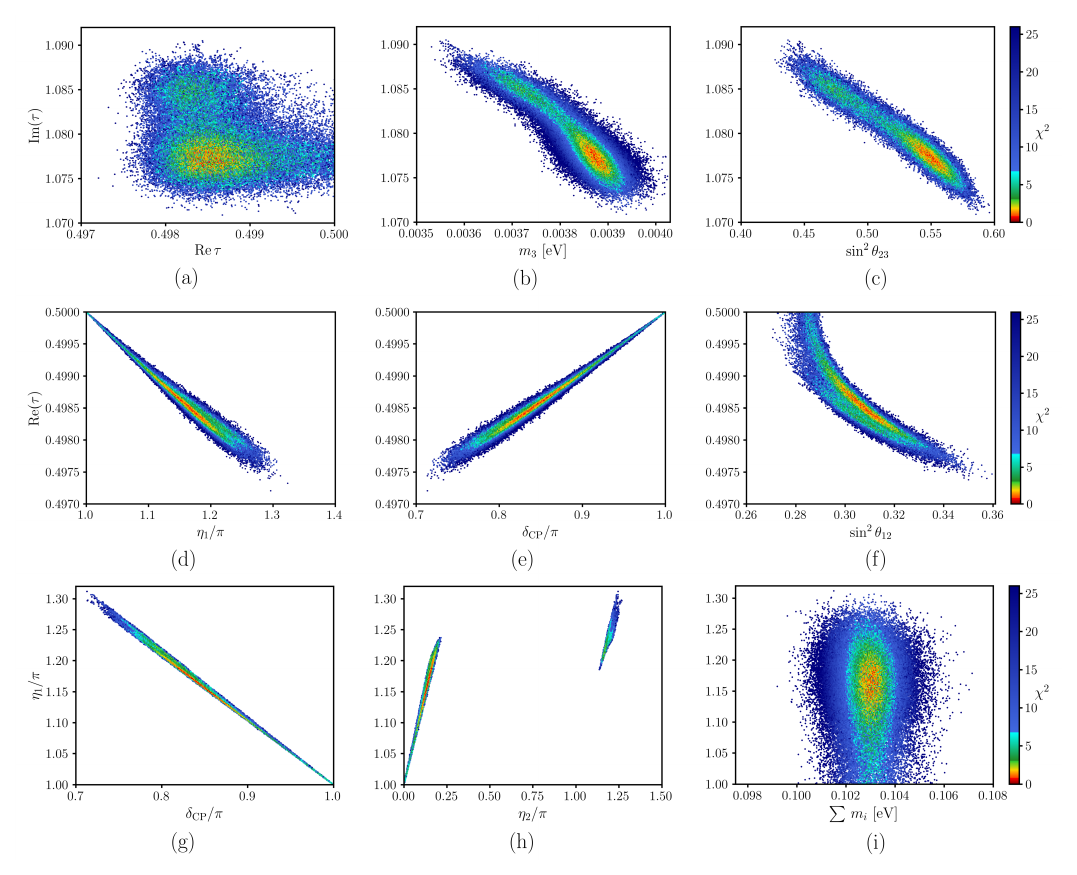}
	\caption{Results of the parameter scan for model C in the IO scenario. The panels show: (a) the allowed region of the modulus field $\tau$; correlations between $\mathrm{Im}(\tau)$ and (b) the lightest neutrino mass $m_3$, (c) the atmospheric mixing angle $\sin^2\theta_{23}$; correlations between $\mathrm{Re}(\tau)$ and (d) the Majorana phase $\eta_1$, (e) the Dirac CP phase $\delta_{\mathrm{CP}}$, (f) the solar mixing angle $\sin^2\theta_{12}$; and correlations of the Majorana phase $\eta_1$ with (g) the Dirac CP phase $\delta_{\mathrm{CP}}$, (h) the Majorana phase $\eta_2$, and (i) the total neutrino mass $\sum m_i$. The color scale indicates the $\chi^2$ value.}
	\label{fig:m3io}
\end{figure}

Figure~\ref{fig:m3io} presents the results of the parameter search for model C in the IO scenario. For the modulus field and its correlations, Fig.~\ref{fig:m3io}(a) shows the allowed region of in its fundamental domain. The viable region is concentrated near the boundary $\mathrm{Re}(\tau) = 0.5$, with some points extending exactly to the boundary. This proximity to the boundary has implications for CP violation, as we shall see. Figures~\ref{fig:m3io}(b) and~\ref{fig:m3io}(c) illustrate the correlations between the imaginary part of the modulus $\mathrm{Im}(\tau)$, and two observables, the lightest neutrino mass $m_3$ and the atmospheric mixing angle $\sin^2\theta_{23}$. The lightest neutrino mass is predicted to a narrow range, $m_3 \in [3.5, 4.0]$ meV, with a best-fit value of $3.875$ meV. The atmospheric mixing angle varies between approximately $0.53$ and $0.56$, showing a slight positive correlation with $\mathrm{Im}(\tau)$.

Figures~\ref{fig:m3io}(d)–~\ref{fig:m3io}(f) illustrate the role played by the real part of the modulus $\mathrm{Re}(\tau)$ in determining the CP-violating phases and the solar mixing angle. As $\mathrm{Re}(\tau)$ approaches the boundary value of $0.5$, several notable trends emerge. The Dirac CP phase $\delta_{\mathrm{CP}}$ (Fig.~\ref{fig:m3io}(e)) tends toward $\pi$, indicating the restoration of CP conservation. This is a consequence of the residual symmetry that persists at the boundary $\mathrm{Re}(\tau) = 0.5$. The Majorana phase $\eta_1$ (Fig.~\ref{fig:m3io}(d)) exhibits a similar behavior, approaching $\pi$ as $\mathrm{Re}(\tau) \to 0.5$. The solar mixing angle $\sin^2\theta_{12}$ (Fig.~\ref{fig:m3io}(f)) shows a negative correlation with $\mathrm{Re}(\tau)$, varying within the range $0.29$–$0.32$. These correlations are a distinctive feature of modular flavor symmetry. The geometry of the modulus space controls the amount of CP violation, with CP being preserved exactly on the boundary of the fundamental domain.

The Majorana phases and their correlations are shown Figs.~\ref{fig:m3io}(g)–~\ref{fig:m3io}(i), which display relationships between the Majorana phases, the Dirac phase, and the total neutrino mass. Figure~\ref{fig:m3io}(g) shows an approximately linear correlation between $\eta_1$ and $\delta_{\mathrm{CP}}$, spanning the ranges $(1.0\pi, 1.3\pi)$ and $(0.7\pi, 1.0\pi)$ respectively. This relationship implies that a measurement of one CP phase would strongly constrain the other. The behavior of the second Majorana phase $\eta_2$ is peculiar. As shown in Fig.~\ref{fig:m3io}(h), $\eta_2$ exhibits a bimodal distribution, with points clustering in two distinct regions: $(0, 0.25)\pi$ and $(1.1, 1.3)\pi$. The region with smaller value of $\chi^2$ corresponds to the lower range, $(0, 0.25)\pi$, where $\eta_2$ approaches zero as $\eta_1 \to \pi$. This correlation between $\eta_1$ and $\eta_2$ is a distinctive prediction of model C in the IO scenario. Figure~\ref{fig:m3io}(i) shows the correlation between $\eta_1$ and the total neutrino mass $\sum m_i$. The total mass varies within the range $\sum m_i \in [0.099, 0.107]$ eV, which is below the cosmological bound. This is a significant improvement over models A and B, whose IO predictions exceeded this limit.
  
While not explicitly shown in all panels, our numerical results indicate that all three mixing angles $\sin^2\theta_{12}$, $\sin^2\theta_{13}$, and $\sin^2\theta_{23}$ span ranges that fully cover the $3\sigma$ experimental intervals, similar to the NO scenario of model C. This broad coverage, combined with the sharp correlations among CP phases and mass scales, makes model C highly predictive yet flexible enough to accommodate the current data.
\begin{figure}[h!]
	\centering
	\includegraphics[width=1\textwidth]{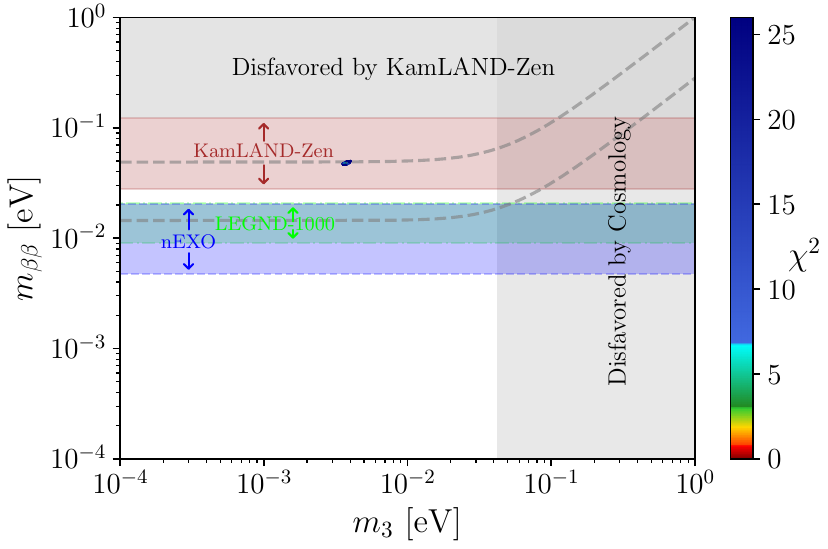}
	\caption{Prediction of model C for the effective Majorana neutrino mass $m_{\beta\beta}$ as a function of the lightest neutrino mass $m_3$ in the IO scenario. The almost pointlike colored region, near the upper boundary of the dashed line, indicates the parameter space favored by the model. Experimental and cosmological constraints are same as the relevant figures in models A and B with IO case.}
	\label{fig:m3io-0vbb}
\end{figure}

Figure~\ref{fig:m3io-0vbb} displays the prediction of model C for $m_{\beta\beta}$ as a function of the lightest neutrino mass $m_3$ in the IO case. The allowed parameter space forms a narrow band with $m_{\beta\beta} \in [0.0462, 0.0502]$ eV. This range lies entirely within the current exclusion limit of the KamLAND-Zen experiment ($0.028$–$0.122$ eV) and, in particular, falls above the projected sensitivity of next-generation $0\nu\beta\beta$ experiments. The entire predicted band lies above the upper ends of LEGEND-1000 and nEXO experiments' sensitivity. This places model C in a  position for experimental testability. The lightest neutrino mass $m_3$, is in the range $3.5$–$4.0$ meV, well below the cosmological exclusion limit indicated by the vertical gray band. This is in contrast to models A and B, whose IO predictions for the lightest neutrino mass were either excluded or in tension with cosmological data.

To highlight the analysis of model C in the case of IO, several remarks are in order. The model provides an excellent fit to the data ($\chi^2_{\text{min}} = 0.2267$), with predictions for all mixing angles covering the full $3\sigma$ experimental ranges. The modulus is concentrated near the boundary $\mathrm{Re}(\tau) = 0.5$, leading to sharp correlations between $\mathrm{Re}(\tau)$ and the CP-violating phases. As $\mathrm{Re}(\tau)$ approaches the boundary, $\delta_{\mathrm{CP}}$ and $\eta_1$ tend to $\pi$, signaling the restoration of CP conservation, a distinctive feature of modular symmetry. The total neutrino mass, $\sum m_i \approx 0.099$–$0.107$ eV, remains below the cosmological bound. The effective Majorana mass is predicted to be $m_{\beta\beta} \approx 0.0462$–$0.0502$ eV, a range that falls within the exclusion regions of KamLAND-Zen and  next-generation $0\nu\beta\beta$ experiments

Having completed the phenomenological analysis of all three models in both mass orderings, we have identified a rich landscape of predictions. Model A favors NO with $\sin^2\theta_{23}$ in the lower octant and $m_{\beta\beta}$ in the range accessible to next-generation experiments. Model B also favors NO but predicts $\sin^2\theta_{23}$ in the upper octant and an extremely small $m_{\beta\beta}$ beyond experimental reach. Model C provides good fits for both orderings, with the IO scenario making particularly sharp predictions for CP phases and placing $m_{\beta\beta}$ within the sensitivity of upcoming $0\nu\beta\beta$ searches. We now turn to the analysis of leptogenesis within these frameworks.

\section{Leptogensis}
\label{sec:lept}

In this section, we investigate the realization of leptogenesis within the models A, B, and C, based on the numerical analysis of neutrino masses and mixing presented in Sec.~\ref{sec:Num}. We adopt the best-fit parameters previously identified and begin by restoring the physical mass scales from the dimensionless quantities used in the fit of neutrino oscillation parameters. We then present the essential theoretical formalism before discussing the numerical results.

The first step is to restore the scaling of the neutrino mass matrix, for models A and B, at the best-fit point, the condition $ \mathcal{O}(m_D/m_{SN}) \sim 1 $ is satisfied~\footnote{ The dimensionless matrices $ m_D $, $ m_{SN} $, and $ m_S $ are respectively defined by $ M_D = \alpha_D v_h m_D $, $ M_{SN} =  \Lambda m_{SN} $, and $ M_S = \alpha_S m_S $, with a relation $\alpha_S = n (\Lambda / \alpha_D v_h)^2$ among the coefficients.}.  We set $\alpha_D v_h / \Lambda = 10^{-3} $, with reference values $\alpha_D v_h = 10$ GeV and $\Lambda = 10$ TeV, which yields $ \alpha_S = 10^{6} \times n$. Here, the factor $(\alpha_D v_h / \Lambda)^2$ also quantifies the magnitude of non-unitary effects in the lepton mixing matrix. The chosen parametrization therefore not only guaranties the correct mass hierarchy necessary for the inverse seesaw mechanism but also ensures that the degree of non-unitarity remains within the stringent limits derived from electroweak precision observables and lepton flavor violation searches~\cite{Antusch:2014woa,Blennow:2016jkn,Fernandez-Martinez:2016lgt}. Taking model A as an example, with $n \approx 0.01$ eV, we obtain $\alpha_S \approx 10^{4}$ eV and $M_S \sim \mathcal{O}(10)$ keV, consistent with the required hierarchy. For model B, analogous values follow from its best-fit parameters. In contrast, for model C, the best-fit point yields $\mathcal{O}(m_D/m_{SN}) \sim 10$, necessitating a different choice of scales. We therefore set $\alpha_D v_h / \Lambda = 10^{-4}$ to maintain the appropriate hierarchy while keeping the heavy neutrino masses in the TeV range. With these scale choices, we now proceed to the leptogenesis analysis.

\subsection{Theoretical framework}
\label{sec:lept-framework}

In the inverse seesaw model, the presence of large Yukawa couplings is a salient characteristic. Such couplings lead to equilibration of lepton flavors in the early Universe, causing the flavor-dependent Boltzmann equations to reduce to a single equation governing the total lepton number. Consequently, although we adopt a flavor-blind approach, this simplification aligns with the typical expectations of the inverse seesaw mechanism and simplifies the numerical analysis of leptogenesis~\cite{AristizabalSierra:2009mq, Blanchet:2010kw}.

Our model contains three right-handed neutrinos $N_i$ and three sterile neutrinos $S_i$ ($i = 1,2,3$), which form three pseudo-Dirac pairs $(N_i, S_i)$. In the basis $( N^c, S)$, the $6 \times 6$ mass matrix for the heavy neutral fermions takes the form
\begin{equation}
M_N = \begin{pmatrix}
0 & M_{SN}^T \\
M_{SN} & M_S
\end{pmatrix}.
\end{equation}
This matrix is diagonalized by a unitary matrix $V$ such that
\begin{equation}
V^T M_N V = \mathrm{diag}\left(M_1, M_2, M_3, M_4, M_5, M_6\right),
\end{equation}
where the eigenvalues are ordered such that $M_1 < M_2 < \cdots < M_6$. In the mass eigenstate basis, the Yukawa coupling matrix elements relevant for leptogenesis are given by
\begin{equation}
h_{i\alpha} = \sum_{j=1}^3  (Y_D)_{\alpha j}V_{ji}^*,
\end{equation}
with $i = 1,\ldots,6$ labeling the heavy neutrino mass eigenstates, $\alpha = e, \mu, \tau$ denoting lepton flavor indices, and $Y_D$ representing the Dirac Yukawa matrix in the flavor basis.

The CP asymmetry parameter $\varepsilon_i$ quantifies the degree of CP violation in the decay of the heavy neutrino $N_i \to \ell H~(\bar{\ell} H^\dagger)$. For a given heavy neutrino $N_i$, it is defined as
\begin{align}
\varepsilon_i &= \frac{\Gamma(N_i \to \ell H) - \Gamma(N_i \to \bar{\ell} H^\dagger)}{\Gamma(N_i \to \ell H) + \Gamma(N_i \to \bar{\ell} H^\dagger)} \\
&= \frac{1}{8\pi (hh^\dagger)_{ii}} \sum_{j\neq i} \mathrm{Im}[(hh^\dagger)^2_{ij}] \, f_{ij},
\label{eq:cp-asym}
\end{align}
where the self-energy correction factor is~\cite{Covi:1996wh, Buchmuller:1997yu, Blanchet:2009kk}
\begin{equation}
f_{ij} = \frac{(M_i^2 - M_j^2) M_i M_j}{(M_i^2 - M_j^2)^2 + (M_i \Gamma_i - M_j \Gamma_j)^2}.
\end{equation}
Here, $\Gamma_i = M_i \left(hh^\dagger\right)_{ii} / 8\pi$ denotes the decay width of the heavy neutrino $M_i$. In scenarios with nearly degenerate masses, this factor can become enhanced, a feature particularly relevant for TeV-scale leptogenesis. Since the lepton asymmetry generated by the decay of heavier particle pairs can be efficiently erased by lepton-number-violating scattering processes involving lighter pairs, we focus on the lightest pseudo-Dirac pair, denoted $(M_1, M_2)$, as the main source of the final asymmetry~\cite{Chakraborty:2021azg}.

The washout parameter $K_i$ characterizes the strength of the washout processes relative to the Hubble expansion rate. It is defined as
\begin{equation}
K_i = \frac{\Gamma_i}{H(T = M_i)}, \qquad
H(T) = 1.66 \sqrt{g_\ast} \frac{T^2}{M_{\mathrm{Pl}}},
\end{equation}
where $H$ is the Hubble parameter, $g_\ast$ is the effective number of relativistic degrees of freedom in the SM plasma at temperature $T$, and $M_{\mathrm{Pl}} = 1.22 \times 10^{19}$ GeV is the Planck mass.

In the inverse seesaw mechanism, the large Yukawa couplings typically lead to very large values of $K_i$, seemingly placing the scenario in the strong washout regime. However, in models with approximately conserved lepton number, such as the inverse seesaw considered here, interactions involving the different sterile states that are usually neglected in the standard type I seesaw framework can become critically important~\cite{Blanchet:2009kk, Shao:2025xav}. These interactions modify the washout dynamics, necessitating the introduction of an effective washout parameter~\cite{Blanchet:2009kk}
\begin{equation}
K^{\mathrm{eff}}_i \equiv \frac{K_i \delta_i^2}{1 + \sqrt{a_i}\,\delta_i + \delta_i^2}
\xrightarrow{\delta_i \ll 1} K_i \delta_i^2.
\end{equation}
Here, $\delta_i = |\Delta M| / \Gamma_i$, with $\Delta M = M_2 - M_1$ being the mass splitting of the lightest pseudo-Dirac pair, and $a_i = (\Gamma_i / M_1)^2$. In the limit of small mass splitting, which is characteristic of the inverse seesaw with a nearly degenerate pair, the effective washout is suppressed by a factor $\delta_i^2$, potentially bringing the scenario into a viable regime despite large nominal $K_i$ values.

The final baryon asymmetry is basically obtained from the CP asymmetry after accounting for washout effects via the efficiency factor $\kappa_i$, together with the sphaleron conversion rate. The efficiency factor, which ranges between 0 and 1, quantifies the fraction of the generated CP asymmetry that survives washout processes and is converted into a net lepton asymmetry. The final baryon asymmetry is given by~\cite{Buchmuller:2004nz}
\begin{equation}
\eta_B = -10^{-2} \sum_{i=1,2} \kappa_i \varepsilon_i,
\end{equation}
where the factor $10^{-2}$ accounts for the conversion of lepton asymmetry to baryon asymmetry via sphaleron processes.

The efficiency factor depends on the washout regime. As in Refs.~\cite{Buchmuller:2004nz, Agashe:2018cuf}, we adopt the following approximate expressions:
\begin{itemize}
  \item Strong washout regime  ($K^{\mathrm{eff}}_i \gg  1$): 
        \begin{equation}
        \kappa_i \approx \frac{2}{z_B K^{\mathrm{eff}}_i} 
                   \left[1 - \exp\left(-\frac{z_B K^{\mathrm{eff}}_i}{2}\right)\right],
        \end{equation}
        where $z_B$  is approximated by
        \begin{equation}
          z_B(K) \approx  1 + \frac{1}{2}\ln\left[1 + \frac{\pi \left(K^{\mathrm{eff}}_i\right)^2}{2^{10}} \left(5\ln\frac{5}{4}+\ln\left(\pi(K^{\mathrm{eff}}_i)^2\right)\right)^5\right].
          \end{equation}
  \item Weak washout regime ($K^{\mathrm{eff}}_i < 1$):
        \begin{equation}
        \kappa_i \approx 
        \begin{cases}
        \displaystyle \frac{1}{4}\, K^{\mathrm{eff}}_i 
        \left(\frac{3\pi}{2} - z_B\right) & \text{for } K_i > 1 ,\\[1.5em]
        \displaystyle \left(\frac{3\pi}{8}\right)^2\, K^{\mathrm{eff}}_i K_i & \text{for } K_i < 1.
        \end{cases}
        \end{equation}
\end{itemize}

With this formalism established, we now proceed to the numerical evaluation of leptogenesis in models A, B and C, utilizing the best-fit parameters obtained from the neutrino masses and mixing analysis.

\subsection{Numerical analysis of leptogenesis}
\label{sec:nu-l}

As discussed in Sec.~\ref{sec:lept-framework}, we adopt the benchmark values $\alpha_D v_h = 10$ GeV and $\Lambda = 10$ TeV in analyzing models A and B to ensure the hierarchy $\alpha_D v_h / \Lambda = 10^{-3}$, required by the inverse seesaw mechanism. For model C, we set $\alpha_D v_h = 1$ GeV and $\alpha_D v_h / \Lambda = 10^{-4}$. In the numerical analysis, we introduce a common scaling factor $r$ that multiplies both $\alpha_D v_h$ and $\Lambda$, therefore adjusting the overall mass scale while preserving their ratio. To avoid unphysically small or large Yukawa couplings, we restrict $r$ to the interval $r \in (0.1, 20)$. By scanning over $r$, we explore the parameter space and compute the resulting baryon asymmetry $\eta_B$ for each model and mass ordering.
\begin{figure}[h!]
	\centering
	\includegraphics[width=1\textwidth]{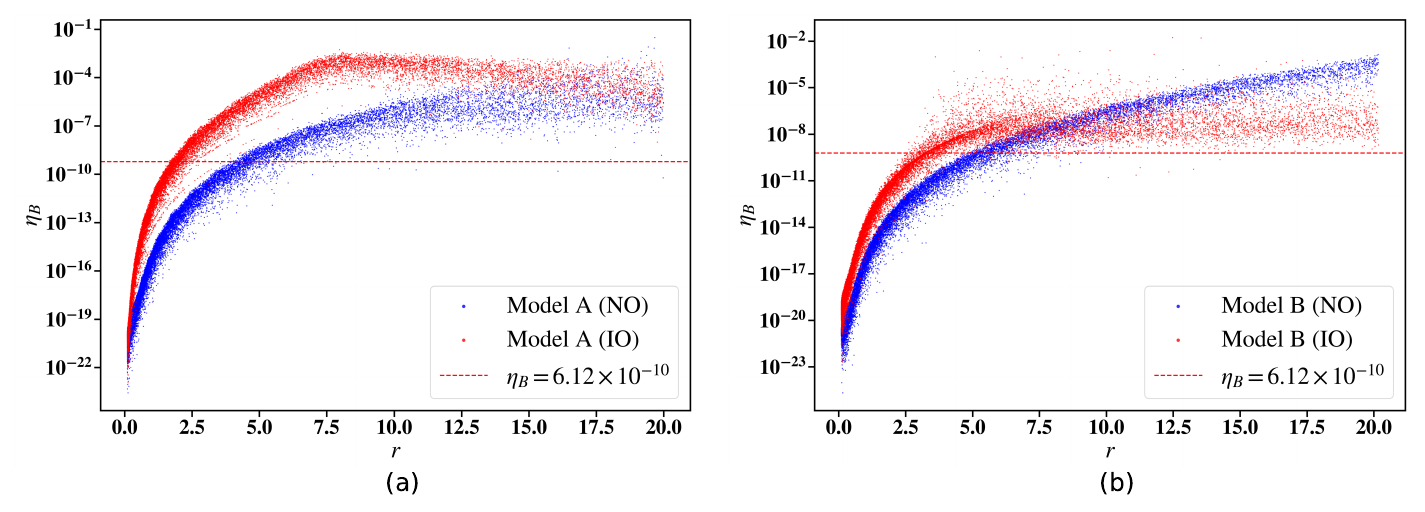}
	\caption{Baryon asymmetry $\eta_B$ as a function of the scaling parameter $r$ for (a) model A and (b) model B. The blue points correspond to the NO scenario, while the red points denote the IO scenario. The horizontal dashed line indicates the experimentally observed value $\eta_B = 6.12 \times 10^{-10}$~\cite{Planck:2018vyg,ParticleDataGroup:2024cfk}. Both models successfully reproduce the observed asymmetry for specific ranges of $r$, with the corresponding heavy neutrino masses located in the TeV regime.}
	\label{fig:mab-lep}
\end{figure}

Figure~\ref{fig:mab-lep} displays the results of our numerical analysis, showing the predicted baryon asymmetry $\eta_B$ as a function of the scaling parameter $r$ for both mass orderings in models A and B. The horizontal dashed line marks the central value of the observed baryon asymmetry $\eta_B = 6.12 \times 10^{-10}$, as reported by the Planck Collaboration~\cite{Planck:2018vyg}. Several essential observations emerge from this figure. For model A in the NO scenario (blue points in Fig.~\ref{fig:mab-lep}(a)), the observed baryon asymmetry is successfully reproduced when the scaling parameter lies in the range $r \in (4.0, 6.0)$. Within this interval, the masses of the lightest pseudo-Dirac pair, which drives the leptogenesis mechanism, fall in the range from $0.792$ TeV to $1.19$ TeV. The masses of the heavier pseudo-Dirac pairs are listed in Table~\ref{tab:lep}. Notably, all relevant heavy neutrinos lie at the TeV scale, making them potentially accessible to current and future collider experiments such as the high-luminosity LHC.
\begin{table}[h!]
  \centering
  \renewcommand{\arraystretch}{2}
  \setlength{\tabcolsep}{5pt}  
  \begin{tabular}{|c|c|c|c|c|c|}
    \hline
    \text{Model} & \text{Order} & $r$ range & $M_1$, $M_2$ [TeV] & $M_3$, $M_4$ [TeV] & $M_5$, $M_6$ [TeV] \\
    \hline  
    \multirow{2}{*}{$\mathbf{A}$} & $\mathbf{NO}$ & $4.0-6.0$ & $0.792-1.19$ & $7.79-11.7$ & $367-550$ \\
    \cline{2-6}
    & $\mathbf{IO}$ & $1.6-2.6$ & $0.838-1.36$ & $2.39-3.88$ & $4.62-7.57$ \\
    \hline 
    \multirow{2}{*}{$\mathbf{B}$} & $\mathbf{NO}$ & $4.6-6.0$ & $1.87-2.24$ & $14.2-18.3$ & $19.0-24.8$ \\
    \cline{2-6}
    & $\mathbf{IO}$ & $2.5-4.0$ & $0.421-0.674$ & $4.68-7.49$ & $328-525$ \\
    \hline 
  \end{tabular}
  \caption{Ranges of the scaling parameter $r$ and the corresponding heavy neutrino masses $M_i$ for which models A and B successfully reproduce the observed baryon asymmetry $\eta_B = 6.12 \times 10^{-10}$. The three pseudo-Dirac pairs are denoted by $(M_1,M_2)$, $(M_3,M_4)$, and $(M_5,M_6)$, with each pair having nearly degenerate masses as characteristic of the inverse seesaw mechanism.}
  \label{tab:lep}  
\end{table}
For model A in the IO (red points in Fig.~\ref{fig:mab-lep}(a)), successful leptogenesis occurs for $r \in (1.6, 2.6)$. The corresponding lightest pseudo-Dirac pair masses range from $0.838$ to $1.36$ TeV. As with the NO case, the heavy neutrino spectrum remains within the TeV regime, reinforcing the testability of this scenario. Model B exhibits similar behavior but with distinct mass scales. For the NO (blue points in Fig.~\ref{fig:mab-lep}(b)), the observed $\eta_B$ is obtained for $r \in (4.6, 6.0)$, yielding the lightest pair masses in the range $1.87$–$2.24$ TeV. This places the relevant heavy neutrinos at the upper end of the TeV scale, potentially accessible to next-generation colliders. Similar phenomenological behavior is observed in model B. For model B in the IO (red points in Fig.~\ref{fig:mab-lep}(b)), successful leptogenesis requires $r \in (2.5, 4.0)$, with the lightest pseudo-Dirac pair masses ranging from $0.421$ to $0.674$ TeV. This lower mass scale improves the prospects for direct detection in existing facilities.

The complete spectrum of heavy neutrino masses for all successful parameter regions is summarized in Table~\ref{tab:lep}. Several features merit comments. The three pseudo-Dirac pairs exhibit a hierarchical structure, with masses spanning from sub-TeV to hundreds of TeV. The lightest pair, responsible for leptogenesis, consistently lies in the TeV range across all scenarios. The mass hierarchies differ significantly between models A and B, as well as between NO and IO scenarios. This variation originates from the different modular weight assignments in the two models, which affect the Yukawa coupling structures and, consequently, the leptogenesis efficiency and its dependence on $r$. In all successful cases, the heavy neutrino masses are sufficiently low to be within the kinematic reach of future colliders, offering promising opportunities for experimental verification.

\begin{figure}[h!]
	\centering
	\includegraphics[width=0.75\textwidth]{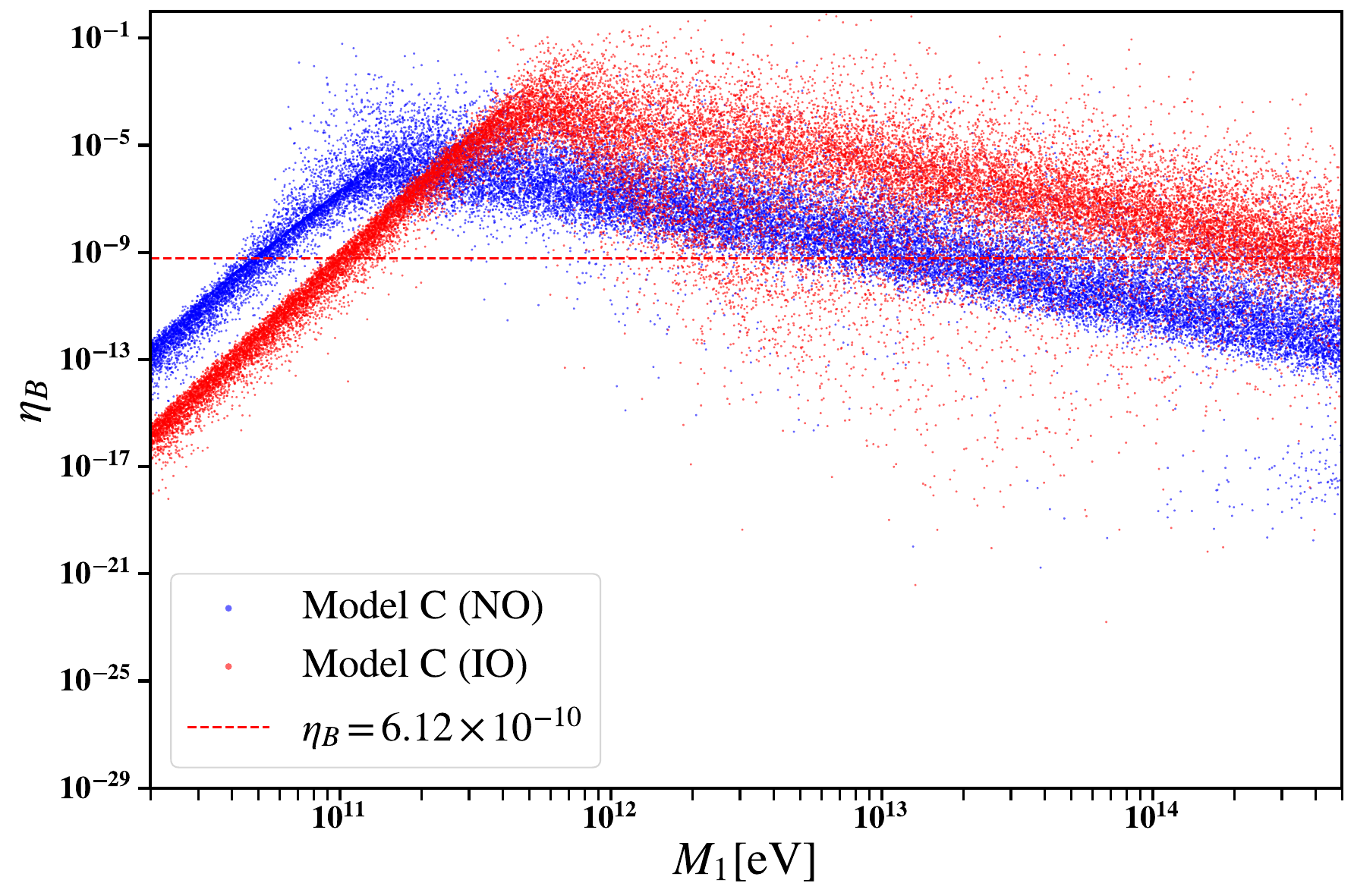}
	\caption{Baryon asymmetry $\eta_B$ as a function of the lightest heavy neutrino mass $M_1$ for model C. Blue points correspond to the NO scenario, while red points denote the IO scenario. The horizontal dashed line indicates the experimentally observed value $\eta_B = 6.12 \times 10^{-10}$.}
	\label{fig:mc-lep}
\end{figure}

Having established the viability of leptogenesis in models A and B, we now turn to the numerical results for model C. Unlike the previous two models, successful leptogenesis in model C does not have a clear regularity for the reference parameter range $r \in (0.1, 20)$. To explore the full predictive capacity of this model, we extended the search to larger values of $r$. Since $r$ is proportional to the heavy neutrino masses, we present the results in terms of the lightest heavy neutrino mass $M_1$, which provides a more direct physical interpretation (see Fig.~\ref{fig:mc-lep}).

For the NO scenario of model C (blue points in Fig.~\ref{fig:mc-lep}), the observed baryon asymmetry is successfully reproduced when $M_1 \approx 0.05$ TeV, corresponding to $r \approx 0.14$. A second viable region appears for $M_1 \gtrsim 3.7$ TeV ($r > 10.2$) on the right side of the plot. Beyond that, $\eta_B$ decreases as $M_1$ increases, until $M_1 \approx 120$ TeV, where the predicted asymmetry falls below the observed value. The IO scenario of model C (red points) exhibits similar behavior. Successful leptogenesis occurs for $M_1 \approx 0.1$ TeV ($r \approx 0.25$), and again for $M_1 \gtrsim 3.5$ TeV ($r > 8.3$). As in the NO case, as $M_1$ increases, the asymmetry gradually falls below the observed value.

In model C, the mass splitting $\Delta M$ of the lightest pseudo-Dirac pair $(N_1, S_1)$ is of order keV, as originating from $M_S \sim \text{keV}$. The decay width of each heavy neutrino satisfies $\Gamma_i \propto M_i (hh^\dagger)_{ii} / (8\pi)$, and is therefore proportional to  $M_i$. Consequently, the parameter$\delta_1 = \frac{\Delta M}{\Gamma_1} \propto \frac{1}{M_1}$, is approximately inversely proportional to $M_1$. When $M_1$ is small, washout is not significantly suppressed and remains strong. Meanwhile, the CP asymmetry parameter $\varepsilon$ increases with $M_1$. Under the combined effect, the baryon asymmetry $\eta_B$ grows with $M_1$. When $M_1$ increases to around $10^{11}$ eV, the effective washout parameter becomes $K_i^{\text{eff}} \approx K_i \delta^2 \quad (\delta \ll 1)$, which is strongly suppressed by $\delta^2$, rendering washout very weak. At the same time, the CP asymmetry no longer increases with $M_1$ (it actually begins to decrease). As a result, $\eta_B$ reaches a maximum and then decreases.

The numerical analysis presented in this section demonstrates that models A and B can successfully account for the observed baryon asymmetry of the Universe via TeV-scale leptogenesis. Both models reproduce the measured value of $\eta_B$ within well-defined ranges of the scaling parameter $r$, with the lightest pseudo-Dirac pair masses consistently lying in the TeV regime. The mass scales and hierarchies differ between models and mass orderings, reflecting the underlying modular symmetry structure and providing distinct experimental targets. The TeV-scale nature of the relevant heavy neutrinos renders these scenarios testable at current and future collider experiments, establishing a promising connection between leptogenesis and direct searches. In contrast, model C exhibits a considerably broader viable parameter space, a distinction that arises from its different assignments of field representations and modular weights, which modify the Yukawa coupling structures and associated washout dynamics. This diversity of phenomenological outcomes underscores the richness of model building within the modular symmetry framework and highlights its capacity to accommodate a wide range of experimental signatures. Taken together, these results confirm that all three models, each rooted in the non-SUSY modular $A^\prime_5$ inverse seesaw mechanism, can successfully generate the observed matter-antimatter asymmetry while remaining consistent with neutrino oscillation data. 

\section{Conclusions}
\label{sec:Con}

In this paper, we have constructed a neutrino mass model based on the non-SUSY modular $A^{\prime}_5$ symmetry framework, without introducing any flavon fields. The lepton flavor structure has been investigated in this context, and leptogenesis has been successfully realized at the TeV energy scale. Neutrino masses are generated via the inverse seesaw mechanism, which entails the introduction of three singlet state $S$ on top of the three right-handed neutrinos $N$, and a scalar singlet $\phi$. The leptogenesis mechanism is driven by the out-of-equilibrium decays of the lightest pseudo-Dirac pair $(N_1, S_1)$, whose masses remain in the TeV range. To maintain modular invariance, the Yukawa couplings are constructed from polyharmonic Maa\ss~form forms, and a gCP symmetry is imposed to reduce the number of free parameters, resulting in a highly constrained predictive framework.

We have performed a detailed numerical analysis of three specific realizations, named models A, B, and C, each of which is characterized by distinct assignments of modular representations and weights. Models A and B share similar irreducible representations for the physical fields under $A^\prime_5$ but differ in their modular weight assignments, allowing us to systematically explore the impact of weight choices on phenomenological predictions. Both models exhibit a preference for the normal mass ordering. Notably, model A predicts the atmospheric mixing angle in the lower octant, while model B favors the upper octant, demonstrating the sensitivity of this observable to the modular weight structure. Model C, which features a different representation assignment and one additional free parameter, provides a better fit to the experimental data in both mass orderings, with best-fit points located well within the $1\sigma$ regions of the global oscillation data. Across all three models, the predicted ranges for neutrino masses, $0\nu\beta\beta$ decay effective Majorana mass $m_{\beta\beta}$, effective electron neutrino mass $m_\beta$, and the Dirac and Majorana phases are provided, offering specific targets for ongoing and future experimental investigations.

The viable parameter spaces identified in this work are compatible with current experimental constraints and yield predictions that can be tested by next-generation facilities. Long-baseline experiments such as DUNE~\cite{DUNE:2020ypp} and Hyper-Kamiokande~\cite{Hyper-Kamiokande:2018ofw} will probe the mixing parameters with unprecedented precision. The JUNO reactor neutrino experiment~\cite{JUNO:2015zny} begins to provide high-precision measurements of neutrino oscillation parameters and plays an important role in determining the neutrino mass ordering. Complementary kinematic searches, such as those carried out by KATRIN~\cite{KATRIN:2022ayy} or Project 8~\cite{Project8:2022wqh} , will continue to constrain the absolute neutrino mass scale. Such high-precision measurements will not only provide rigorous tests of the models presented here but may also reveal new physical features beyond the current understanding.

Our analysis shows interesting correlations between physical parameters, particularly between mixing angles and the Dirac CP phase in specific regions of the modulus space $\tau$. These correlations, also observed in other modular symmetry studies~\cite{Zhang:2025dsa}, reflect the underlying group-theoretic structure and provide interconnected predictions that can be experimentally scrutinized. Furthermore, the allowed regions of the modulus field $\tau$ in all three models are located at or near the boundary of the fundamental domain, or in the vicinity of $\mathrm{Re}(\tau)=0$. This pattern, which has also been noted in the literature~\cite{Feruglio:2022koo,Li:2025kcr,Granelli:2025lds}, reflects intrinsic characteristics of modular-symmetric models or is related to the stability of the modulus VEV, a question that warrants further theoretical investigation.

Modular symmetry has emerged as a powerful paradigm for deciphering the lepton flavor structure. Its core advantage lies in the ability to naturally explain flavor hierarchies and mixing patterns with a minimal number of parameters through group-theoretic constraints. In contrast to traditional formulations that rely on SUSY framework, the non-holomorphic modular symmetry adopted in this work breaks free from this limitation, offering a framework for flavor physics in non-SUSY scenarios. The results presented here demonstrate that modular symmetry can simultaneously account for lepton flavor mixing and the baryon asymmetry of the Universe via TeV-scale leptogenesis. Successful reproduction of the observed baryon asymmetry $\eta_B$ with heavy neutrino masses in the TeV range provides direct support for the phenomenological viability of modular symmetry at experimentally accessible energies and establishes a compelling connection with current and future collider experiments.

Through the construction of non-SUSY lepton models under the modular group $A^{\prime}_5$, this work has elucidated the dual role of modular symmetry in generating neutrino masses and mixing, while simultaneously explaining the matter-antimatter asymmetry of the Universe. The models yield verifiable phenomenological signals at the TeV scale, offering concrete targets for experimental investigation and providing crucial support for modular symmetry as a compelling framework for BSM flavor physics. Future research directions may include the extension of this framework to the quark sector, the exploration of other finite modular groups, and a more detailed investigation of collider signatures and their detectability at upcoming facilities.

\section*{Acknowledgements}
This work was supported by the National Natural Science Foundation of the People’s Republic of China under Grant No. 12565016.


\appendix

\section{\texorpdfstring{$A^\prime_5$}{A5prime} Modular Symmetry}
\label{app:Modular}

Since this work is formulated within the framework of non-holomorphic modular invariant theories, we provide here a brief overview of the relevant mathematical background. The complete modular group $\Gamma = SL(2, \mathbb{Z})$ is an important discrete subgroup of $SL(2, \mathbb{R})$, which consists of $2 \times 2$ matrices $\gamma = \begin{pmatrix} a & b \\ c & d \end{pmatrix}$ with integer entries satisfying $ad - bc = 1$. It acts on the upper half-plane $\mathcal{H} = \{\tau \in \mathbb{C} \mid \mathrm{Im}(\tau) > 0\}$ via linear fractional transformations:
\begin{equation} \tau \mapsto \gamma\tau = \frac{a\tau + b}{c\tau + d}, \label{eq:modular_transformation} \end{equation}
where $ \tau $ is the complex modulus. The modular group $\Gamma$ is generated by two fundamental elements, $S$ and $T$, with matrix representations
\begin{equation}
S = \begin{pmatrix} 0 & 1 \\ -1 & 0 \end{pmatrix}, \qquad
T = \begin{pmatrix} 1 & 1 \\ 0 & 1 \end{pmatrix}.
\end{equation}
These generators satisfy the defining relations $S^4 = (ST)^3 = \mathbf{I}$, where $\mathbf{I}$ is the $2 \times 2$ identity matrix, together with the commutation relation $S^2 T = T S^2$.

The principal congruence subgroup of level $N$, denoted $\Gamma(N)$, is defined as
\begin{equation}
\Gamma(N) = \left\{ \begin{pmatrix} a & b \\ c & d \end{pmatrix} \in SL(2, \mathbb{Z}) \;\Bigg|\;
\begin{pmatrix} a & b \\ c & d \end{pmatrix} \equiv \begin{pmatrix} 1 & 0 \\ 0 & 1 \end{pmatrix} \pmod{N} \right\},
\end{equation}
with $N \in \mathbb{Z}^+$. The homogeneous finite modular group $ \Gamma^{\prime}_N $ is obtained as the quotient $\Gamma^{\prime}_N \equiv \Gamma / \Gamma(N)$ of the modular group $ \Gamma $ and its principal congruence subgroup $ \Gamma(N) $ of level $N$. For the inhomogeneous finite modular group $\Gamma_N$, when $N > 2$, $\Gamma^{\prime}_N$ is its double cover group, satisfying $\Gamma^{\prime}_N \cong \Gamma_N / \{\pm \mathbf{I}\}$. At low levels, $\Gamma^{\prime}_N$ coincides with the well-known double covers of permutation and alternating groups $\Gamma^\prime_3 \cong T^\prime$, $\Gamma^\prime_4 \cong S^\prime_4$, and $\Gamma^\prime_5 \cong A^{\prime}_5$. This correspondence provides an origin for discrete flavor symmetries within the modular framework.

In this work, we employ the homogeneous finite modular group $\Gamma^{\prime}_5 \cong A^{\prime}_5$ at the level $N = 5$. This group is the unique non-trivial central extension of the alternating group $A_5$ (the rotational symmetry group of the regular icosahedron) by a central element of order two. Geometrically, it corresponds to the spin double cover icosahedral or dodecahedral symmetry in the space of unit quaternions. The group $A^{\prime}_5$ can be generated by two elements $S$ and $T$ satisfying the defining relations
\begin{equation}
S^4 = T^5 = (ST)^3 = \mathbf{I}, \qquad S^2 T = T S^2.
\end{equation}

The group $A^{\prime}_5$ has nine irreducible representations: $\mathbf{1}$, $\hat{\mathbf{2}}$, $\hat{\mathbf{2}}^{\prime}$, $\mathbf{3}$, $\mathbf{3}^{\prime}$, $\mathbf{4}$, $\hat{\mathbf{4}}^{\prime}$, $\mathbf{5}$, and $\hat{\mathbf{6}}$. Representations marked with a hat are unique to the double cover $A^{\prime}_5$ and do not have an analog in $A_5$, while those without a hat are common to both groups. The multiplication rules for the irreducible representations relevant to this work are given below:
\begin{equation}
\begin{aligned}  
& \hat{\mathbf{2}}\otimes \mathbf{3'}=\hat{\mathbf{2}}'\otimes \mathbf{3}=\hat{\mathbf{6}}, \quad\hat{\mathbf{2}}'\otimes \mathbf{3'}=\hat{\mathbf{2}}'\oplus \hat{\mathbf{4}}', \quad\mathbf{3}\otimes \mathbf{3}=\mathbf{1_s}\oplus \mathbf{3_a}\oplus \mathbf{5_s},\\
&\mathbf{3}\otimes \mathbf{3'}=\mathbf{4}\oplus \mathbf{5},\quad\hat{\mathbf{4}}'\otimes \hat{\mathbf{4}}'=\mathbf{1_a}\oplus \mathbf{3_s}\oplus \mathbf{3'_s}\oplus \mathbf{4_s}\oplus \mathbf{5_a},\\ 
& \hat{\mathbf{6}}\otimes \hat{\mathbf{6}}=\mathbf{1_a}\oplus \mathbf{3_{1,s}}\oplus \mathbf{3_{2,s}}\oplus \mathbf{3'_{1,s}}\oplus \mathbf{3'_{2,s}}\oplus \mathbf{4_s}\oplus \mathbf{4_a}\oplus \mathbf{5_{1,s}}\oplus \mathbf{5_{2,a}}\oplus \mathbf{5_{3,a}}. 
\end{aligned}
\end{equation}
The subscripts $\mathbf{s}$ and $\mathbf{a}$ denote symmetric and antisymmetric combinations, respectively. For a complete list of multiplication rules and the associated Clebsch-Gordan coefficients, we refer to Refs.~\cite{Ding:2023htn, Yao:2020zml}.

Under the constraints of modular symmetry, Yukawa couplings are promoted to modular forms, which are holomorphic functions of the modulus $\tau$ that transform covariantly under the modular group. Traditionally, modular flavor symmetry has been formulated within SUSY frameworks. In the non-SUSY context adopted here, the modular forms are generalized to polyharmonic Maa\ss~forms, which satisfy the Laplace condition~\cite{Qu:2024rns}
\begin{equation}
	\Delta_k Y(\tau) = 0,
\end{equation}
where $\Delta_k$ is the weight-$k$ Laplace operator defined as  
\begin{equation}
	\Delta_k = -y^2\left(\frac{\partial^2}{\partial x^2} + \frac{\partial^2}{\partial y^2}\right) + iky\left(\frac{\partial}{\partial x} + i\frac{\partial}{\partial y}\right) = -4y^2\frac{\partial}{\partial \tau}\frac{\partial}{\partial \bar{\tau}} + 2iky\frac{\partial}{\partial \bar{\tau}},
\end{equation}
with $\tau = x + iy$. Polyharmonic Maa\ss~forms $Y^{(k)}(\tau)$ of level $N$ and weight $k$ satisfy the modular transformation property
\begin{equation}
Y(\gamma \tau) = (c\tau + d)^k Y(\tau), \quad \forall \gamma \in \Gamma(N).
\end{equation}

The space of such forms, denoted $M_k(\Gamma(N))$, is finite dimensional and furnishes representations of the modular group. Modular multiplets $Y_{\mathbf{r}}(\tau)$ transforming under irreducible representations $\rho_{\mathbf{r}}$ of the finite modular group satisfy~\cite{Ding:2023htn}
\begin{equation}
Y_{\mathbf{r}}(\gamma \tau) = (c\tau + d)^k \rho_{\mathbf{r}}(\gamma) Y_{\mathbf{r}}(\tau).
\end{equation}
To ensure modular invariance of the full Lagrangian, matter fields are assigned modular weights $-k$ that cancel the weights carried by the Yukawa couplings. In this work, we employ the $A^\prime_5$ modular group with integer modular weights in the range $-4 \leq k \leq 3$. The explicit expressions for the modular forms used in our analysis, together with their derivations, can be found in Ref.~\cite{Li:2025kcr}.

%
%

\bibliographystyle{RAN}
\bibliography{biblio}

\begin{thebibliography}{10}
\providecommand{\url}[1]{\texttt{#1}}
\providecommand{\urlprefix}{URL }
\providecommand{\eprint}[2][]{\url{#2}}

\bibitem{ATLAS:2012yve}
G.~Aad et~al. (ATLAS), \emph{{Observation of a new particle in the search for
  the Standard Model Higgs boson with the ATLAS detector at the LHC}},
  \MYhref[journalLinks]{http://dx.doi.org/10.1016/j.physletb.2012.08.020}{Phys.
  Lett. B
  }\MYhref[journalLinks]{http://dx.doi.org/10.1016/j.physletb.2012.08.020}{\textbf{716}
  (2012) 1--29},
  \MYhref[eprintLinks]{http://arxiv.org/abs/1207.7214}{{\ttfamily
  arXiv:1207.7214 [hep-ex]}}.

\bibitem{CMS:2012qbp}
S.~Chatrchyan et~al. (CMS), \emph{{Observation of a New Boson at a Mass of 125
  GeV with the CMS Experiment at the LHC}},
  \MYhref[journalLinks]{http://dx.doi.org/10.1016/j.physletb.2012.08.021}{Phys.
  Lett. B
  }\MYhref[journalLinks]{http://dx.doi.org/10.1016/j.physletb.2012.08.021}{\textbf{716}
  (2012) 30--61},
  \MYhref[eprintLinks]{http://arxiv.org/abs/1207.7235}{{\ttfamily
  arXiv:1207.7235 [hep-ex]}}.

\bibitem{Kajita:2016cak}
T.~Kajita, \emph{{Nobel Lecture: Discovery of atmospheric neutrino
  oscillations}},
  \MYhref[journalLinks]{http://dx.doi.org/10.1103/RevModPhys.88.030501}{Rev.
  Mod. Phys.
  }\MYhref[journalLinks]{http://dx.doi.org/10.1103/RevModPhys.88.030501}{\textbf{88}
  (2016) 3 030501}.

\bibitem{McDonald:2016ixn}
A.~B. McDonald, \emph{{Nobel Lecture: The Sudbury Neutrino Observatory:
  Observation of flavor change for solar neutrinos}},
  \MYhref[journalLinks]{http://dx.doi.org/10.1103/RevModPhys.88.030502}{Rev.
  Mod. Phys.
  }\MYhref[journalLinks]{http://dx.doi.org/10.1103/RevModPhys.88.030502}{\textbf{88}
  (2016) 3 030502}.

\bibitem{Mohapatra:1986aw}
R.~N. Mohapatra, \emph{{Mechanism for Understanding Small Neutrino Mass in
  Superstring Theories}},
  \MYhref[journalLinks]{http://dx.doi.org/10.1103/PhysRevLett.56.561}{Phys.
  Rev. Lett.
  }\MYhref[journalLinks]{http://dx.doi.org/10.1103/PhysRevLett.56.561}{\textbf{56}
  (1986) 561--563}.

\bibitem{Mohapatra:1986bd}
R.~N. Mohapatra and J.~W.~F. Valle, \emph{{Neutrino Mass and Baryon Number
  Nonconservation in Superstring Models}},
  \MYhref[journalLinks]{http://dx.doi.org/10.1103/PhysRevD.34.1642}{Phys. Rev.
  D
  }\MYhref[journalLinks]{http://dx.doi.org/10.1103/PhysRevD.34.1642}{\textbf{34}
  (1986) 1642}.

\bibitem{Deppisch:2004fa}
F.~Deppisch and J.~W.~F. Valle, \emph{{Enhanced lepton flavor violation in the
  supersymmetric inverse seesaw model}},
  \MYhref[journalLinks]{http://dx.doi.org/10.1103/PhysRevD.72.036001}{Phys.
  Rev. D
  }\MYhref[journalLinks]{http://dx.doi.org/10.1103/PhysRevD.72.036001}{\textbf{72}
  (2005) 036001},
  \MYhref[eprintLinks]{http://arxiv.org/abs/hep-ph/0406040}{{\ttfamily
  arXiv:hep-ph/0406040}}.

\bibitem{Dev:2009aw}
P.~S.~B. Dev and R.~N. Mohapatra, \emph{{TeV Scale Inverse Seesaw in SO(10) and
  Leptonic Non-Unitarity Effects}},
  \MYhref[journalLinks]{http://dx.doi.org/10.1103/PhysRevD.81.013001}{Phys.
  Rev. D
  }\MYhref[journalLinks]{http://dx.doi.org/10.1103/PhysRevD.81.013001}{\textbf{81}
  (2010) 013001},
  \MYhref[eprintLinks]{http://arxiv.org/abs/0910.3924}{{\ttfamily
  arXiv:0910.3924 [hep-ph]}}.

\bibitem{CentellesChulia:2020dfh}
S.~Centelles~Chuli{\'a}, R.~Srivastava and A.~Vicente, \emph{{The inverse
  seesaw family: Dirac and Majorana}},
  \MYhref[journalLinks]{http://dx.doi.org/10.1007/JHEP03(2021)248}{JHEP
  }\MYhref[journalLinks]{http://dx.doi.org/10.1007/JHEP03(2021)248}{\textbf{03}
  (2021) 248}, \MYhref[eprintLinks]{http://arxiv.org/abs/2011.06609}{{\ttfamily
  arXiv:2011.06609 [hep-ph]}}.

\bibitem{Planck:2018vyg}
N.~Aghanim et~al. (Planck), \emph{{Planck 2018 results. VI. Cosmological
  parameters}},
  \MYhref[journalLinks]{http://dx.doi.org/10.1051/0004-6361/201833910}{Astron.
  Astrophys.
  }\MYhref[journalLinks]{http://dx.doi.org/10.1051/0004-6361/201833910}{\textbf{641}
  (2020) A6}, [Erratum: Astron.Astrophys. 652, C4 (2021)],
  \MYhref[eprintLinks]{http://arxiv.org/abs/1807.06209}{{\ttfamily
  arXiv:1807.06209 [astro-ph.CO]}}.

\bibitem{ParticleDataGroup:2024cfk}
S.~Navas et~al. (Particle Data Group), \emph{{Review of particle physics}},
  \MYhref[journalLinks]{http://dx.doi.org/10.1103/PhysRevD.110.030001}{Phys.
  Rev. D
  }\MYhref[journalLinks]{http://dx.doi.org/10.1103/PhysRevD.110.030001}{\textbf{110}
  (2024) 3 030001}.

\bibitem{Sakharov:1967dj}
A.~D. Sakharov, \emph{{Violation of CP Invariance, C asymmetry, and baryon
  asymmetry of the universe}},
  \MYhref[journalLinks]{http://dx.doi.org/10.1070/PU1991v034n05ABEH002497}{Pisma
  Zh. Eksp. Teor. Fiz.
  }\MYhref[journalLinks]{http://dx.doi.org/10.1070/PU1991v034n05ABEH002497}{\textbf{5}
  (1967) 32--35}.

\bibitem{Fukugita:1986hr}
M.~Fukugita and T.~Yanagida, \emph{{Baryogenesis Without Grand Unification}},
  \MYhref[journalLinks]{http://dx.doi.org/10.1016/0370-2693(86)91126-3}{Phys.
  Lett. B
  }\MYhref[journalLinks]{http://dx.doi.org/10.1016/0370-2693(86)91126-3}{\textbf{174}
  (1986) 45--47}.

\bibitem{Pilaftsis:1997jf}
A.~Pilaftsis, \emph{{CP violation and baryogenesis due to heavy Majorana
  neutrinos}},
  \MYhref[journalLinks]{http://dx.doi.org/10.1103/PhysRevD.56.5431}{Phys. Rev.
  D
  }\MYhref[journalLinks]{http://dx.doi.org/10.1103/PhysRevD.56.5431}{\textbf{56}
  (1997) 5431--5451},
  \MYhref[eprintLinks]{http://arxiv.org/abs/hep-ph/9707235}{{\ttfamily
  arXiv:hep-ph/9707235}}.

\bibitem{Pilaftsis:2003gt}
A.~Pilaftsis and T.~E.~J. Underwood, \emph{{Resonant leptogenesis}},
  \MYhref[journalLinks]{http://dx.doi.org/10.1016/j.nuclphysb.2004.05.029}{Nucl.
  Phys. B
  }\MYhref[journalLinks]{http://dx.doi.org/10.1016/j.nuclphysb.2004.05.029}{\textbf{692}
  (2004) 303--345},
  \MYhref[eprintLinks]{http://arxiv.org/abs/hep-ph/0309342}{{\ttfamily
  arXiv:hep-ph/0309342}}.

\bibitem{Feruglio:2017spp}
F.~Feruglio, \emph{{Are neutrino masses modular forms?}}, pages 227--266 (2019)
  \MYhref[eprintLinks]{http://arxiv.org/abs/1706.08749}{{\ttfamily
  arXiv:1706.08749 [hep-ph]}}.

\bibitem{Feruglio:2019ybq}
F.~Feruglio and A.~Romanino, \emph{{Lepton flavor symmetries}},
  \MYhref[journalLinks]{http://dx.doi.org/10.1103/RevModPhys.93.015007}{Rev.
  Mod. Phys.
  }\MYhref[journalLinks]{http://dx.doi.org/10.1103/RevModPhys.93.015007}{\textbf{93}
  (2021) 1 015007},
  \MYhref[eprintLinks]{http://arxiv.org/abs/1912.06028}{{\ttfamily
  arXiv:1912.06028 [hep-ph]}}.

\bibitem{Ding:2023htn}
G.-J. Ding and S.~F. King, \emph{{Neutrino mass and mixing with modular
  symmetry}},
  \MYhref[journalLinks]{http://dx.doi.org/10.1088/1361-6633/ad52a3}{Rept. Prog.
  Phys.
  }\MYhref[journalLinks]{http://dx.doi.org/10.1088/1361-6633/ad52a3}{\textbf{87}
  (2024) 8 084201},
  \MYhref[eprintLinks]{http://arxiv.org/abs/2311.09282}{{\ttfamily
  arXiv:2311.09282 [hep-ph]}}.

\bibitem{Qu:2024rns}
B.-Y. Qu and G.-J. Ding, \emph{{Non-holomorphic modular flavor symmetry}},
  \MYhref[journalLinks]{http://dx.doi.org/10.1007/JHEP08(2024)136}{JHEP
  }\MYhref[journalLinks]{http://dx.doi.org/10.1007/JHEP08(2024)136}{\textbf{08}
  (2024) 136}, \MYhref[eprintLinks]{http://arxiv.org/abs/2406.02527}{{\ttfamily
  arXiv:2406.02527 [hep-ph]}}.

\bibitem{Qu:2025ddz}
B.-Y. Qu, J.-N. Lu and G.-J. Ding, \emph{{Non-holomorphic modular flavor
  symmetry and odd weight polyharmonic Maa{\ss} form}},
  \MYhref[journalLinks]{http://dx.doi.org/10.1007/JHEP11(2025)140}{JHEP
  }\MYhref[journalLinks]{http://dx.doi.org/10.1007/JHEP11(2025)140}{\textbf{11}
  (2025) 140}, \MYhref[eprintLinks]{http://arxiv.org/abs/2506.19822}{{\ttfamily
  arXiv:2506.19822 [hep-ph]}}.

\bibitem{Kumar:2024uxn}
B.~Kumar and M.~K. Das, \emph{{Study of neutrino phenomenology and
  0$\nu\beta\beta$ decay using polyharmonic Maass forms}},
  \MYhref[journalLinks]{http://dx.doi.org/10.1142/S0217751X25500903}{Int. J.
  Mod. Phys. A
  }\MYhref[journalLinks]{http://dx.doi.org/10.1142/S0217751X25500903}{\textbf{40}
  (2025) 23 2550090},
  \MYhref[eprintLinks]{http://arxiv.org/abs/2405.10586}{{\ttfamily
  arXiv:2405.10586 [hep-ph]}}.

\bibitem{Nomura:2024atp}
T.~Nomura and H.~Okada, \emph{{Type-II seesaw of a non-holomorphic modular
  $A_4$ symmetry}},
  \MYhref[journalLinks]{http://dx.doi.org/10.1016/j.physletb.2025.139763}{Phys.
  Lett. B
  }\MYhref[journalLinks]{http://dx.doi.org/10.1016/j.physletb.2025.139763}{\textbf{868}
  (2025) 139763},
  \MYhref[eprintLinks]{http://arxiv.org/abs/2408.01143}{{\ttfamily
  arXiv:2408.01143 [hep-ph]}}.

\bibitem{Nomura:2024nwh}
T.~Nomura and H.~Okada, \emph{{Zee model in a non-holomorphic modular $A_4$
  symmetry}},
  \MYhref[journalLinks]{http://dx.doi.org/10.1016/j.physletb.2025.139618}{Phys.
  Lett. B
  }\MYhref[journalLinks]{http://dx.doi.org/10.1016/j.physletb.2025.139618}{\textbf{867}
  (2025) 139618},
  \MYhref[eprintLinks]{http://arxiv.org/abs/2412.18095}{{\ttfamily
  arXiv:2412.18095 [hep-ph]}}.

\bibitem{Kobayashi:2025hnc}
T.~Kobayashi, H.~Okada and Y.~Orikasa, \emph{{Zee-Babu model in a
  non-holomorphic modular $A_4$ symmetry and modular stabilization}}  (2025),
  \MYhref[eprintLinks]{http://arxiv.org/abs/2502.12662}{{\ttfamily
  arXiv:2502.12662 [hep-ph]}}.

\bibitem{Loualidi:2025tgw}
M.~A. Loualidi, M.~Miskaoui and S.~Nasri, \emph{{Nonholomorphic $A_4$ modular
  invariance for fermion masses and mixing in SU(5) GUT}},
  \MYhref[journalLinks]{http://dx.doi.org/10.1103/PhysRevD.112.015008}{Phys.
  Rev. D
  }\MYhref[journalLinks]{http://dx.doi.org/10.1103/PhysRevD.112.015008}{\textbf{112}
  (2025) 1 015008},
  \MYhref[eprintLinks]{http://arxiv.org/abs/2503.12594}{{\ttfamily
  arXiv:2503.12594 [hep-ph]}}.

\bibitem{Kumar:2025bfe}
B.~Kumar and M.~K. Das, \emph{{Leptogenesis, 0$\nu\beta\beta$ and lepton flavor
  violation in modular left-right asymmetric model with polyharmonic Maass
  forms}},
  \MYhref[journalLinks]{http://dx.doi.org/10.1007/JHEP09(2025)071}{JHEP
  }\MYhref[journalLinks]{http://dx.doi.org/10.1007/JHEP09(2025)071}{\textbf{09}
  (2025) 071}, \MYhref[eprintLinks]{http://arxiv.org/abs/2504.21701}{{\ttfamily
  arXiv:2504.21701 [hep-ph]}}.

\bibitem{Nomura:2025ovm}
T.~Nomura, H.~Okada and X.-Y. Wang, \emph{{A radiative neutrino mass model with
  leptoquarks under non-holomorphic modular A$_{4}$ symmetry}},
  \MYhref[journalLinks]{http://dx.doi.org/10.1007/JHEP09(2025)163}{JHEP
  }\MYhref[journalLinks]{http://dx.doi.org/10.1007/JHEP09(2025)163}{\textbf{09}
  (2025) 163}, \MYhref[eprintLinks]{http://arxiv.org/abs/2504.21404}{{\ttfamily
  arXiv:2504.21404 [hep-ph]}}.

\bibitem{Nomura:2025raf}
T.~Nomura and H.~Okada, \emph{{Neutrino mass model at a three-loop level from a
  non-holomorphic modular $A_4$ symmetry}}  (2025),
  \MYhref[eprintLinks]{http://arxiv.org/abs/2506.02639}{{\ttfamily
  arXiv:2506.02639 [hep-ph]}}.

\bibitem{Zhang:2025dsa}
X.~Zhang and Y.~Reyimuaji, \emph{{Inverse seesaw model in nonholomorphic
  modular $A_4$ flavor symmetry}},
  \MYhref[journalLinks]{http://dx.doi.org/10.1103/PhysRevD.112.075050}{Phys.
  Rev. D
  }\MYhref[journalLinks]{http://dx.doi.org/10.1103/PhysRevD.112.075050}{\textbf{112}
  (2025) 7 075050},
  \MYhref[eprintLinks]{http://arxiv.org/abs/2507.06945}{{\ttfamily
  arXiv:2507.06945 [hep-ph]}}.

\bibitem{Priya:2025wdm}
Priya, L.~Singh, B.~C. Chauhan and S.~Verma, \emph{{Type-III Seesaw in
  Non-Holomorphic Modular Symmetry and Leptogenesis}}  (2025),
  \MYhref[eprintLinks]{http://arxiv.org/abs/2508.05047}{{\ttfamily
  arXiv:2508.05047 [hep-ph]}}.

\bibitem{Kumar:2025nut}
B.~Kumar and M.~K. Das, \emph{{Neutrino phenomenology and Dark matter in a
  left-right asymmetric model with non-holomorphic modular $A_{4}$ group}}
  (2025), \MYhref[eprintLinks]{http://arxiv.org/abs/2509.01205}{{\ttfamily
  arXiv:2509.01205 [hep-ph]}}.

\bibitem{Nanda:2025lem}
S.~K. Nanda, M.~R. Devi and S.~Patra, \emph{{Non-Holomorphic $A_4$ Modular
  Symmetry in Type-I Seesaw: Implications for Neutrino Masses and
  Leptogenesis}}  (2025),
  \MYhref[eprintLinks]{http://arxiv.org/abs/2509.22108}{{\ttfamily
  arXiv:2509.22108 [hep-ph]}}.

\bibitem{Jangid:2025thp}
S.~Jangid and H.~Okada, \emph{{A radiative seesaw model in a non-invertible
  selection rule with the assistance of a non-holomorphic modular $A_4$
  symmetry}}  (2025),
  \MYhref[eprintLinks]{http://arxiv.org/abs/2510.17292}{{\ttfamily
  arXiv:2510.17292 [hep-ph]}}.

\bibitem{Gao:2025jlw}
X.-Y. Gao and C.-C. Li, \emph{{Minimal lepton models with non-holomorphic
  modular $A_{4}$ symmetry}}  (2025),
  \MYhref[eprintLinks]{http://arxiv.org/abs/2512.07158}{{\ttfamily
  arXiv:2512.07158 [hep-ph]}}.

\bibitem{Okada:2025jjo}
H.~Okada and Y.~Orikasa, \emph{{A radiative seesaw in a non-holomorphic modular
  $S_3$ flavor symmetry}}  (2025),
  \MYhref[eprintLinks]{http://arxiv.org/abs/2501.15748}{{\ttfamily
  arXiv:2501.15748 [hep-ph]}}.

\bibitem{Ding:2024inn}
G.-J. Ding, J.-N. Lu, S.~T. Petcov and B.-Y. Qu, \emph{{Non-holomorphic modular
  S$_{4}$ lepton flavour models}},
  \MYhref[journalLinks]{http://dx.doi.org/10.1007/JHEP01(2025)191}{JHEP
  }\MYhref[journalLinks]{http://dx.doi.org/10.1007/JHEP01(2025)191}{\textbf{01}
  (2025) 191}, \MYhref[eprintLinks]{http://arxiv.org/abs/2408.15988}{{\ttfamily
  arXiv:2408.15988 [hep-ph]}}.

\bibitem{Li:2024svh}
C.-C. Li, J.-N. Lu and G.-J. Ding, \emph{{Non-holomorphic modular A$_{5}$
  symmetry for lepton masses and mixing}},
  \MYhref[journalLinks]{http://dx.doi.org/10.1007/JHEP12(2024)189}{JHEP
  }\MYhref[journalLinks]{http://dx.doi.org/10.1007/JHEP12(2024)189}{\textbf{12}
  (2024) 189}, \MYhref[eprintLinks]{http://arxiv.org/abs/2410.24103}{{\ttfamily
  arXiv:2410.24103 [hep-ph]}}.

\bibitem{Li:2025kcr}
C.-C. Li and G.-J. Ding, \emph{{Lepton models from non-holomorphic
  $A^{\prime}_{5}$ modular flavor symmetry}}  (2025),
  \MYhref[eprintLinks]{http://arxiv.org/abs/2509.15183}{{\ttfamily
  arXiv:2509.15183 [hep-ph]}}.

\bibitem{Nasri:2026nbf}
S.~Nasri, L.~Singh, Tapender and S.~Verma, \emph{{Dark-Portal Leptogenesis in a
  Non-Holomorphic Modular Scoto-Seesaw Model}}  (2026),
  \MYhref[eprintLinks]{http://arxiv.org/abs/2601.06435}{{\ttfamily
  arXiv:2601.06435 [hep-ph]}}.

\bibitem{Herrero-Brocal:2023czw}
A.~Herrero-Brocal and A.~Vicente, \emph{{The majoron coupling to charged
  leptons}},
  \MYhref[journalLinks]{http://dx.doi.org/10.1007/JHEP01(2024)078}{JHEP
  }\MYhref[journalLinks]{http://dx.doi.org/10.1007/JHEP01(2024)078}{\textbf{01}
  (2024) 078}, \MYhref[eprintLinks]{http://arxiv.org/abs/2311.10145}{{\ttfamily
  arXiv:2311.10145 [hep-ph]}}.

\bibitem{Vicente:2026zen}
A.~Vicente, \emph{{Neutrino mass models:~A short review with emphasis on the
  majoron}}, in \emph{{NuPhys2026}: {Prospect in Neutrino Physics}} (2026)
  \MYhref[eprintLinks]{http://arxiv.org/abs/2604.15995}{{\ttfamily
  arXiv:2604.15995 [hep-ph]}}.

\bibitem{TWIST:2014ymv}
R.~Bayes et~al. (TWIST), \emph{{Search for two body muon decay signals}},
  \MYhref[journalLinks]{http://dx.doi.org/10.1103/PhysRevD.91.052020}{Phys.
  Rev. D
  }\MYhref[journalLinks]{http://dx.doi.org/10.1103/PhysRevD.91.052020}{\textbf{91}
  (2015) 5 052020},
  \MYhref[eprintLinks]{http://arxiv.org/abs/1409.0638}{{\ttfamily
  arXiv:1409.0638 [hep-ex]}}.

\bibitem{COMET:2018auw}
R.~Abramishvili et~al. (COMET), \emph{{COMET Phase-I Technical Design Report}},
  \MYhref[journalLinks]{http://dx.doi.org/10.1093/ptep/ptz125}{PTEP
  }\MYhref[journalLinks]{http://dx.doi.org/10.1093/ptep/ptz125}{\textbf{2020}
  (2020) 3 033C01},
  \MYhref[eprintLinks]{http://arxiv.org/abs/1812.09018}{{\ttfamily
  arXiv:1812.09018 [physics.ins-det]}}.

\bibitem{Xing:2022rob}
T.~Xing et~al., \emph{{Search for Majoron at the COMET experiment*}},
  \MYhref[journalLinks]{http://dx.doi.org/10.1088/1674-1137/ac9897}{Chin. Phys.
  C
  }\MYhref[journalLinks]{http://dx.doi.org/10.1088/1674-1137/ac9897}{\textbf{47}
  (2023) 1 013108},
  \MYhref[eprintLinks]{http://arxiv.org/abs/2209.12802}{{\ttfamily
  arXiv:2209.12802 [hep-ex]}}.

\bibitem{Avila:2025qsc}
I.~M. {\'A}vila et~al., \emph{{Dark matter as the source of neutrino mass:
  Theory overview and experimental prospects}},
  \MYhref[journalLinks]{http://dx.doi.org/10.1016/j.physrep.2026.02.003}{Phys.
  Rept.
  }\MYhref[journalLinks]{http://dx.doi.org/10.1016/j.physrep.2026.02.003}{\textbf{1173}
  (2026) 1--81},
  \MYhref[eprintLinks]{http://arxiv.org/abs/2506.24027}{{\ttfamily
  arXiv:2506.24027 [hep-ph]}}.

\bibitem{Wang:2024qhe}
Z.~Wang, Y.~Reyimuaji and N.~Yalikun, \emph{{$Z_4$ symmetric inverse seesaw
  model for neutrino masses and FIMP dark matter}},
  \MYhref[journalLinks]{http://dx.doi.org/10.1103/3tvj-qmld}{Phys. Rev. D
  }\MYhref[journalLinks]{http://dx.doi.org/10.1103/3tvj-qmld}{\textbf{112}
  (2025) 5 055041},
  \MYhref[eprintLinks]{http://arxiv.org/abs/2412.15672}{{\ttfamily
  arXiv:2412.15672 [hep-ph]}}.

\bibitem{Novichkov:2019sqv}
P.~P. Novichkov, J.~T. Penedo, S.~T. Petcov and A.~V. Titov, \emph{{Generalised
  CP Symmetry in Modular-Invariant Models of Flavour}},
  \MYhref[journalLinks]{http://dx.doi.org/10.1007/JHEP07(2019)165}{JHEP
  }\MYhref[journalLinks]{http://dx.doi.org/10.1007/JHEP07(2019)165}{\textbf{07}
  (2019) 165}, \MYhref[eprintLinks]{http://arxiv.org/abs/1905.11970}{{\ttfamily
  arXiv:1905.11970 [hep-ph]}}.

\bibitem{Novichkov:2018ovf}
P.~P. Novichkov, J.~T. Penedo, S.~T. Petcov and A.~V. Titov, \emph{{Modular
  S$_{4}$ models of lepton masses and mixing}},
  \MYhref[journalLinks]{http://dx.doi.org/10.1007/JHEP04(2019)005}{JHEP
  }\MYhref[journalLinks]{http://dx.doi.org/10.1007/JHEP04(2019)005}{\textbf{04}
  (2019) 005}, \MYhref[eprintLinks]{http://arxiv.org/abs/1811.04933}{{\ttfamily
  arXiv:1811.04933 [hep-ph]}}.

\bibitem{Esteban:2020cvm}
I.~Esteban et~al., \emph{{The fate of hints: updated global analysis of
  three-flavor neutrino oscillations}},
  \MYhref[journalLinks]{http://dx.doi.org/10.1007/JHEP09(2020)178}{JHEP
  }\MYhref[journalLinks]{http://dx.doi.org/10.1007/JHEP09(2020)178}{\textbf{09}
  (2020) 178}, \MYhref[eprintLinks]{http://arxiv.org/abs/2007.14792}{{\ttfamily
  arXiv:2007.14792 [hep-ph]}}.

\bibitem{Esteban:2024eli}
I.~Esteban et~al., \emph{{NuFit-6.0: updated global analysis of three-flavor
  neutrino oscillations}},
  \MYhref[journalLinks]{http://dx.doi.org/10.1007/JHEP12(2024)216}{JHEP
  }\MYhref[journalLinks]{http://dx.doi.org/10.1007/JHEP12(2024)216}{\textbf{12}
  (2024) 216}, \MYhref[eprintLinks]{http://arxiv.org/abs/2410.05380}{{\ttfamily
  arXiv:2410.05380 [hep-ph]}}.

\bibitem{Xing:2007fb}
Z.-z. Xing, H.~Zhang and S.~Zhou, \emph{{Updated Values of Running Quark and
  Lepton Masses}},
  \MYhref[journalLinks]{http://dx.doi.org/10.1103/PhysRevD.77.113016}{Phys.
  Rev. D
  }\MYhref[journalLinks]{http://dx.doi.org/10.1103/PhysRevD.77.113016}{\textbf{77}
  (2008) 113016},
  \MYhref[eprintLinks]{http://arxiv.org/abs/0712.1419}{{\ttfamily
  arXiv:0712.1419 [hep-ph]}}.

\bibitem{FlavorPy}
A.~Baur, \emph{{FlavorPy}} (2024),
  \urlprefix\url{https://doi.org/10.5281/zenodo.11060597}.

\bibitem{Katrin:2024tvg}
M.~Aker et~al. (Katrin), \emph{{Direct neutrino-mass measurement based on 259
  days of KATRIN data}}  (2024),
  \MYhref[eprintLinks]{http://arxiv.org/abs/2406.13516}{{\ttfamily
  arXiv:2406.13516 [nucl-ex]}}.

\bibitem{KamLAND-Zen:2024eml}
S.~Abe et~al. (KamLAND-Zen), \emph{{Search for Majorana Neutrinos with the
  Complete KamLAND-Zen Dataset}}  (2024),
  \MYhref[eprintLinks]{http://arxiv.org/abs/2406.11438}{{\ttfamily
  arXiv:2406.11438 [hep-ex]}}.

\bibitem{Vagnozzi:2017ovm}
S.~Vagnozzi et~al., \emph{{Unveiling $\nu$ secrets with cosmological data:
  neutrino masses and mass hierarchy}},
  \MYhref[journalLinks]{http://dx.doi.org/10.1103/PhysRevD.96.123503}{Phys.
  Rev. D
  }\MYhref[journalLinks]{http://dx.doi.org/10.1103/PhysRevD.96.123503}{\textbf{96}
  (2017) 12 123503},
  \MYhref[eprintLinks]{http://arxiv.org/abs/1701.08172}{{\ttfamily
  arXiv:1701.08172 [astro-ph.CO]}}.

\bibitem{DESI:2024mwx}
A.~G. Adame et~al. (DESI), \emph{{DESI 2024 VI: cosmological constraints from
  the measurements of baryon acoustic oscillations}},
  \MYhref[journalLinks]{http://dx.doi.org/10.1088/1475-7516/2025/02/021}{JCAP
  }\MYhref[journalLinks]{http://dx.doi.org/10.1088/1475-7516/2025/02/021}{\textbf{02}
  (2025) 021}, \MYhref[eprintLinks]{http://arxiv.org/abs/2404.03002}{{\ttfamily
  arXiv:2404.03002 [astro-ph.CO]}}.

\bibitem{Chebat:2025kes}
D.~Chebat et~al., \emph{{Cosmological neutrino mass: a frequentist overview in
  light of DESI}},
  \MYhref[journalLinks]{http://dx.doi.org/10.1088/1475-7516/2026/01/041}{JCAP
  }\MYhref[journalLinks]{http://dx.doi.org/10.1088/1475-7516/2026/01/041}{\textbf{01}
  (2026) 041}, \MYhref[eprintLinks]{http://arxiv.org/abs/2507.12401}{{\ttfamily
  arXiv:2507.12401 [astro-ph.CO]}}.

\bibitem{LEGEND:2021bnm}
N.~Abgrall et~al. (LEGEND), \emph{{The Large Enriched Germanium Experiment for
  Neutrinoless $\beta\beta$ Decay}: {LEGEND-1000 Preconceptual Design Report}}
  (2021), \MYhref[eprintLinks]{http://arxiv.org/abs/2107.11462}{{\ttfamily
  arXiv:2107.11462 [physics.ins-det]}}.

\bibitem{nEXO:2021ujk}
G.~Adhikari et~al. (nEXO), \emph{{nEXO: neutrinoless double beta decay search
  beyond 10$^{28}$ year half-life sensitivity}},
  \MYhref[journalLinks]{http://dx.doi.org/10.1088/1361-6471/ac3631}{J. Phys. G
  }\MYhref[journalLinks]{http://dx.doi.org/10.1088/1361-6471/ac3631}{\textbf{49}
  (2022) 1 015104},
  \MYhref[eprintLinks]{http://arxiv.org/abs/2106.16243}{{\ttfamily
  arXiv:2106.16243 [nucl-ex]}}.

\bibitem{GAMBITCosmologyWorkgroup:2020rmf}
P.~St\"ocker et~al. (GAMBIT Cosmology Workgroup), \emph{{Strengthening the
  bound on the mass of the lightest neutrino with terrestrial and cosmological
  experiments}},
  \MYhref[journalLinks]{http://dx.doi.org/10.1103/PhysRevD.103.123508}{Phys.
  Rev. D
  }\MYhref[journalLinks]{http://dx.doi.org/10.1103/PhysRevD.103.123508}{\textbf{103}
  (2021) 12 123508},
  \MYhref[eprintLinks]{http://arxiv.org/abs/2009.03287}{{\ttfamily
  arXiv:2009.03287 [astro-ph.CO]}}.

\bibitem{Lorenz:2018fzb}
C.~S. Lorenz, L.~Funcke, E.~Calabrese and S.~Hannestad, \emph{{Time-varying
  neutrino mass from a supercooled phase transition: current cosmological
  constraints and impact on the $\Omega_m$-$\sigma_8$ plane}},
  \MYhref[journalLinks]{http://dx.doi.org/10.1103/PhysRevD.99.023501}{Phys.
  Rev. D
  }\MYhref[journalLinks]{http://dx.doi.org/10.1103/PhysRevD.99.023501}{\textbf{99}
  (2019) 2 023501},
  \MYhref[eprintLinks]{http://arxiv.org/abs/1811.01991}{{\ttfamily
  arXiv:1811.01991 [astro-ph.CO]}}.

\bibitem{Antusch:2014woa}
S.~Antusch and O.~Fischer, \emph{{Non-unitarity of the leptonic mixing matrix:
  Present bounds and future sensitivities}},
  \MYhref[journalLinks]{http://dx.doi.org/10.1007/JHEP10(2014)094}{JHEP
  }\MYhref[journalLinks]{http://dx.doi.org/10.1007/JHEP10(2014)094}{\textbf{10}
  (2014) 094}, \MYhref[eprintLinks]{http://arxiv.org/abs/1407.6607}{{\ttfamily
  arXiv:1407.6607 [hep-ph]}}.

\bibitem{Blennow:2016jkn}
M.~Blennow et~al., \emph{{Non-Unitarity, sterile neutrinos, and Non-Standard
  neutrino Interactions}},
  \MYhref[journalLinks]{http://dx.doi.org/10.1007/JHEP04(2017)153}{JHEP
  }\MYhref[journalLinks]{http://dx.doi.org/10.1007/JHEP04(2017)153}{\textbf{04}
  (2017) 153}, \MYhref[eprintLinks]{http://arxiv.org/abs/1609.08637}{{\ttfamily
  arXiv:1609.08637 [hep-ph]}}.

\bibitem{Fernandez-Martinez:2016lgt}
E.~Fernandez-Martinez, J.~Hernandez-Garcia and J.~Lopez-Pavon, \emph{{Global
  constraints on heavy neutrino mixing}},
  \MYhref[journalLinks]{http://dx.doi.org/10.1007/JHEP08(2016)033}{JHEP
  }\MYhref[journalLinks]{http://dx.doi.org/10.1007/JHEP08(2016)033}{\textbf{08}
  (2016) 033}, \MYhref[eprintLinks]{http://arxiv.org/abs/1605.08774}{{\ttfamily
  arXiv:1605.08774 [hep-ph]}}.

\bibitem{AristizabalSierra:2009mq}
D.~Aristizabal~Sierra, M.~Losada and E.~Nardi, \emph{{Lepton Flavor
  Equilibration and Leptogenesis}},
  \MYhref[journalLinks]{http://dx.doi.org/10.1088/1475-7516/2009/12/015}{JCAP
  }\MYhref[journalLinks]{http://dx.doi.org/10.1088/1475-7516/2009/12/015}{\textbf{12}
  (2009) 015}, \MYhref[eprintLinks]{http://arxiv.org/abs/0905.0662}{{\ttfamily
  arXiv:0905.0662 [hep-ph]}}.

\bibitem{Blanchet:2010kw}
S.~Blanchet, P.~S.~B. Dev and R.~N. Mohapatra, \emph{{Leptogenesis with TeV
  Scale Inverse Seesaw in SO(10)}},
  \MYhref[journalLinks]{http://dx.doi.org/10.1103/PhysRevD.82.115025}{Phys.
  Rev. D
  }\MYhref[journalLinks]{http://dx.doi.org/10.1103/PhysRevD.82.115025}{\textbf{82}
  (2010) 115025},
  \MYhref[eprintLinks]{http://arxiv.org/abs/1010.1471}{{\ttfamily
  arXiv:1010.1471 [hep-ph]}}.

\bibitem{Covi:1996wh}
L.~Covi, E.~Roulet and F.~Vissani, \emph{{CP violating decays in leptogenesis
  scenarios}},
  \MYhref[journalLinks]{http://dx.doi.org/10.1016/0370-2693(96)00817-9}{Phys.
  Lett. B
  }\MYhref[journalLinks]{http://dx.doi.org/10.1016/0370-2693(96)00817-9}{\textbf{384}
  (1996) 169--174},
  \MYhref[eprintLinks]{http://arxiv.org/abs/hep-ph/9605319}{{\ttfamily
  arXiv:hep-ph/9605319}}.

\bibitem{Buchmuller:1997yu}
W.~Buchmuller and M.~Plumacher, \emph{{CP asymmetry in Majorana neutrino
  decays}},
  \MYhref[journalLinks]{http://dx.doi.org/10.1016/S0370-2693(97)01548-7}{Phys.
  Lett. B
  }\MYhref[journalLinks]{http://dx.doi.org/10.1016/S0370-2693(97)01548-7}{\textbf{431}
  (1998) 354--362},
  \MYhref[eprintLinks]{http://arxiv.org/abs/hep-ph/9710460}{{\ttfamily
  arXiv:hep-ph/9710460}}.

\bibitem{Blanchet:2009kk}
S.~Blanchet, T.~Hambye and F.-X. Josse-Michaux, \emph{{Reconciling leptogenesis
  with observable $\mu \to e \gamma$ rates}},
  \MYhref[journalLinks]{http://dx.doi.org/10.1007/JHEP04(2010)023}{JHEP
  }\MYhref[journalLinks]{http://dx.doi.org/10.1007/JHEP04(2010)023}{\textbf{04}
  (2010) 023}, \MYhref[eprintLinks]{http://arxiv.org/abs/0912.3153}{{\ttfamily
  arXiv:0912.3153 [hep-ph]}}.

\bibitem{Chakraborty:2021azg}
I.~Chakraborty, H.~Roy and T.~Srivastava, \emph{{Resonant leptogenesis in (2,2)
  inverse see-saw realisation}},
  \MYhref[journalLinks]{http://dx.doi.org/10.1016/j.nuclphysb.2022.115780}{Nucl.
  Phys. B
  }\MYhref[journalLinks]{http://dx.doi.org/10.1016/j.nuclphysb.2022.115780}{\textbf{979}
  (2022) 115780},
  \MYhref[eprintLinks]{http://arxiv.org/abs/2106.08232}{{\ttfamily
  arXiv:2106.08232 [hep-ph]}}.

\bibitem{Shao:2025xav}
Y.~Shao and Z.-h. Zhao, \emph{{Linear seesaw leptogenesis before and after
  electroweak symmetry breaking}},
  \MYhref[journalLinks]{http://dx.doi.org/10.1103/85vc-x819}{Phys. Rev. D
  }\MYhref[journalLinks]{http://dx.doi.org/10.1103/85vc-x819}{\textbf{112}
  (2025) 11 115033},
  \MYhref[eprintLinks]{http://arxiv.org/abs/2509.14524}{{\ttfamily
  arXiv:2509.14524 [hep-ph]}}.

\bibitem{Buchmuller:2004nz}
W.~Buchmuller, P.~Di~Bari and M.~Plumacher, \emph{{Leptogenesis for
  pedestrians}},
  \MYhref[journalLinks]{http://dx.doi.org/10.1016/j.aop.2004.02.003}{Annals
  Phys.
  }\MYhref[journalLinks]{http://dx.doi.org/10.1016/j.aop.2004.02.003}{\textbf{315}
  (2005) 305--351},
  \MYhref[eprintLinks]{http://arxiv.org/abs/hep-ph/0401240}{{\ttfamily
  arXiv:hep-ph/0401240}}.

\bibitem{Agashe:2018cuf}
K.~Agashe et~al., \emph{{Natural Seesaw and Leptogenesis from Hybrid of
  High-Scale Type I and TeV-Scale Inverse}},
  \MYhref[journalLinks]{http://dx.doi.org/10.1007/JHEP04(2019)029}{JHEP
  }\MYhref[journalLinks]{http://dx.doi.org/10.1007/JHEP04(2019)029}{\textbf{04}
  (2019) 029}, \MYhref[eprintLinks]{http://arxiv.org/abs/1812.08204}{{\ttfamily
  arXiv:1812.08204 [hep-ph]}}.

\bibitem{DUNE:2020ypp}
B.~Abi et~al. (DUNE), \emph{{Deep Underground Neutrino Experiment (DUNE), Far
  Detector Technical Design Report, Volume II: DUNE Physics}}  (2020),
  \MYhref[eprintLinks]{http://arxiv.org/abs/2002.03005}{{\ttfamily
  arXiv:2002.03005 [hep-ex]}}.

\bibitem{Hyper-Kamiokande:2018ofw}
K.~Abe et~al. (Hyper-Kamiokande), \emph{{Hyper-Kamiokande Design Report}}
  (2018), \MYhref[eprintLinks]{http://arxiv.org/abs/1805.04163}{{\ttfamily
  arXiv:1805.04163 [physics.ins-det]}}.

\bibitem{JUNO:2015zny}
F.~An et~al. (JUNO), \emph{{Neutrino Physics with JUNO}},
  \MYhref[journalLinks]{http://dx.doi.org/10.1088/0954-3899/43/3/030401}{J.
  Phys. G
  }\MYhref[journalLinks]{http://dx.doi.org/10.1088/0954-3899/43/3/030401}{\textbf{43}
  (2016) 3 030401},
  \MYhref[eprintLinks]{http://arxiv.org/abs/1507.05613}{{\ttfamily
  arXiv:1507.05613 [physics.ins-det]}}.

\bibitem{KATRIN:2022ayy}
M.~Aker et~al. (KATRIN), \emph{{KATRIN: status and prospects for the neutrino
  mass and beyond}},
  \MYhref[journalLinks]{http://dx.doi.org/10.1088/1361-6471/ac834e}{J. Phys. G
  }\MYhref[journalLinks]{http://dx.doi.org/10.1088/1361-6471/ac834e}{\textbf{49}
  (2022) 10 100501},
  \MYhref[eprintLinks]{http://arxiv.org/abs/2203.08059}{{\ttfamily
  arXiv:2203.08059 [nucl-ex]}}.

\bibitem{Project8:2022wqh}
A.~A. Esfahani et~al. (Project 8), \emph{{The Project 8 Neutrino Mass
  Experiment}}, in \emph{{Snowmass 2021}} (2022)
  \MYhref[eprintLinks]{http://arxiv.org/abs/2203.07349}{{\ttfamily
  arXiv:2203.07349 [nucl-ex]}}.

\bibitem{Feruglio:2022koo}
F.~Feruglio, \emph{{Universal Predictions of Modular Invariant Flavor Models
  near the Self-Dual Point}},
  \MYhref[journalLinks]{http://dx.doi.org/10.1103/PhysRevLett.130.101801}{Phys.
  Rev. Lett.
  }\MYhref[journalLinks]{http://dx.doi.org/10.1103/PhysRevLett.130.101801}{\textbf{130}
  (2023) 10 101801},
  \MYhref[eprintLinks]{http://arxiv.org/abs/2211.00659}{{\ttfamily
  arXiv:2211.00659 [hep-ph]}}.

\bibitem{Granelli:2025lds}
A.~Granelli et~al., \emph{{Modular-symmetry-protected seesaw}},
  \MYhref[journalLinks]{http://dx.doi.org/10.1007/JHEP12(2025)035}{JHEP
  }\MYhref[journalLinks]{http://dx.doi.org/10.1007/JHEP12(2025)035}{\textbf{12}
  (2025) 035}, \MYhref[eprintLinks]{http://arxiv.org/abs/2505.21405}{{\ttfamily
  arXiv:2505.21405 [hep-ph]}}.

\bibitem{Yao:2020zml}
C.-Y. Yao, X.-G. Liu and G.-J. Ding, \emph{{Fermion masses and mixing from the
  double cover and metaplectic cover of the $A_5$ modular group}},
  \MYhref[journalLinks]{http://dx.doi.org/10.1103/PhysRevD.103.095013}{Phys.
  Rev. D
  }\MYhref[journalLinks]{http://dx.doi.org/10.1103/PhysRevD.103.095013}{\textbf{103}
  (2021) 9 095013},
  \MYhref[eprintLinks]{http://arxiv.org/abs/2011.03501}{{\ttfamily
  arXiv:2011.03501 [hep-ph]}}.

\end{thebibliography}

\end{document}